\documentclass[12pt,nofootinbib,superscriptaddress]{article}

\pdfoutput=1

\usepackage{graphicx}
\usepackage{amsmath,amssymb} 
\usepackage{latexsym}
\usepackage{mathrsfs}
\usepackage{color}
\usepackage{slashed}
\definecolor{red}{rgb}{1,0,0}
\usepackage{cancel}
\usepackage{comment}
\usepackage{cite,mathtools}
\usepackage[normalem]{ulem}

\usepackage{caption}
\usepackage{subcaption}
\definecolor{red}{rgb}{1,0,0}
\usepackage[english]{babel}

\usepackage{boldline}
\usepackage{gensymb}

\newcommand{\be}{\begin{equation}}
\newcommand{\ee}{\end{equation}}
\newcommand{\bee}{\begin{equation*}}
\newcommand{\eee}{\end{equation*}}
\newcommand{\bea}{\begin{eqnarray}}
\newcommand{\eea}{\end{eqnarray}}
\newcommand{\bean}{\begin{eqnarray*}}
\newcommand{\eean}{\end{eqnarray*}}

\def\bea{\begin{eqnarray}} \def\eea{\end{eqnarray}}
\def\be{\begin{equation}} \def\ee{\end{equation}}

\newcommand{\promille}{%
  \relax\ifmmode\promillezeichen
        \else\leavevmode\(\mathsurround=0pt\promillezeichen\)\fi}
\newcommand{\promillezeichen}{%
  \kern-.05em%
  \raise.5ex\hbox{\the\scriptfont0 0}%
  \kern-.15em/\kern-.15em%
  \lower.25ex\hbox{\the\scriptfont0 00}}

\definecolor{commam}{rgb}{0.2,0.5,1.0}
\definecolor{myred}{rgb}{0.8, 0.3, 0.3}
\definecolor{myblue}{rgb}{0, 0, 0.6}
\definecolor{mygreen}{rgb}{0.04, 0.7, 0.5}

 \def\be   {\begin{equation}}   \def\ee   {\end{equation}}
 \def\ba   {\begin{array}}      \def\ea   {\end{array}}
 \def\bea  {\begin{eqnarray}}   \def\eea  {\end{eqnarray}}
 \def\bean {\begin{eqnarray*}}  \def\eean {\end{eqnarray*}}
 
 \def\bry{\begin{array}}
 \def\ery{\end{array}}

\newcommand{\GeV}{\,\mathrm{GeV}}

\usepackage[colorlinks,linkcolor=black,citecolor=blue,urlcolor=blue,linktocpage]{hyperref}
\makeatletter
\def\section{\@startsection {section}{1}{\z@}{-3.5ex plus -1ex minus
 -.2ex}{2.3ex plus .2ex}{\large\bf}}
\def\subsection{\@startsection{subsection}{2}{\z@}{-3.25ex plus -1ex
minus -.2ex}{1.5ex plus .2ex}{\normalsize\bf}}
\makeatother
\makeatletter

\@addtoreset{equation}{section}

\makeatother
\begin{document}

\setcounter{page}{0}
\thispagestyle{empty}
%\pagestyle{empty}

%%%%%%%%%%%%%%%%%%%%%%%%%%%%%%%%%%%%%%%%%%%%%%%%%%%%%%%%%%%%%%%%%%%%%%%%%%%%%%%
\begin{flushright}
%astro-ph/yymmnnn
%CERN-PH-TH/2012-368\\
DESY 17-229
%\today
\end{flushright}

\vskip 8pt

\begin{center}
{\bf \LARGE {Electroweak Phase Transition and}}\\
\vskip 10pt
{\bf \LARGE {Baryogenesis in Composite Higgs Models}}
\end{center}

\vskip 16pt

\begin{center}
 { \bf  Sebastian Bruggisser$^{a}$, Benedict von Harling$^a$, Oleksii Matsedonskyi$^{a}$ and G\'eraldine Servant$^{a,b}$}
 \end{center}

\vskip 14pt

\begin{center}
\centerline{$^{a}${\it DESY, Notkestrasse 85, 22607 Hamburg, Germany}}
\centerline{$^{b}${\it II.~Institute of Theoretical Physics, University of Hamburg, 22761 Hamburg, Germany}}

\vskip .1cm
\centerline{\tt sebastian.bruggisser@desy.de, benedict.von.harling@desy.de,}
\centerline{\tt oleksii.matsedonskyi@desy.de, geraldine.servant@desy.de}
\end{center}

\vskip 10pt

\begin{abstract}
\vskip 3pt
\noindent

We present a comprehensive study of the electroweak phase transition
in composite Higgs models, where the Higgs arises from a new,
strongly-coupled sector which confines near the TeV scale. 
This work extends our study in Ref.~\cite{Bruggisser:2018mus}. 
We describe the confinement phase transition in terms of the dilaton, the pseudo-Nambu-Goldstone boson of broken conformal
invariance of the composite Higgs sector.  From the analysis of the joint Higgs-dilaton 
potential we conclude that in this scenario the electroweak phase transition can naturally be first-order, allowing for electroweak baryogenesis. 
We then extensively discuss possible options to generate a sufficient amount of CP violation -- another key ingredient of baryogenesis -- from quark Yukawa couplings which vary during the phase transition.  
For one such an option, with a varying charm quark Yukawa coupling, we perform a full numerical analysis of tunnelling in the Higgs-dilaton potential and determine regions of parameter space which
allow for successful baryogenesis. This scenario singles out the light dilaton region while satisfying all experimental bounds. We discuss future tests. Our results bring new opportunities and strong motivations for electroweak baryogenesis.

\end{abstract}

\newpage

\tableofcontents

\vskip 13pt

\newpage

\section{Introduction}

Models in which the Higgs boson 
is a composite pseudo-Nambu-Goldstone boson (PNGB) and the Standard Model (SM) fermions are partially composite offer a very popular alternative to Supersymmetry for solving the hierarchy problem and are prime targets at the LHC (see \cite{Panico:2015jxa} for a review). These models  feature a new sector with a strong dynamics which confines around the TeV scale. This sector possesses an approximate global symmetry $G$, which is spontaneously broken at the condensation scale  to a subgroup $H$. The lightest Goldstone boson associated with this breaking is identified with the Higgs boson, which can then be naturally light. If one wants to ensure a custodial symmetry to suppress oblique corrections to electroweak precision tests, the minimal possible coset is $SO(5)/SO(4)$, while larger groups can be viable too.

In these composite Higgs models, the Higgs potential is generated due to an explicit breaking of the Goldstone symmetry.
This explicit breaking comes from an elementary sector consisting of fermions and gauge bosons which do not respect the global symmetry $G$. It is communicated to the composite sector from which the Higgs originates through  elementary/composite interactions present in the theory. These interactions are also responsible for the Yukawa couplings by inducing mixing between the elementary and composite fermions. The size of the Yukawa couplings is then determined by the degree of compositeness of the states that are identified with the SM fermions. The Higgs potential in composite Higgs models is thus intimately tied with the Yukawa couplings. 
This framework therefore appears to be an ideal laboratory to study the connection between electroweak symmetry breaking and flavour physics.

Despite the fame of these models, the electroweak phase transition in composite Higgs models has not been studied in much detail yet. In particular, the phase transition is relevant for the question whether composite Higgs models can allow for electroweak baryogenesis and thereby explain the baryon asymmetry of the universe. In the SM, this scenario fails because the electroweak phase transition is not first-order and the amount of CP violation is also not enough. As was shown in \cite{Grojean:2004xa,Bodeker:2004ws,Delaunay:2007wb,Grinstein:2008qi}, dimension-6 operators involving the Higgs which are expected to arise from the strong sector can make the phase transition first-order and also provide a new source of CP violation. Alternatively, this can be achieved if an additional singlet changes its \emph{vev} during the electroweak phase transition. Such a singlet can arise as an extra PNGB in non-minimal composite Higgs models with global symmetry breaking patterns such as  $SO(6) / SO(5)$ or $SO(7) / SO(6)$. The former pattern was studied with regard to electroweak baryogenesis in \cite{Espinosa:2011eu}, while the latter was considered in \cite{Chala:2016ykx}. 
In all these studies, however,
the confinement scale of the strong sector $f$ has been taken to be constant which implicitly assumes that the confinement phase transition happens well before the electroweak phase transition. 
This assumption is not always justified. In~\cite{Bruggisser:2018mus}, we have studied the joint dynamics of the Higgs and the order parameter $f$ of the strong sector and found that both transitions can naturally happen simultaneously.
This opens a rich range of possibilities for the nature of the electroweak phase transition and for electroweak baryogenesis in composite Higgs models. 
In this paper, we will extend this analysis.

In order to ensure a large separation between the UV scale (e.g.~the Planck scale) and the confinement scale, the strong sector should be near a conformal fixed point for most of its evolution when running down to lower energies. This (nearly) conformal invariance is spontaneously broken when the strong sector confines. Provided that the explicit breaking of the conformal invariance is small, the spectrum of composite states then contains an associated light PNGB, the so-called dilaton. The \emph{vev} of this field sets the confinement scale $f$ and this field can thus be thought of as an order parameter for the confinement of the strong sector. 

The analyses in~\cite{Bruggisser:2018mus} and in this paper are based on the combined potential for the Higgs and the dilaton.  
To this end, we rely on a four-dimensional effective field theory (EFT) describing the lowest lying degrees of freedom, such as the SM particles, the Higgs boson, and the dilaton. We present a universal simplified description, suitable to parametrise the IR physics resulting from different possible explicit UV complete constructions.
Our ignorance about the details of the strong sector in the confined phase is parametrised by the minimal set of coupling constants and masses, in the spirit of~\cite{Giudice:2007fh,Chala:2017sjk}. The crucial ingredients of our framework are the fundamental symmetries which the strong sector is expected to feature, including the spontaneously broken global symmetry $G$ and its surviving subgroup $H$, as well as an approximate conformal symmetry above the confinement scale. We also include temperature corrections to the potential and then study the phase transition in the two-field potential for the Higgs and dilaton.

The scalar potential for the Higgs arises at one-loop, and, because of the PNGB nature of the Higgs,
is a trigonometric function of the ratio $h/f$.
It takes the generic form \cite{Panico:2015jxa}
\be
\label{eq:TunedHiggsPotential}
V [h] \,\sim \, f^4 \left[ \alpha  \sin^2 (h/f) + \beta \sin^4 (h/f) \right],
\ee
where the two terms are loop-induced by the same dynamics and $\alpha$ is expected to be at least as large as $\beta$~\cite{Panico:2012uw}. However, 
in order to obtain the correct electroweak symmetry breaking scale $h=v$ and the correct Higgs mass, $\alpha$ and $\beta$ have to fulfil the relations
\be
\label{eq:TunedHiggsRelations}
\alpha \, = \, -2 \beta  \sin^2 (v/f) \ , \quad \, \, m_h^2 \, \approx \, 8  f^2 \sin^2 (v/f) \, \beta \, .
\ee
This means that $|\alpha/\beta| =  2 \sin^2 (v/f)$ which in turn needs to be suppressed since  electroweak precision tests and Higgs coupling measurements constrain $\sin^2 (v/f) \lesssim 0.1...0.2$ \cite{Grojean:2013qca}. Therefore, contrary to the generic estimate $\alpha \gtrsim \beta$,  $\alpha$ has to be suppressed with respect to $\beta$, which requires some accidental cancellation. This is the irreducible tuning of composite Higgs models \cite{Csaki:2017eio}.

Note that when expanding the potential (\ref{eq:TunedHiggsPotential}) for small $h/f$ to match the coefficients of the Higgs potential in the EFT approach~\cite{Grojean:2004xa}, we find that for a constant $f$, no first-order phase transition can follow from a negative quartic coupling in the tree-level potential as discussed in~\cite{Grojean:2004xa}.
In this paper, we will take into account the fact that $f, \alpha$ and $\beta$ are dynamical and show the implications for the nature of the electroweak phase transition.
We denote the dilaton as $\chi$.
As the confinement scale $f$ is set by $\chi$, we in particular replace $f \rightarrow \chi$ (times a proportionality factor to be determined later) in Eq.~\eqref{eq:TunedHiggsPotential} in order to derive the joint potential for the Higgs and dilaton. 
This part of the potential is then minimized for $h \propto \chi$ (however with $h \ll \chi$, as currently $v\ll f$) which leads to a valley in the two-field potential along this direction.  If the phase transition happens only at temperatures below the electroweak scale and temperature corrections to the potential are correspondingly small, this valley can attract the tunnelling trajectory and both $h$ and $\chi$ can obtain their \emph{vevs} simultaneously. Since the confinement phase transition can be naturally very strongly first-order, this makes the electroweak phase transition first-order too and thereby solves the first problem that electroweak baryogenesis faces in the SM. In particular, we then do not need to invoke higher-dimensional operators or extra scalars for this purpose.

While the Higgs potential appears tuned today, it can become detuned in the early universe when the dilaton \emph{vev} differs from its value today $ \sim f$. In particular, the sizes of the elementary-composite mixings mentioned above depend on the confinement scale and thus on $\chi$.  
Since they affect various loop corrections, and therefore the values of $\alpha$ and $\beta$, we generically expect  the tuned relations in Eq.~\eqref{eq:TunedHiggsRelations} to be fulfilled only near the minimum of the potential. 
The resulting detuning away from the minimum can lead to an additonal valley either along the direction $h \sim 0$ in the potential, or along $h \sim \chi$. 
Again this valley can attract the tunnelling trajectory and affect the relation between $h$ and $\chi$ during the phase transition. For a deep valley along $h \sim 0$, the Higgs \emph{vev} vanishes during most of the transition which is not suitable for electroweak baryogenesis. On the other hand, for $h \sim \chi$ the Higgs \emph{vev} is on average larger than for the tuned case discussed above.
This behaviour can, in particular, be helpful for generating enough CP asymmetry for electroweak baryogenesis as we discuss below.  

\begin{figure}[t]
\centering
\includegraphics[width=7cm]{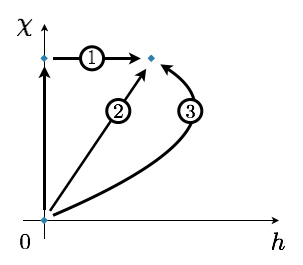}
\caption{\small \it Sketch of different trajectories for the phase transition in composite Higgs models. The field $\chi$ sets the size of the strong-sector condensate and $h$ is the Higgs vev. The blue points correspond to (meta)stable vacua of the theory, and the black lines show possible phase transition trajectories.
\label{fig:phtr}}
\end{figure}

We summarize the various ways in which the phase transition can occur in composite Higgs models in Fig.~\ref{fig:phtr}.
The trajectory (1) corresponds to a two-step transition, where the confinement phase transition happens in the first step and electroweak symmetry is only broken subsequently in a second step. 
This chain of events was implicitly assumed in the previous works \cite{Espinosa:2011eu,Chala:2016ykx,Delaunay:2007wb,Grinstein:2008qi,Grojean:2004xa,Bodeker:2004ws} (in \cite{Espinosa:2011eu,Chala:2016ykx} the additional scalar also changes its \emph{vev} in the second step). 
For the other two options presented in Fig.~\ref{fig:phtr} the electroweak phase transition is instead tied to the confinement phase transition.
The linear trajectory (2), where $h$ scales linearly with $\chi$ all the way between the two minima, can occur in models where the Higgs potential is not detuned away from the minimum of the potential. On the other hand, the composite Higgs models studied in this paper can have a trajectory like (3) for which $ h \sim \chi $ during most of the phase transition.

Apart from affecting the tunnelling trajectory,
the variation of the mixings during the phase transition can also provide additional CP violation. This can then solve the second problem for electroweak baryogenesis in the SM. Indeed, varying mixings lead to varying Yukawa couplings which have previously been shown \cite{Bruggisser:2017lhc} to result in a new source of CP violation.
This new source is not in conflict with bounds from e.g.~electric dipole moments since it is active only during the phase transition. 
For instance, the CP violation can be large if at least one light quark  mixing increases to the size of the top quark mixing when the dilaton \emph{vev} is sent to zero. 
The CP violation then arises from the interplay between the top quark mixing and one light quark mixing. 
The dependence of the mixings on the dilaton \emph{vev} is mostly determined by the scaling dimension of the composite operators to which the elementary fermions are linearly coupled.  For constant scaling dimensions one can expect the mixings to either stay approximately constant too or decrease when the dilaton \emph{vev} is sent to zero. 
The required growing behaviour instead can be obtained if the scaling dimension of the corresponding operator becomes energy-dependent. Such an energy-dependence can well occur in composite Higgs models. Alternatively, a sufficient amount of CP violation can be generated by the top quark mixings alone if they have a phase which varies with the dilaton \emph{vev}.

Our approach enables to test a broad range of possible UV complete theories. Together, the analyses in~\cite{Bruggisser:2018mus} and in this paper make the following progress:
\begin{itemize}
\item First general analysis of the confinement and electroweak phase transitions in PNGB composite Higgs models.
\item First analysis of electroweak baryogenesis in PNGB composite Higgs models during the combined confinement and electroweak phase transition, including the analysis of the new CP-violating source from varying quark mixings.
\item Full numerical calculation of the two-field tunnelling for electroweak baryogenesis in PNGB composite Higgs models. This is especially important as the Higgs \emph{vev} has a non-trivial dependence on the \emph{vev} of the strong condensate. 
\end{itemize}
Compared to\cite{Bruggisser:2018mus} which focussed on varying top mixings, we will in particular extend the analysis of the CP-violating source and systematically list the cases where varying quark mixings can generate the CP asymmetry.
For one promising case, with varying charm mixings (which is both quantitatively and qualitatively different from  varying top mixings), we will perform a scan of the parameter space, and determine the parameter region for which the confinement and electroweak phase transitions happen simultaneously and are first-order, and the amount of CP violation is sufficient for explaining the baryon asymmetry of the universe.
Furthermore, we will present a detailed analysis of phenomenological implications of our scenario for collider physics, such as Higgs coupling measurements and dilaton searches. 
Overall, this analysis follows a series of papers on the impact of Yukawa coupling variation for electroweak baryogenesis \cite{Bruggisser:2018mus,Baldes:2016rqn,Baldes:2016gaf,vonHarling:2016vhf,Bruggisser:2017lhc,Servant:2018xcs}.

The structure of this paper is the following. In Sec.~\ref{sec:review} we review the basic ingredients needed to construct our EFT, including the generation of the SM fermion masses, the Higgs mass and conformal symmetry breaking. These key ingredients are combined together in Sec.~\ref{sec:zteft} to provide a single framework for studying the phase transition. In Sec.~\ref{sec:fteft} we add finite temperature effects. Sec.~\ref{sec:cpv} is devoted to the analysis of the CP-violating source associated with the varying quark mixings. The numerical analysis of the phase transition and the resulting baryon asymmetry is described and the main results are presented in Sec.~\ref{sec:res}. Finally, we discuss experimental tests of our scenario in Sec.~\ref{ExperimentalTests}, including the current bounds on flavour-changing neutral currents, CP-violating Higgs couplings and gravitational wave signals, and conclude in Sec.~\ref{sec:conc}. An appendix reviews the formalism to calculate the tunnelling path and action in potentials for one and more fields.

\section{Review of key concepts}
\label{sec:review}

In this section, we will summarize the main concepts which we will adopt for our description of the phase transition and electroweak baryogenesis in composite Higgs models. These concepts represent a typical (though not the only possible) picture of composite Higgs models and their flavour structure, and have been motivated in a large amount of literature, of which we will only point out a few representative works. For general reviews of composite Higgs models~\cite{Kaplan:1983fs}, we refer the reader to~\cite{Contino:2010rs,Panico:2015jxa,Bellazzini:2014yua}.

\subsection{Standard Model masses from anomalous dimensions}
\label{sec:PartialCompositeness}

We now review the basics of fermion mass generation in composite Higgs models and state the notations. Let us begin with the flavour-diagonal case, \emph{i.e.}~without mixing between different SM generations. The SM fermion masses are generated from couplings
\be\label{eq:uvpc}
y_i  \bar q_i {\cal O}_i
\ee 
of elementary fermions $q_i$ to operators from the strong sector ${\cal O}_i$. 
The dimensionless coefficients $y_i$ are assumed to be of order one in the far UV, where the couplings are generated. The renormalisation group (RG) evolution then changes them when running down to the confinement scale. This is driven mostly by anomalous dimensions of the operators ${\cal O}_i$, which remain sizeable over a wide energy range due to an approximate conformal symmetry. The RG equation reads (see \emph{e.g.}~\cite{Contino:2010rs})
\be\label{eq:yrun}
\frac{\partial y_i}{\partial \log \mu} \, = \, \gamma_i y_i \, + \,  c_i \frac{y_i^3}{g_\star^2} \, + \, \dots \,,
\ee
where $\gamma_i$ is the scaling dimension of ${\cal O}_i$ minus 5/2 (which is constrained as $\gamma_i \geq -1$ by virtue of the unitarity bound on fermionic CFT operators), $g_\star\sim 1...4\pi$ is the typical coupling of the strong sector (we will give more meaning to it later),
and the ellipsis stands for terms suppressed by higher powers of $y_i^2/(4 \pi)^2$ and $g_\star^2/(4 \pi)^2$.
This running stops at the confinement scale, 
\be
\sim g_\star f,
\ee
where the CFT disappears and the operators ${\cal O}_i$ can excite bound states of the strong dynamics, the fermionic partners. At energies below $\sim g_\star f$, Eq.~(\ref{eq:uvpc}) is mapped onto mixings between these composite fermions, which we denote as $\psi_i$, and the elementary fermions $q_i$,
\be\label{eq:irpc}
y_i \, f \bar q_i U \psi_i \,,
\ee 
where $U$ is the Goldstone matrix and the couplings $y_i$ are now defined at the confinement scale. As the strong sector spontaneously breaks the global symmetry,
\be
G\to H ,
\ee
the Goldstones are introduced as a compensator between the elementary fermions transforming in $G$ and composite multiplets of $H$ (the elementary fermions generically do not fill complete multiplets of $G$ but can be given some transformation properties under $G$ in order to write down the original operator~(\ref{eq:uvpc})). The Higgs multiplet comes as a part of these Goldstones:
\be\label{eq:u}
U \sim \exp[i h/f] \,.
\ee
Since the interaction~(\ref{eq:irpc}) leads to the mixing of elementary and composite fermions, this mechanism is known as partial compositeness. By integrating out the partners, we obtain the Yukawa couplings
\be
\lambda_i \, \sim \, {y_{Li} y_{Ri} \over g_\psi} \,,
\label{eq:yukawavsmixing}
\ee
where $y_{Li}$ and $y_{Ri}$ are the mixing parameters of left- and right-handed fermions, respectively, and we have parametrized the masses of the partners as 
\be
m_\psi \, = \, g_\psi f . 
\ee
The SM fermion mass hierarchy is then explained by order-one differences in the anomalous dimensions of the operators in Eq.~\eqref{eq:uvpc}.  
\begin{figure}[t]
\centering
\includegraphics[width=10cm]{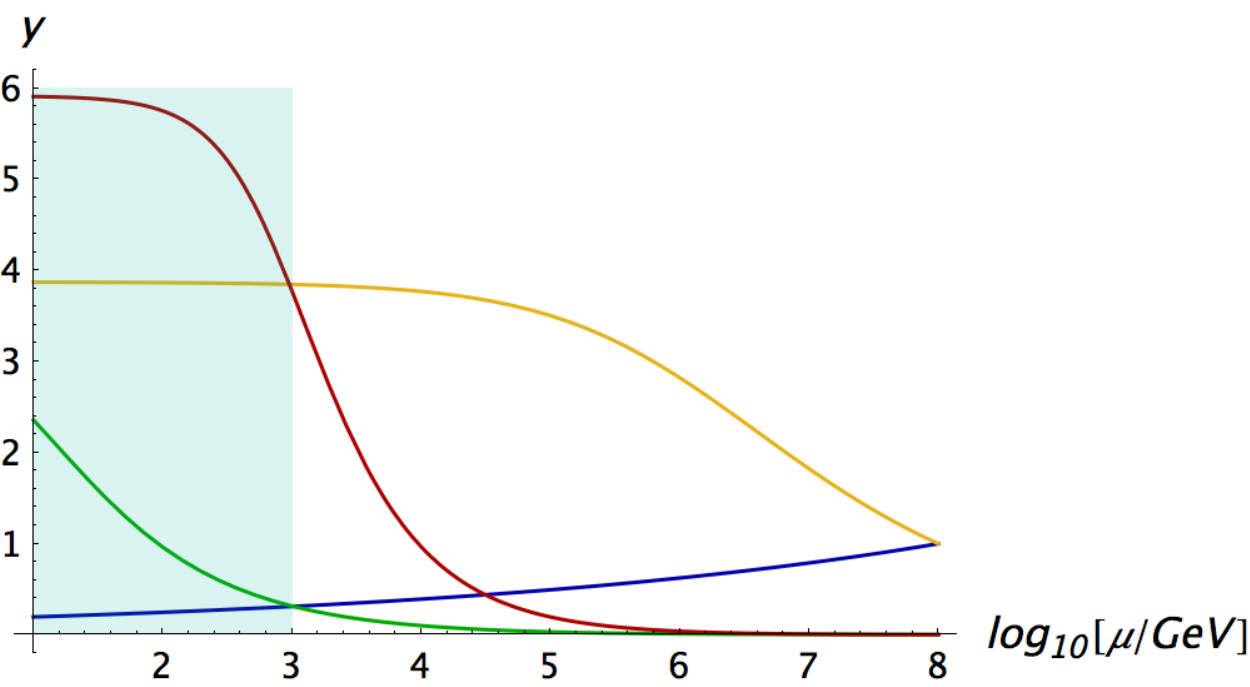}
\caption{\small \it Examples of the running of the mixing parameters (\ref{eq:solmixing}) (the effective Yukawa couplings follow similar trends), for positive $\gamma$ and an order-one initial value (blue), a small negative $\gamma$ and a small initial value (green), a negative $\gamma$ and a small initial value (red), a negative $\gamma$ and an order-one initial value (yellow). The green area shows the range of energies relevant for our analysis. The behaviour of the mixings outside the green region is not relevant for our analysis and can differ from what we show. For instance, all the initial values could be of order one, but the sign of the anomalous dimension could change with energy.}
\label{fig:yuk1}
\end{figure}

The RG equation (\ref{eq:yrun}) only defines the running down to the condensation scale, where the CFT description breaks down. We will later link this condensation scale to the \emph{vev} of the dilaton $\chi$, the PNGB of the spontaneously broken conformal invariance of the strong sector.
When the dilaton \emph{vev} changes, the condensation scale changes accordingly and the mixing parameters $y_i$ in turn change following the RG. 
This allows us to use the RG equation (\ref{eq:yrun}) to obtain the dependence of $y_i$ on the dilaton {\it vev} $\chi$,~\footnote{The general solution for the case where the anomalous dimension $\gamma$ is itself scale-dependent was discussed in Ref.~\cite{vonHarling:2016vhf}.}
\be 
\label{eq:solmixing}
y_i[\chi] \, = \, y_{0,i} \left(\frac{\chi}{\chi_0}\right)^{\gamma_i} \, \left(1 \, + \, \frac{c_i  y_{0,i}^2}{\gamma_i  g_\star^2} \left[1- \left(\frac{\chi}{\chi_0}\right)^{2 \gamma_i} \right] \right)^{-1/2}, 
\ee
where we have fixed the integration constant by requiring that for the dilaton \emph{vev} today $\chi=\chi_0$ the mixing is $y_i=y_{0,i}$. 
More generally, throughout the paper sub/superscripts `$0$' will refer to the present-day values of the parameters. 
Let us discuss some special cases of this general solution. 
A positive anomalous dimension $\gamma_i$ leads to a mixing $y_i$ which is decaying towards the IR. If then $\gamma_i  g_\star^2\gtrsim c_i  y_{0,i}^2$, the second term in the {\it r.h.s.} of Eq.~(\ref{eq:yrun}) is never important and Eq.~\eqref{eq:solmixing} simplifies to
\be
y_i[\chi] \, \simeq \, y_{0,i} \, \left(\frac{\chi}{\chi_0}\right)^{\gamma_i} .
\ee 
This behaviour of the mixing parameter can be used to obtain small Yukawas for the light SM fermions. Small Yukawas are irrelevant for the purpose of our analysis, however, as they only contribute negligibly to the Higgs potential (see next section) or CP violation (see Sec.~\ref{sec:cpv}). A negative anomalous dimension, on the other hand, can make the mixing sizeable in the IR. 
Once $y_i$ is large, the second term on the {\it r.h.s.} of Eq.~(\ref{eq:yrun}) becomes important. For $c_i$ being positive, this causes $y_i$ to run to the fixed point
 \be
y_i\, \simeq \,  \sqrt{-\gamma_i/c_i} \,g_\star \, .
 \ee
In Fig.~\ref{fig:yuk1}, we show several examples of the running according to Eq.~(\ref{eq:yrun}). 
The behaviour corresponding to the red and yellow line is suitable to obtain respectively a varying or an almost constant top Yukawa at the energies of interest. The behaviour corresponding to the green and blue line, on the other hand, can yield respectively \emph{e.g.}~a sizeably growing or a decaying charm Yukawa. As will be shown later, a varying top or a growing charm Yukawa are essential for creating a sufficient amount of CP violation during the phase transition.

\medskip

We will now turn to the more realistic case with flavour mixing, which as we will see later can be crucial for otaining enough CP violation. The couplings (\ref{eq:uvpc}) in the UV and (\ref{eq:irpc}) in the IR then respectively generalize to
\be\label{eq:uvirpcflav}
y_{ij} \, \bar q_i {\cal O}_j \quad \, \text{and} \quad \, y_{ij} \, f \bar q_i U \psi_j\,.
\ee 
Differences in the anomalous dimensions of the operators ${\cal O}_j$ lead in the IR to large hierarchies between entries in the matrix $y_{ij}$ with different index $j$. Aside from this, the structure of $y_{ij}$ can be either arbitrary (``anarchic") with different entries $i$ varying by order-one factors, or it can have a certain pattern dictated by flavour symmetries (see \emph{e.g.}~\cite{Csaki:2008eh}). Integrating out the composite partners, Eq.~\eqref{eq:yukawavsmixing} for the Yukawa couplings generalizes to
\be
\lambda_{ij} \, \sim \,  (y_L)_{ik} \,(g_\psi)^{-1}_{kl} \, (y_R)_{lj}^\dagger \,.  
\ee
It can be shown that for a matrix $y_{ij}$ with hierarchical entries with respect to the index $j$, the approximate equality $y_{ij} \simeq V_{ij} y_{jj}$ holds, where $V_{ij}$ is a unitary matrix (see \emph{e.g.}~\cite{Panico:2015jxa}). Therefore after performing unitary rotations with these $V_{ij}$ on the left- and right-handed SM fermions, we obtain
\be
\lambda_{ij} \, \sim \, (y_L)_{ii} \,(g_\psi)^{-1}_{ij} \, (y_R)_{jj}^\dagger  \,.
\ee
Even if the matrix $(g_\psi)_{ij}$ is anarchic, this leads to hierarchically different Yukawa eigenvalues 
\be
\label{eq:yukawaexpression}
\lambda_i \, \sim \,  \frac{(y_{L})_{ii} (y_{R})_{ii}} {g_\psi}\, ,
\ee
where $g_\psi$ now stands for a typical entry of the matrix.
From Eq.~(\ref{eq:yukawaexpression}), it is now clear that the 
Yukawa couplings of the SM fermions depend on the scale $\chi$. This dependence will crucially impact the dynamics of the electroweak phase transition if it is to happen at the same time as the confinement phase transition.

\subsection{Higgs shift symmetry breaking}
\label{sec:hssb}

If we want to ensure a custodial symmetry in order to suppress oblique corrections to electroweak precision observables, the minimally sufficient global symmetry of the strong sector $G$ is $SO(5)$~\cite{Agashe:2004rs}. If this symmetry were only broken spontaneously, the Higgs potential would vanish as expected for an exact Goldstone boson. However the SM fermions do not fill out complete multiplets of $G$, while only part of $G$ is gauged by the SM gauge bosons. The elementary-composite couplings of Eq.~(\ref{eq:uvpc}) therefore involve operators $\cal O$ which transform under $G$, and fermions $q$ which do not. This is an explicit breaking of $G$ which leads to a potential for the Higgs boson. Similarly the gauge interactions also explicitly break $G$ which gives an additional, though subdominant contribution to the potential (see below). 
The contribution from Eq.~(\ref{eq:uvpc}) can not be too large either, however, as the Higgs mass should be suppressed with respect to the masses of the rest of the strong-sector resonances. The general form of the Higgs potential therefore can be considered as an expansion in an adimensional quantity $y/g_\star$ parametrising the relative strength of the explicit breaking,   
\be\label{eq:vhzero}
V_h \, = \, g_\star^2 f^4 \frac{y^2}{(4\pi)^2}  \sum_i \left(\frac{y}{g_\star}\right)^{p_i} {\cal I}_i\left(\frac{h}{f}\right)\, ,
\ee
where $p_i\geq0$ and ${\cal I}_i$ are trigonometric functions. This expression is intuitively clear from the following arguments. The loop factor suppression follows from the fact that the effective potential is generated at one loop with an elementary state running in the loop. The factor $g_\star^2 f^4$ reproduces the correct overall dimension of the potential (see Sec.~\ref{sec:match}) and is composed of the two characteristic parameters of the strong sector -- the typical coupling and the value of the strong condensate. Finally, that the ${\cal I}_i$ are trigonometric functions is expected from the fact that the Higgs enters the theory in the form of the Goldstone matrix~(\ref{eq:u}).

The same mixings (\ref{eq:uvpc}) and (\ref{eq:irpc}) which are responsible for the Yukawa couplings \eqref{eq:yukawavsmixing}
thus also produce the Higgs potential.  
Given that the current value of the top quark Yukawa coupling dominates over the rest, we can expect that the top quark mixings set the size of the Higgs potential at present times. A more quantitative description of the potential is postponed to Sec.~\ref{sec:hpot}.

In the following, for the sake of decreasing the overall number of parameters, we will set the typical strong-sector coupling $g_\star$ and the coupling determining the mass of the fermionic partners $g_\psi$ equal. 
The rational behind this is that the fermionic resonances are expected to give a sizeable contribution to the potential, which therefore by dimensional analysis has to depend on their masses and couplings.   
This assumption is however not completely flawless, as some of the explicit models show a preference for a sizeable mass gap between the top quark partners and the rest of the composite resonances implying 
~\cite{Matsedonskyi:2012ym,Redi:2012ha,Marzocca:2012zn,Pomarol:2012qf}
\be
g_\psi<g_\star \ .
\ee
We will comment more on this assumption in  Sec.~\ref{sec:cpv}, as its discussion requires a dedicated analysis of the energy dependence of the mixings.

The value of the Goldstone decay constant $f$ determines by how much our model is deformed with respect to the SM and in particular sets the scale by which higher-dimensional operators are suppressed.
It is therefore forced to be somewhat larger than the Higgs \emph{vev}, with the currently preferred value being $f \sim 1$~TeV. However, the potential of the type~(\ref{eq:vhzero}) generically has a minimum at either $h=0$ or $h\sim f$. This can be seen by taking a concrete example of the trigonometric function ${\cal I}_i$, \emph{e.g.}~$\sin^2[h/f]$, which does not allow for minima at $0<h \ll f$~\footnote{This can be translated to the inability to generate a Higgs quartic coupling which is less suppressed than the quadratic term. Expanding the trigonometric functions appearing in the leading order of $y/g_\star$, we obtain both the quadratic term $f^2 h^2$ and the quartic term $h^4$, with an overall coefficient of the same size. This generically gives $h\sim f$ in the minimum. More elaborate constructions can however alleviate this tuning, see \emph{e.g.}~recent attempts in~\cite{Batell:2017kho,Csaki:2017eio}.}.
To obtain $h\ll f$ as required therefore necessitates a tuning, of the order $\sim v^2/f^2$ (a more quantitative way to estimate the tuning will be given in Sec.~\ref{sec:prelimAnalysis}). 

The Higgs potential also receives additional contributions, in particular from interactions with SM gauge bosons. Their effect, though generically expected to be subdominant, can become important in the region where the leading contribution of the SM fermions is tuned to smaller values~\cite{Panico:2012uw}. We however do not need to consider this source of shift symmetry breaking separately. As we will discuss in more detail in Sec.~\ref{sec:hpot}, we will fix the current Higgs potential by the observed Higgs mass and \emph{vev}, and will not distinguish the separate contributions to it. The Higgs potential at different energies, which we will instead consider in more details, is expected to become detuned, with the large fermionic contributions dominating over the contributions from gauge bosons.     

The Higgs also induces the masses of the electroweak gauge bosons which are given by
\be\label{eq:mW}
m_W^2 \, \sim \, g_W^2 f^2 \sin^2 (v/f) \, \equiv \, g_W^2 v_{\text{SM}}^2\, , 
\ee
where $v$ is the Higgs \emph{vev} today, $v_{\text{SM}} = 246$~GeV and $W$ stands for both the $W$ and $Z$ boson.
Notice that this differs from the corresponding expression  in the SM (and $v$ similarly slightly differs from $v_{\text{SM}}$).
From this, we also see why the composite Higgs couplings deviate by order $v_\text{SM}^2/f^2$ from the corresponding couplings in the SM. Varying the mass term for the gauge bosons with respect to the Higgs, we obtain a trilinear vertex
\be\label{eq:vhWW}
\Gamma_{hWW} \, \sim \, \frac{\delta}{\delta h \, \delta W_\mu \, \delta W^\mu} (m_W^2 W_\mu W^\mu)|_{h=v} \, = \, 2 f \sin (v/f) \cos (v/f) \, = \, 2 \, v_{\text{SM}} \cos(v/f)
\ee  
which deviates by a factor of $\cos(v/f) \sim 1-\frac{v^2}{2f^2}$ from the tree-level result in the SM.

\subsection{Conformal symmetry breaking}
\label{sec:DilatonPotential}

We will assume that the confinement phase transition and the formation of the strong condensate can be described as the transition of a single field $\chi$ from zero to some finite \emph{vev}. All mass scales of the strong sector, including $f$ discussed in the previous section, will be linked to the \emph{vev} of $\chi$. In this section we will discuss the main factors determining the dynamics of $\chi$ in isolation, \emph{i.e.}~we will momentarily neglect the Higgs.

In order to reproduce the observed flavour structure of the SM using partial compositeness, large anomalous dimensions for the operators which mix with the fundamental fermions are required. Furthermore, these large anomalous dimensions need to persist over a wide range of energies from some high UV scale 
down to roughly the TeV scale, where the strong sector confines. This can be achieved if the strong sector exhibits an approximate conformal invariance and remains strongly coupled over this energy range (see \emph{e.g.}~\cite{Contino:2010rs}). It is not possible to break the conformal symmetry purely spontaneously and we instead need to add an explicit source of conformal-symmetry breaking \cite{cpr,Coradeschi:2013gda,Bellazzini:2013fga,Chacko:2012sy,Megias:2014iwa}.
To this end, we introduce the strong sector operator 
\be
\epsilon {\cal O}_\epsilon
\ee
whose scaling dimension differs from four. This results in a scale dependence of the renormalized dimensionless parameter $\epsilon$ which to lowest order satisfies the RG equation
\be\label{eq:epsrun}
\frac{\partial \epsilon}{\partial \log \mu} \, \simeq \, \gamma_\epsilon  \epsilon\, ,
\ee
where $\gamma_\epsilon$ is the scaling dimension of ${\cal O}_\epsilon$ minus $4$ (which is constrained as $\gamma_\epsilon \geq -3$ by virtue of the unitarity bound on scalar CFT operators). In order to break conformal invariance in the IR, we choose $\gamma_\epsilon$ negative. Provided that $\gamma_\epsilon$ is small in absolute value and $\epsilon$ at the UV scale is somewhat small too, $\epsilon$ only slowly grows when running towards the IR. The (nearly) conformal invariance is then maintained for a large energy range. Eventually, however, $\epsilon$ and thus the distortion induced by $\mathcal{O}_\epsilon$ grows so large that the strong sector condenses and conformal invariance becomes spontaneously broken.
We further assume that $\gamma_\epsilon$ remains small even at the condensation scale. In this case, the explicit breaking of the conformal invariance is weak compared to the spontaneous breaking by the non-vanishing condensate. There is then an associated light PNGB, the so-called dilaton~\footnote{Despite the fact that the dilaton is allowed to be rather light in our construction, it can not be used as the SM Higgs impostor~\cite{Megias:2016jcw}.}, which we identify with the field $\chi$ introduced earlier. 
The breaking of conformal invariance is reflected by a non-trivial potential for the dilaton given by
\be\label{eq:vcft}
V_\chi \, = \, c_\chi g_\chi^2 \chi^4 \, - \, \epsilon[\chi] \chi^4 \, + \, \dots \, ,
\ee
where all mass dimensions are set by $\chi$ which is the only mass source in the theory and $\epsilon$ is run down to the scale $\chi$ using the RG equation \eqref{eq:epsrun}. 
The first term in the potential respects scale invariance, $x_\mu \to x_\mu \lambda, \chi \to \chi/\lambda$, and therefore is not $\epsilon$-suppressed. For this reason we chose to normalise it with a generic dilaton coupling $g_\chi$ with a power that follows from dimensional analysis (see Sec.~\ref{sec:match}). The constant $c_\chi$ is of order one and
the ellipsis stand for contributions in higher order of $\epsilon[\chi]/g_\chi^2$. 
The first term alone would not allow for a global minimum with a non-vanishing $\chi$, which confirms the need for explicit conformal-symmetry breaking. Without loss of generality we assume that the leading term in the potential which breaks conformal invariance is proportional to the first power of $\epsilon[\chi]$. The non-trivial dependence of $\epsilon[\chi]$ on the condensation scale, and hence on $\chi$, allows for a global minimum of the potential. We denote this minimum as $\chi_0$, the dilaton \emph{vev} today. Minimizing the potential and using the RG equation \eqref{eq:epsrun}, we can determine $\gamma_\epsilon$ and $\epsilon[\chi_0]$, in terms of $\chi_0$ and the dilaton mass $m_\chi$:  
\bea
\label{EpsilonChi0Rel}
V_\chi^\prime[\chi_0] = 0 & \;\;\, \, \Rightarrow \;\; \, \, & \epsilon[\chi_0]=c_\chi \frac{g_\chi^2}{1+\gamma_\epsilon/4}\,, \\
V_\chi^{\prime\prime}[\chi_0]=m_\chi^2 & \;\; \, \, \Rightarrow \;\; \, \,& \gamma_\epsilon = - \frac 1 {4c_\chi} \frac{m_\chi^2}{g_\chi^2 \chi_0^2}\,.
\label{GammaEpsilonRel}
\eea

We can take $\epsilon[\chi]$ as an independent breaking source (assuming that its microscopic description can be provided in the UV-complete theory) and fix the scaling dimension and boundary condition to generate the desired $\chi_0$ and $m_\chi$.
On the other hand, we can associate $\epsilon[\chi]$ with the conformal symmetry breaking due to partial compositeness that is already present in our model. Indeed, loops involving composite and elementary fermions generate
\be\label{eq:epsy}
\epsilon[\chi] \, \sim \, g_\star^2 N_c \frac{y[\chi]^2}{(4 \pi)^2}\,,\;\;\, \, \, \gamma_\epsilon \, = \, 2 \gamma_y\,,
\ee   
where $N_c=3$ is the number of QCD colors. If no other type of conformal-symmetry breaking is allowed in the theory, the presence of a large $y[\chi_0] \sim 4 \pi/\sqrt{N_c}$ seems then to be necessary to generate the non-vanishing condensate. This large mixing, however, could lead to a large breaking of the shift symmetry of the Higgs, 
which consequently would no longer be an approximate Goldstone boson. To solve this issue one may for instance consider a scenario where the large mixing is associated with the elementary right-handed top which is then chosen to transforms as a singlet under $G$, so that its large mixing would not break the Higgs shift symmetry. We will leave a further study of this option for future work. However, the terms in Eq.~\eqref{eq:epsy} will give an additional, albeit generically subdominant, contribution to the dilaton potential as we discuss in Sec.~\ref{sec:hpot}.

The cancellation of an NDA-sized scale-invariant quartic, $c_\chi\sim1$, requires the scale-invariance breaking sources to exit the perturbative region. From Eq.~\eqref{EpsilonChi0Rel}, we see that then $\epsilon[\chi_0]\sim g_\chi^2$. Since $\epsilon$ grows further towards $\chi=0$, the description of the dynamics of the phase transition becomes less robust within our approach. In order to proceed, we will assume a moderate suppression of the scale-invariant quartic, $c_\chi<1$,  so that cancelling it does not require $\epsilon[\chi_0] \sim g_\chi^2$. To prevent $\epsilon$ from exiting the perturbative regime at lower energies we cut its growth by accounting for the next-to-leading order term in the RG equation,
\be
\label{epsilonRGeq}
\frac{\partial \epsilon}{\partial \log \mu} \, \simeq \, \gamma_\epsilon \epsilon \, + \, c_\epsilon \frac{\epsilon^2}{g_\chi^2}\,,
\ee
with the order-one coefficient $c_\epsilon$ taken positive.
As in Eq.~\eqref{EpsilonChi0Rel}, we can fix the integration constant by the requirement that the potential is minimized at $\chi=\chi_0$. The solution to the RG equation then reads
\begin{equation}
\label{epsilonexpression}
\epsilon [\chi] \, = \, \frac{8 \, c_\chi g_\chi^2 \gamma_\epsilon (\chi/\chi_0)^{\gamma_\epsilon}}{\gamma_\epsilon \left(4 + \gamma_\epsilon + \sqrt{16 \,  c_\epsilon c_\chi + (4 + \gamma_\epsilon)^2} \right) + 8 \,  c_\epsilon c_\chi \left(1 - (\chi/\chi_0)^{\gamma_\epsilon} \right)}\, . 
\end{equation}
As in Eq.~\eqref{GammaEpsilonRel}, we could further trade the scaling dimension $\gamma_\epsilon$ for the dilaton mass $m_\chi$.

Note that this has a form reminiscent of the Goldberger-Wise potential \cite{Goldberger:1999uk} which arises in certain 5D duals of confining theories. However, for $\gamma_\epsilon<0$  the corresponding Goldberger-Wise potential has a barrier at zero temperature with crucial implications for the strength of the phase transition (see \cite{vonHarling:2017yew}), while our dilaton potential \eqref{eq:vcft} with \eqref{epsilonexpression} has no such barrier. 
We will show that a strong first-order phase transition can nevertheless follow.

\section{Zero-temperature effective field theory}
\label{sec:zteft}

\subsection{Scalar potential: Matching two descriptions}
\label{sec:match}

In the previous section, we have reviewed the effective potentials of the Higgs, in the one coupling-one mass scenario, and the dilaton. We will now combine both potentials, relating the mass scale of the problem to the dilaton \emph{vev}.
This will provide us with a unified framework allowing for the description of the confinement phase transition and the electroweak phase transition together. 
It is most natural to assume that the Higgs is a meson-like state of the underlying confining theory, in analogy to the QCD pions associated with chiral symmetry breaking. 
For the dilaton, instead, one can argue for it being either meson-like (see \emph{e.g.} the lattice studies observing a light meson-like state~\cite{Aoki:2014oha,Appelquist:2016viq} in $N_c=3, N_f=8$ theories) or glueball-like (as expected in theories dual to a warped extra dimension such as~\cite{Randall:2006py}).  

In the case where both the Higgs and the dilaton are meson-like states, we can consistently assume that they are characterised by approximately the same typical coupling, hence we set
\be
g_\star \, = \, g_\chi .
\ee
A glueball-like dilaton would instead behave very differently. In large-$N$ confining theories, which we will use as a reference, an interaction involving $l$ glueballs and $k$ mesons scales like~\cite{Witten:1979kh} 
\be\label{eq:largeN}
N^{-l-k/2+1} \, \,(k\neq0)  \, \, \,  \text{or} \, \, \,  N^{-l+2} \, \, (k=0)\,.
\ee 
This suggests defining the coupling of the glueball-like states as 
\be\label{eq:e2}
g_\chi \, \equiv \, 4\pi/N
\ee
and for the meson-like states as
\be\label{eq:e3}
g_\star (g_\chi) \, \equiv \, 4 \pi/\sqrt N.
\ee
The factors of $4 \pi$  are chosen to ensure that one recovers a generic strongly-coupled theory in the limit $N\to1$~\footnote{Notice, however, that for instance in explicit 5D constructions this normalization can differ by a factor of order a few.}. Let us now find the general form of the effective potential involving simultaneously the Higgs $h$ and the dilaton $\chi$, forgetting about Goldstone symmetries for the moment. In the absence of explicit mass scales, we can only construct the effective potential out of the fields $\chi$ and $h$ and the couplings $g_\chi$ and $g_\star$. A generic term in the effective potential can thus be written as 
\be\label{eq:e1}
g_\star^\alpha g_\chi^\beta \, h^\gamma \chi^\delta\,.
\ee
Relations between the powers $\alpha,\beta,\gamma$ and $\delta$ can be obtained by dimensional analysis. 
Keeping units of length $L$, time $T$ and mass $M$ (\emph{i.e.}~not working in natural units where $\hbar=c=1$), one concludes that the dimensions of the potential, fields, couplings and derivatives are (see \emph{e.g.}~\cite{Panico:2015jxa,Chala:2017sjk})
\be 
[V] \, = \, \frac{[\hbar]}{L^4}\, , \quad \, [\chi] \, = \, [h] \, = \, \frac{[\hbar]^{1/2}}{L}\, , \quad \, [g_\chi]\,= \, [g_\star] \, = \, \frac{1}{[\hbar]^{1/2}}\, , \quad \, [\partial_\mu] \, = \, \frac{1}{L} \, ,
\ee
where $[\hbar]= M L^2/ T$. From this, one finds the relations $\gamma + \delta =4$ and $\alpha+\beta=2$. Imposing also the large-$N$ scaling from Eqs.~(\ref{eq:largeN}), (\ref{eq:e2}), (\ref{eq:e3}) gives an additional relation and  
fixes the term in Eq.~(\ref{eq:e1}) to 
\be\label{eq:e4}
g_\star^{2}\, (g_\chi \chi/g_\star)^{4} \left(\frac{h}{g_\chi \chi/g_\star}\right)^\gamma  \;\;(\gamma\neq 0) \quad \text{or}  \quad g_\chi^2 \chi^4 \;\;(\gamma=0) \, . 
\ee
This applies to both a glueball-like or a meson-like dilaton, depending on which $N$-scaling we choose for $g_\chi$. 

We are now in a position to derive the dependence of the Higgs potential on the dilaton, restoring the full Goldstone symmetry. The dilaton \emph{vev} is the only source of mass in the theory. We therefore need to replace the Goldstone decay constant $f$, which balances the Higgs in the functions  ${\cal I}_i$ of Eq.~(\ref{eq:vhzero}), by the dilaton \emph{vev} $\chi$ times a proportionality factor. In order to determine this factor, we can write the functions ${\cal I}_i$ as power series in $h/\chi$ and match with Eq.~\eqref{eq:e4}. This gives
\be\label{eq:vgen}
V_{h} \, =  g_\star^2 \left({g_\chi} \chi /{g_\star} \right)^4 \frac{y^2}{(4\pi)^2} \sum_{i} \left(\frac{y}{g_\star}\right)^{p_i} {\cal I}_i \left[\frac{h}{{g_\chi} \chi /{g_\star}}\right]\,.
\ee
Note that this has still the right dimensions since after restoring factors of $\hbar$, the loop factor is $y^2\hbar/(4 \pi)^2$ which is dimensionless.   
From this expression we can in particular read off the relation between the current Goldstone decay constant $f$ and the current dilaton \emph{vev} $\chi_0$,
\be
f \, = \, g_\chi \chi_0/g_\star\,.
\label{eq:fversuschi}
\ee
In order to obtain the combined Higgs-dilaton potential, we then need to add Eqs.~\eqref{eq:vcft} and \eqref{eq:vgen}. Notice that the first term in Eq.~\eqref{eq:vcft} has the correct power of the coupling $g_\chi$ as follows from Eq.~\eqref{eq:e4}.

\subsection{Canonical variables}

To complete our EFT we need to include the kinetic terms of the Higgs and dilaton.
For consistency of our description, these kinetic terms need to be invariant under the shift symmetries. It is precisely this invariance which allows us to fix the form of the Higgs potential in terms of trigonometric functions and suppression factors proportional to the symmetry breaking sources~(\ref{eq:vhzero}). In case of an isolated Higgs and a constant scale $f$ it is sufficient to choose the Higgs kinetic term as
\be\label{eq:hkin1}
{\cal L}_{\text{kin}} \, = \, \frac 1 2 (\partial_\mu h)^2
\ee 
to respect the symmetry $h \to h + \text{const.}$ (which is weakly broken by the scalar potential) and the smaller (gauge) symmetry $h \to h + 2 \pi f k\, (k \in \mathbb{Z})$ (which is an exact symmetry of the full Lagrangian). The presence of the gauge symmetry follows from the fact that the Higgs parametrizes the phase of the Goldstone matrix, $U\sim \exp [i h/f]$, and phases rotated by $2 \pi$ are not distinguishable.  However, we can not trivially apply the same kinetic term in case of a dynamical scale $f \to g_\chi \chi/ g_\star$. Although the scalar potential~(\ref{eq:vgen}) does respect the symmetry $h \to h + 2 \pi g_\chi \chi/ g_\star$, the kinetic term~(\ref{eq:hkin1}) does not.
The simplest way to derive the invariant kinetic terms is to switch to the dimensionless Goldstone boson $\theta$, which substitutes the argument $g_\star h / g_\chi \chi$ in the trigonometric functions of the scalar potential~(\ref{eq:vgen}). Using dimensional analysis as discussed in Sec.~\ref{sec:match}~\footnote{We recall that $\theta \propto h$ transforms as an electroweak doublet. Therefore at the level of dimension-four operators we can not write down a kinetic mixing term. Higher-order operators will be discussed in Sec.~\ref{sec:prelimAnalysis}.}, we then find
\be
{\cal L}_{\text{kin}} \, = \, \frac 1 2 \frac{g_\chi^2}{g_\star^2} \chi^2 (\partial_\mu \theta)^2 + \frac 1 2 (\partial_\mu \chi)^2\,,
\ee
which trivially respects the gauge symmetry $\theta\to\theta+2 \pi k $. The constant prefactor of the kinetic term for $\theta$ has been chosen such that at a constant dilaton \emph{vev}, we can switch to the field 
\be
\label{eq:hdef}
h \, = \, \theta g_\chi \chi/g_\star 
\ee 
and reproduce the Higgs potential~(\ref{eq:vhzero}).
For later convenience, we also introduce the field
\be
\hat{h} \, \equiv \, \theta g_\chi \chi_0/g_\star = \theta f 
\ee
which has a canonically normalized kinetic term at $\chi=\chi_0$. On the other hand, $h$ necessarily has a mixed kinetic term $\sim \partial_\mu \chi \partial^\mu h$ needed to insure invariance under $h \to h + 2 \pi g_\chi \chi/ g_\star$.    
 
In the following, when considering the phase transition, it will be convenient to use field variables which always remain canonically normalized,
\be
\label{chi1chi2def}
\chi_1 \, \equiv \, \chi \sin [g_\chi \theta/g_\star]\, , \, \,\; \; \; \;\;\chi_2 \, \equiv \, \chi \cos [g_\chi \theta/g_\star]\, ,
\ee 
so that
\be
{\cal L}_{\text{kin}} \, = \, \frac 1 2 (\partial_\mu \chi_1)^2 \, + \, \frac 1 2 (\partial_\mu \chi_2)^2\,.
\ee 
We can express the scalar potential~(\ref{eq:vgen}) in terms of these new field variables by using the inverse relations
\be
\label{eq:dictionary}
\chi \, = \,  (\chi_1^2+\chi_2^2)^{1/2}, \, \, \, \, \;\; \;\;\;
 {g_\star h}/{g_\chi \chi} =  \theta \, = \, (g_\star/g_\chi)\arcsin[\chi_1/(\chi_1^2+\chi_2^2)^{1/2}]\,.
\ee

\subsection{Parametric form of the scalar potential}
\label{sec:hpot}

We are now ready to add the final details to the zero-temperature potential given by Eqs.~\eqref{eq:vcft} and \eqref{eq:vgen}, namely to take into account the tuning which is necessary to obtain the observed Higgs mass and \emph{vev} today. From this we will also see
how the varying mixings affect the potential at values of $\chi$ different from the value $\chi_0$ today.  
Let us begin with discussing the Higgs potential today.
For the functions $\mathcal{I}_i$, we choose a parametrisation which can be matched onto the most commonly used models~\cite{Panico:2012uw}
\be\label{eq:vh0}
V_h^0 \, = \, \alpha^0 \sin^2 \theta \, + \, \beta^0 \sin^4 \theta\, .
\ee
The coefficients $\alpha^0$ and $\beta^0$ are fixed by the observed Higgs mass and \emph{vev} as
\be
\alpha^0 \, = \,  -2 \beta^0 \sin^2 (v/f) \, \simeq \, -{1 \over 4} f^2  m_h^2\, , \;\;\;\;\;\;\; \beta^0 \, \simeq \, {1\over 8} m_h^2 f^2 /\sin^2 (v/f) \, .
\ee
As mentioned in the introduction, reproducing the SM Higgs parameters requires a certain amount of tuning. This manifests itself in the fact that $\alpha^0$ and $\beta^0$ typically sizeably deviate from generic NDA estimates. 
However, the model parameters generically vary if we change $\chi$ from its \emph{vev} today $\chi_0$ as happens during the phase transition. For example, we have seen in Sec.~\ref{sec:PartialCompositeness} that the mixings $y$ can substantially change with $\chi$. We expect that such a variation leads to a detuning of the potential and that the potential becomes completely generic away from the current minimum. 
Choosing the trigonometric functions as in Eq.~(\ref{eq:vh0}), the leading-order estimate in Eq.~(\ref{eq:vgen}) for the Higgs potential generated by $n_f$ quarks reads
\be\label{eq:vnda}
V_h^{\text{NDA}}[y] \, = \,  \alpha^{\text{NDA}}[y] \sin^2 \theta \, + \,  \beta^{\text{NDA}}[y] \sin^4\theta 
\ee
with the coefficients given by 
\be
\alpha^{\text{NDA}}[y] = c_\alpha \sum_{n_f} \, g_\star^2 {N_c \,y^{2}[\chi] \over (4 \pi)^2}  \left(\frac{g_\chi}{g_\star} \chi \right)^4  \quad \text {and} \quad  
\beta^{\text{NDA}}[y] = c_\beta \sum_{n_f} \, g_\star^2 {N_c \, y^{2}[\chi] \over (4 \pi)^2}  \left(\frac{g_\chi}{g_\star} \chi \right)^4 \left({y\over g_\star}\right)^{p_\beta} .
\ee
Here $y[\chi]$ is given by Eq.~\eqref{eq:solmixing},
$c_\alpha$ and $c_\beta$ are free parameters of our EFT whose absolute values are expected to be of order one and $N_c$ is the number of SM QCD colors enhancing the quark loops.
A  non-zero $p_\beta$ means that the leading contribution to the coefficient $\beta$ is suppressed with respect to the naive estimate~\cite{Matsedonskyi:2012ym,Panico:2012uw}. In the known composite Higgs models one can have $p_\beta=0,2$. For instance, contributions with $p_\beta=0$ arise from fermionic composite operators in Eq.~(\ref{eq:uvpc}) transforming in $G$ representations with dimension ${\bf r}_{L/R}={\bf 14}$ and $p_\beta=2$ is generated from operators with ${\bf r}_{L/R}={\bf 5},{\bf 10}$~\cite{Panico:2012uw}.   

Altogether, we can approximate the Higgs potential for arbitrary $\chi$ as 
\be
\label{eq:Vhpotential}
V_h \, = \, \left(\frac{g_\chi\chi}{g_\star f}\right)^4 \left(\alpha^0 \sin^2\theta  +  \beta^0 \sin^4\theta \right) \, + \, \left(V_h^{\text{NDA}}[y] - V_h^{\text{NDA}}[y_0]\right)
\ee
with $\theta$ in terms of $h$ and $\chi$ given in Eq.~\eqref{eq:dictionary}. This interpolates between the tuned potential in Eq.~\eqref{eq:vh0} for the dilaton \emph{vev} $\chi=\chi_0$ and the mixings $y=y_0$ today and the detuned potential in Eq.~\eqref{eq:vnda} for $\chi\neq \chi_0$ and $y \neq y_0$.
In this expression we have also accounted for the fact that $V_h^{0}$ at fixed $y=y_0$ scales as the fourth power of $\chi$.

For the Higgs-independent dilaton potential, we will similarly include a $y$-dependent term which we expect to be generated from the mixings as discussed around Eq.~\eqref{eq:epsy}. Altogether this gives
\be
\label{eq:Vchipotential}
V_\chi \, = \, c_{\chi} g_\chi^2 \chi^4 \, - \, \epsilon[\chi] \chi^4 \, + \, c_{\chi y} \sum_{n_f} \,g_\star^2  \frac{N_c \, y^2[\chi]}{(4 \pi)^2} \left(\frac{g_\chi}{g_\star} \chi \right)^4,
\ee
where the coefficients $c_{\chi}$ and $c_{\chi y}$ are generically of order one.  
The $y$-dependent term is expected to be related to the Higgs potential. Since this dependence can only be extracted in explicit models which we do not discuss, however, we will limit ourselves to the assumption that there are no correlations between the $c$-coefficients controlling different terms of the scalar potential.   
Note that the $y$-dependent term can produce a barrier between $\chi=0$ and $\chi_0$ at zero temperature if $c_{\chi y}>0$, analogously to the Goldberger-Wise potential for the dilaton which was used in the studies of the phase transition of Randall-Sundrum models in~\cite{Creminelli:2001th,Randall:2006py,Nardini:2007me,Konstandin:2010cd,vonHarling:2017yew,Dillon:2017ctw}.

\subsection{Key properties of the two-field potential and preliminary analysis}\label{sec:prelimAnalysis}

\begin{figure}[t]
\centering
\includegraphics[width=6.5cm]{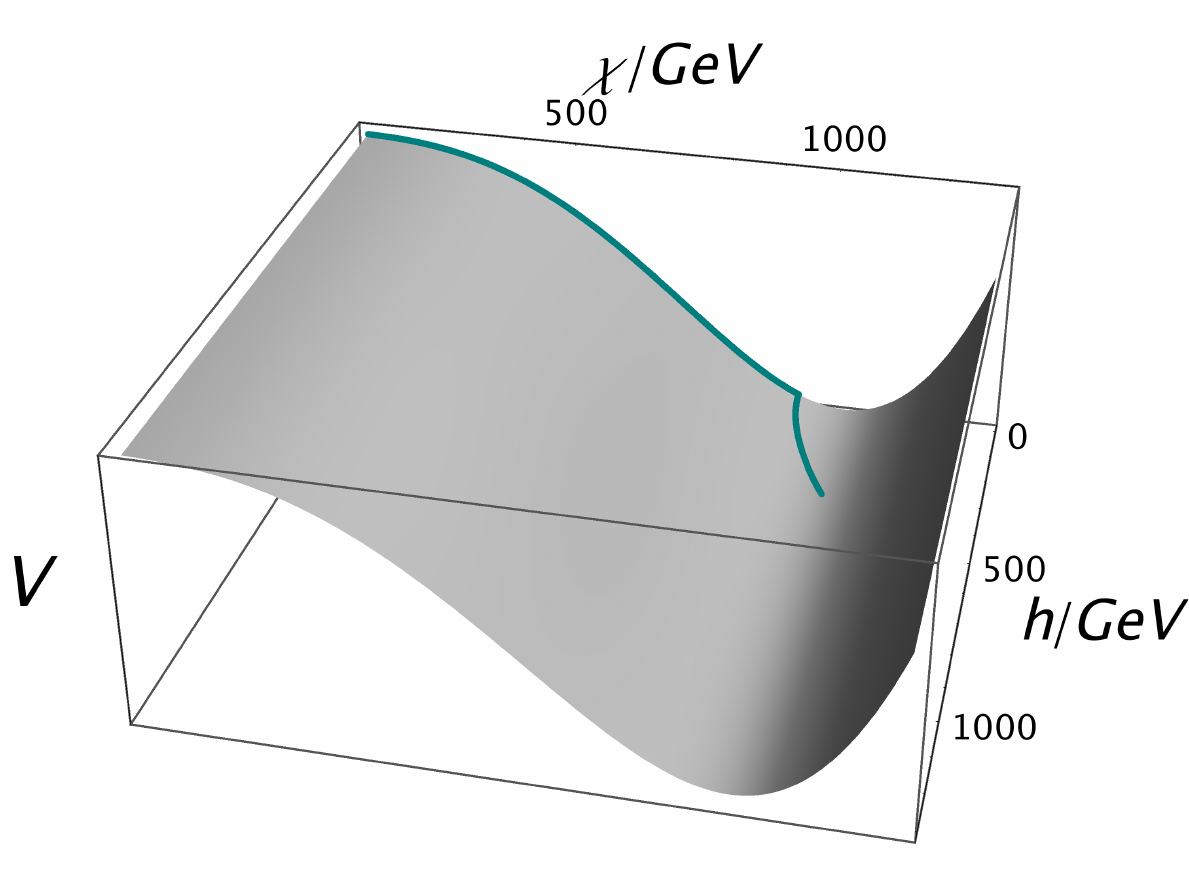}
\hspace{1cm}
\includegraphics[width=6.5cm]{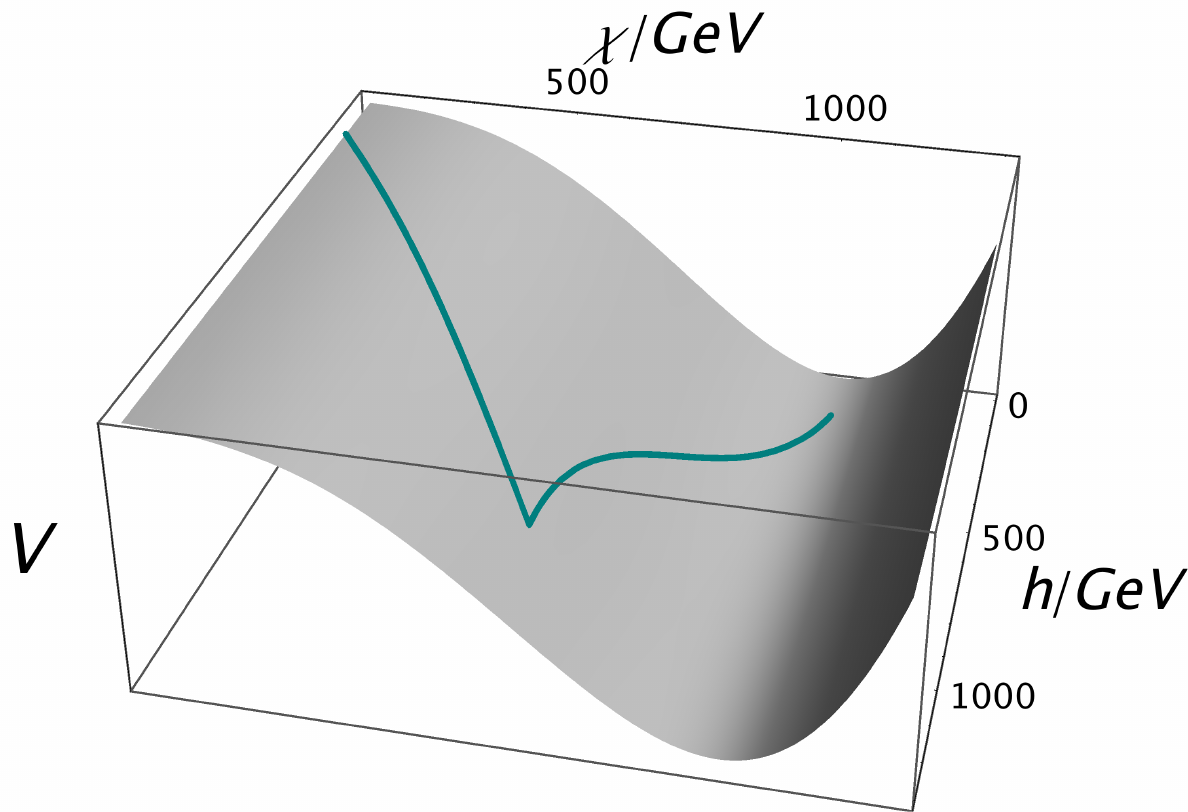}
\caption{ \small \it Examples of potentials with a valley along the direction $h=0$ (left plot) and with a valley along $h\sim\chi$ (right plot). The green line shows the $\chi$-dependent minimum of the Higgs potential. Since the Higgs potential is loop-suppressed with respect to the dilaton potential, and we assume order-one values of the mixings $y$, the valleys are not very pronounced.}
\label{fig:simppot}
\end{figure}

\begin{figure}[t]
\centering
\includegraphics[width=8.cm]{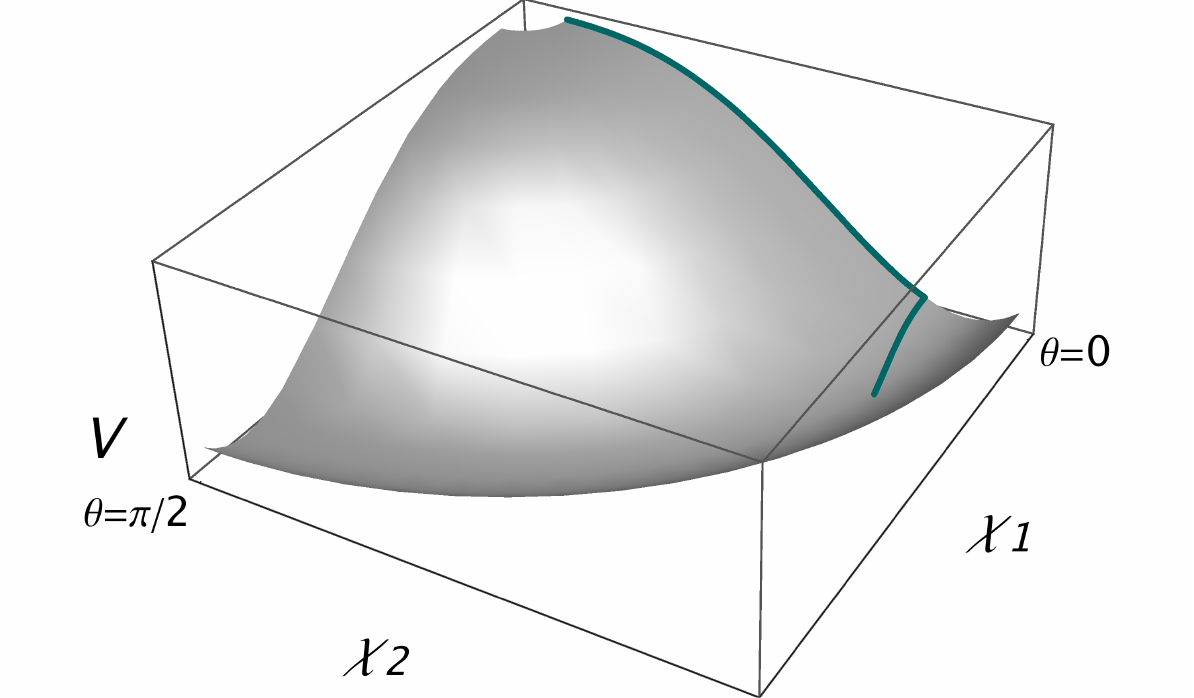}
\includegraphics[width=8.cm]{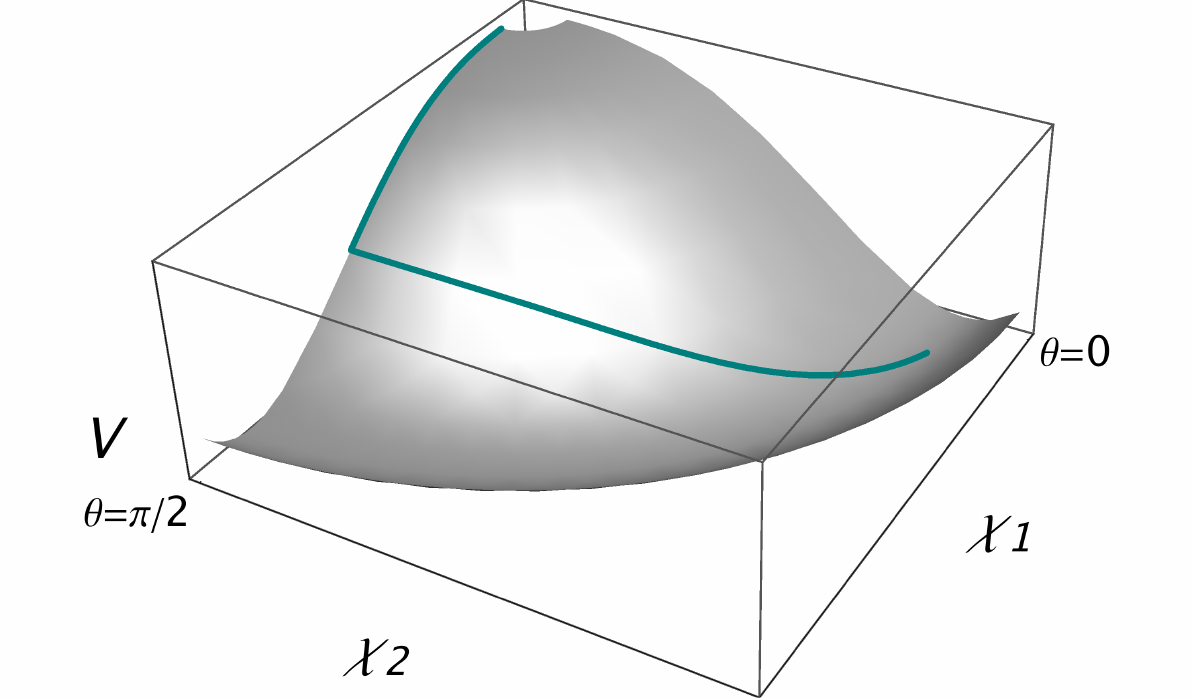}
\caption{ \small \it Same as in Fig.~\ref{fig:simppot} but in canonical variables $\chi_{1,2}$.}
\label{fig:simppot2}
\end{figure}

We can now perform a preliminary analysis of the scalar potential. 
The Higgs potential away from $\chi=\chi_0$ is typically dominated by its detuned part, $(V_h^{\text{NDA}}[y] - V_h^{\text{NDA}}[y_0])$, as the NDA contributions are larger than what is needed to reproduce the potential today at $\chi=\chi_0$ and $y= y_0$. This detuned part admits minima which are typically at $\theta=0$ or $\theta=\pi/2$ for fixed $\chi<\chi_0$, depending on the signs of the coefficients $\alpha^{\text{NDA}}$ and $\beta^{\text{NDA}}$. Sketches of the two-dimensional potential for these two cases are shown in 
Figs.~\ref{fig:simppot} and \ref{fig:simppot2}.
The minima with respect to $\theta$ result in corresponding valleys in the two-dimensional potential which will attract the tunnelling trajectory during a first-order phase transition. If this attraction is large, which is expected for large mixings $y$ and a resulting deep valley, the tunnelling trajectory  can come very close to this valley. 

Along a valley $\theta \propto h=0$, the electroweak gauge bosons generically are massless which leaves the sphaleron processes active during most of the phase transition and is thus not suitable for electroweak baryogenesis. We are therefore instead interested in a valley along $\theta = \pi/2$. 
If the mixings $y$ are growing towards the IR, such a valley is present if
\be\label{eq:cab1}
c_\alpha +  c_\beta (y/g_\star)^{p_\beta} \, < \, 0 \, .
\ee
On the other hand, we need the SM fermions to be massive as well, in order to source the CP violation. For this, the condition~(\ref{eq:cab1}) is necessary but not sufficient. Indeed, in many explicit composite Higgs models the fermion masses are proportional to $\sin \theta \cos \theta$ and thus also vanish at $\theta= \pi/2$.
Therefore if the potential features a valley along ${\theta= \pi/2}$ which is deep enough to make the tunnelling trajectory follow it closely, electroweak baryogenesis can be spoiled as well. The exact tunnelling path depends on a number of parameters and does not necessarily follow the valley.
If this does happen, on the other hand, we can choose parameters such that the valley is along a direction $0 < \theta < \pi/2$ as we will see later. Alternatively, we can use the fact that the cosine in the expression for the mass is absent when we couple the elementary fermions to composite operators in the following $SO(5)$ representations: ${\bf r}_L + {\bf r}_R={\bf5}+{\bf1}, {\bf5}+{\bf10}, {\bf10}+{\bf5}, {\bf5}+{\bf14}, {\bf14}+{\bf5}, {\bf10}+{\bf10}$~\cite{Pomarol:2012qf}. Interestingly, this selection criterium disfavours the minimal embeddings with $\bf 1$ and $\bf 5$ representations only (taking into account that the allowed ${\bf5}+{\bf1}$ has difficulties in reproducing a realistic Higgs potential~\cite{Panico:2012uw}). The preferred models therefore can have a rich spectrum of composite fermions as a distinctive phenomenological feature~\cite{Pappadopulo:2013vca,Matsedonskyi:2014lla}. 
In summary of this discussion, we present in 
Figs.~\ref{fig:simppot} and~\ref{fig:simppot2} 
two cases, with a valley along $\theta=0$ and along $\theta=\pi/2$. 

Another important comment concerns the overall tuning of the model. As was already mentioned, the NDA estimates of the Higgs potential typically give too large a Higgs mass and either $h \sim0$ or $h\sim f$ for the Higgs \emph{vev}. Obtaining $h \sim v\ll f$ then requires some tuning of the Higgs potential, which can be estimated to be of order 
\be
\xi \, \equiv v^2/f^2 \, .
\ee
Current experimental constraints coming from various observables indicate that \cite{Grojean:2013qca}
 \be
 \xi\, \lesssim \, 0.1...0.2 \, .
 \ee
In order to estimate the overall tuning of the Higgs sector (which includes the  tuning in $\xi$), we take the product of ratios of the required values of the Higgs potential coefficients over the values that are generically expected,
\be
\text{tuning} \, \sim \, \frac{\alpha^0}{\alpha^{\text{NDA}}[y_0(\text{top}),f]} \frac{\beta^0}{\beta^{\text{NDA}}[y_0(\text{top}),f]}\, .
\ee 
Here we only include the contribution to $\alpha$ and $\beta$ from the top quark, as the other quark mixings are expected to be small at present times. We emphasize that the amount of fine-tuning that is required in the model with sizeable variation of the mixings is not different from the fine-tuning in the usual composite Higgs models.

We are now also in the position to discuss the important phenomenological question of Higgs-dilaton mixing, in the true minimum of the scalar potential. This mixing can alter the Higgs couplings compared to the SM predictions, in addition to the usual universal composite Higgs deviations proportional to $v^2/f^2$ (\emph{cf.}~Eq.~(\ref{eq:vhWW})). 
We will work in the basis ($\hat{h}$,$\chi$) where possible kinetic mixings are redefined away, therefore all the mixing effects are contained in the scalar potential~(\ref{eq:Vhpotential}) and are determined by
the mixing mass
\be
\begin{split}
\label{eq:vmix1}
m_{\hat{h} \chi}^2 \, &= \, \frac{\delta}{\delta \hat{h}\, \delta \chi} V_{h}\bigl|_{\hat{h}=v,\chi=\chi_0}   \\
\, & =  \, \partial_\chi \left(\frac{g_\chi\chi}{g_\star f}\right)^4 \partial_{\hat{h}} \, (\alpha^0 \sin^2\theta  +  \beta^0 \sin^4\theta)\, + \, \partial_{\hat{h}} \partial_\chi \, (V_h^{\text{NDA}}[y] - V_h^{\text{NDA}}[y_0])\,.
\end{split}
\ee
Note that the first term in the second line vanishes, since the first derivative of the scalar potential vanishes in the minimum:
\be
\partial_{\hat{h}} \, (\alpha^0 \sin^2\theta  +  \beta^0 \sin^4\theta)|_{\hat{h}=v,\chi=\chi_0} \, \sim \, \partial_{\hat{h}} \, V_h|_{\hat{h}=v,\chi=\chi_0} \, = \, 0\,.
\ee 
Mixing between the Higgs and the dilaton can still arise from the second term in the second line which gives
\be
\begin{split}
\label{eq:hchimix2}
m_{\hat{h} \chi}^2  &=  f^2 \frac{N_c g_\star g_\chi}{(4\pi)^2 } \, \{4 \, c_\alpha  y  \beta_y  \sin \theta_0 \cos \theta_0 \, + \, 
4 \, (2+p_\beta) \, c_\beta \, (y^{1+p_\beta}/g_\star^{p_\beta}) \,  \beta_y \, \sin^3 \theta_0 \cos \theta_0 \} \\
 &\simeq   
f^2 \frac{N_c g_\star g_\chi}{(4\pi)^2}  \,  \{4 \,  c_\alpha  y  \beta_y  \sin \theta_0 \}\,,
\end{split}
\ee
where $\theta_0=v/f$, $\beta_y$ is the $\beta$-function in the RG equation \eqref{eq:yrun} and we have neglected higher powers of $ \theta_0$. The mixing mass thus becomes the larger, the more the mixing parameters $y$ vary with energy. 
In the limit of a large dilaton mass $m_{\chi}$, the Higgs-dilaton mixing angle $\delta$ is given by $\tan \delta \sim m_{\hat{h} \chi}^2/m_{\chi}^2$. As follows from~\cite{Chala:2017sjk}, the effect of this type of mixing is most pronounced in the Higgs-photon and Higgs-gluon couplings, where it can easily outrun the universal composite Higgs effects of order $v^2/f^2$. Several other operators relevant for Higgs physics can however also be affected, if the dilaton mass becomes small enough. Once observed, the deviations of the Higgs couplings can become an important test of the scenarios with sizeably varying Yukawa interactions as we discuss in more detail in Sec.~\ref{ExperimentalTests}.
While performing the numerical analysis of the phase transition, we will keep track of the mass mixing and require that $\delta \lesssim 0.1$ to ensure that experimental constraints on the Higgs couplings are fulfilled.

\section{Description at finite temperature}
\label{sec:fteft}

We can distinguish two qualitatively different types of finite-temperature effects: those leading us out of the applicability of the EFT, for temperatures and dilaton \emph{vevs} such that $g_\chi \chi\lesssim T$, and those which do not, for $g_\chi \chi \gtrsim T$. The latter are simply accounted for by the standard thermal corrections. The former can only be properly accounted for in the complete UV description up to the energy scale $\mu \sim T$, including the heavier bound states and eventually their deconfined constituents.  In the following we will use a limited knowledge about this UV theory to determine the main relevant features of the phase transition. This discussion is in many aspects analogous to that in~\cite{Creminelli:2001th,Randall:2006py,Nardini:2007me,Konstandin:2010cd,vonHarling:2017yew,Dillon:2017ctw} for the phase transition in 5D dual models.

For $\chi=0$, the strong sector that gives rise to the Higgs is in its deconfined and (nearly) conformal phase. By dimensional analysis and large-$N$ counting, it is clear that the free energy in this phase scales as $F_{\text{CFT}}[\chi=0]\simeq -c N^2 T^4$, where the constant $c$ depends on the number of \emph{d.o.f.} per color in the strong sector. For definiteness, we will use the result for $\mathcal{N}=4$ $SU(N)$ super-Yang-Mills  (including a factor $3/4$ due to strong coupling which can be calculated from the AdS dual \cite{Gubser:1996de}). The free energy then reads
\be\label{eq:fcft}
F_{\text{CFT}}[\chi=0]\, \simeq \, -\frac{\pi^2 N^2}{8} T^4\, .
\ee
We expect that any realistic strong sector would require an approximately similar number of \emph{d.o.f.}~per color as $\mathcal{N}=4$ $SU(N)$ super-Yang-Mills and that the free energy would thus not differ much from the relation that we use.
An additional contribution ${F_{\text{SM}} \simeq - \pi^2 g_{\rm SM} T^4/90}$ arises from the elementary SM fields, with $g_{\rm SM}\simeq 100$ being the total number of \emph{d.o.f.}~of the SM. We will neglect this contribution for now assuming $N\gg1$ (it will be accounted for in our numerical study). We also assume that the conformal symmetry breaking effects are sufficiently small to be negligible.

\begin{figure}[t]
\centering
\includegraphics[width=10cm]{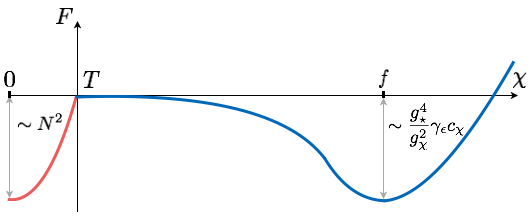}
\caption{\small \it Schematic shape of the free energy as a function of  $\chi$, in the ``hot''  region with $g_\chi \chi\lesssim T$ (in red) and in the ``cold'' region with $g_\chi \chi\gtrsim T$ (in blue).}
\label{fig:F}
\end{figure}

We next need to determine the potential at finite temperatures in the regime $\chi>0$. For $\chi>T/g_\chi$, the thermal corrections to the potential from composite states are calculable and small. Again momentarily neglecting the thermal corrections from the SM, the free energy in this regime is well approximated by
\be\label{eq:feft}
F_{\text{conf.}}[\chi>T/g_\chi]\,  \simeq \, V_h + V_\chi \, ,
\ee
where $V_h$ and $V_\chi$ are given in Eqs.~\eqref{eq:Vhpotential} and \eqref{eq:Vchipotential}. The relative energy of the two phases described by Eqs.~\eqref{eq:fcft} and \eqref{eq:feft} is fixed by the fact that both should match in the limit $T\to0$ and $\chi\to0$. Given the absence of robust knowledge about the temperature corrections in the opposite regime $0\lesssim \chi \lesssim T/g_\chi$, we could neglect this part of the potential and glue \eqref{eq:fcft} to \eqref{eq:feft} at $\chi=T/g_\chi$ using a step-function transition. 
The tunnelling rate which is obtained for such an approximation to the free energy would of course only give an estimate. It is however expected not to be significantly different from the exact result. 
To see this, recall that we are interested in regions of parameter space where the nucleation temperature is below the weak scale, and therefore 
significantly below the scale $\chi_0\sim f \gg v$ of the true minimum. Whatever features the bounce action has from the region $0 \lesssim \chi \lesssim T/g_\chi$, they are characterized by the only relevant scale $T \ll v$, and are therefore expected to be smaller than the contribution from the region under control $T/g_\chi \lesssim \chi \lesssim \chi_0$. An additional factor making the latter region more important is the fact that the dilaton potential in this region is characterised by the behaviour $\sim \chi^4 (g_\chi^2-\epsilon)$ with $\epsilon\simeq g_\chi^2$ and slowly varying. The height of the potential barrier is therefore almost constant over the large interval of $\chi$ starting from the maximal point at $\chi=T/g_\chi$. A sketch of the described behaviour of the free energy is shown in Fig.~\ref{fig:F}.

The critical temperature at which the transition becomes energetically allowed is given by the temperature at which the free energies in the deconfined and confined phases become equal. Neglecting the Higgs-dependent part in (\ref{eq:feft}), the relevant free energy in the confined phase is given by the global minimum of the dilaton potential at $\chi_0= g_\star f/g_\chi$,
\be\label{eq:vchimin}
V_{\chi}^{\text{min}} \, \simeq \,   {\gamma_\epsilon \over 4} c_\chi g_\chi^2 \chi_0^4 \, = \,  {\gamma_\epsilon \over 4} c_\chi \frac{g_\star^4}{g_\chi^2} f^4\, .
\ee 
Equating with \eqref{eq:fcft}, the critical temperature follows as 
\be
\label{eq:CriticalTemperature}
T_c \, \simeq \, 2 \left(\frac{g_\star^2}{4 \pi g_\chi N}\right)^{1/2} (2 \gamma_\epsilon c_\chi)^{1/4}f \, = \, 2 (2 \gamma_\epsilon c_\chi)^{1/4}f \times 
\begin{cases}
N^{-3/4}\,, \quad  &\text{for}\,\, g_\chi  =  4\pi/\sqrt{N} \\
N^{-1/2}\,, \quad  & \text{for}\,\, g_\chi  =  4\pi/N \, .
\end{cases}
\ee
This in particular shows that the strength of the phase transition grows with increasing $N$. 
The thermal corrections from the SM fields which we have neglected so far increase the relative depth of the minimum in the deconfined phase at $\chi=0$ and thereby decrease the critical temperature.

\medskip

Instead of using the step-function behaviour described above, we will model a smooth transition of the free energy in the regime between $\chi \sim 0$ and $\chi  \sim T/g_\chi$. It is reasonable to expect that some order parameter of this hot phase, which characterises the size of the strong-sector condensate and the breaking of electroweak symmetry and which we will also call $\chi$, experiences a continuous evolution from $\chi\sim 0$ to $\chi\sim T/g_\chi$, where it smoothly transits into the dilaton of our EFT. This assumption allows us to introduce the Higgs as a variable which defines the direction of the $\chi$ condensate in the $G$ space. Despite the fact that it is generally difficult to argue for the existence of composite states in the hot plasma, as at some point they get ``dissolved'' to their elementary constituents, the PNGB Higgs is rather a collective excitation of the $\chi$ condensate with a given quantum number and hence can be considered in the hot phase as well, once we assume the existence of $\chi$ in that regime.
As in~\cite{Randall:2006py}, we will model the behaviour of $\chi$ in the regime $0\lesssim \chi \lesssim T/g_\chi$ by adding to the zero-temperature potential ((\ref{eq:feft}) extrapolated to $\chi<T/g_\chi$) the temperature correction from (besides the SM fields) $N^2$ CFT \emph{d.o.f.}~to which we assign a mass $g_\chi \chi$. The thus defined thermal correction reads 
\begin{equation}\label{eq:v1Tloop}
	\Delta V^{\text{1-loop}}_T \, = \, \sum_{\text{bosons}}\frac{n  T^4}{2\pi^2}J_b\left[\frac{m^2}{T^2}\right]-\sum_{\text{fermions}}\frac{n T^4}{2\pi^2}J_f\left[\frac{m^2}{T^2}\right] ,
\end{equation}
where the sum runs over both CFT and SM {\it d.o.f.}, $n$ is the number of {\it d.o.f.}~for each particle species and
the masses $m$ depend on $\chi$ and/or $h$. Furthermore, the functions $J_b$ and $J_f$ are given by
\begin{equation}
J_b[x] \, = \int_0^\infty dk~k^2 \log\left[1-e^{-\sqrt{k^2+x}}\right] \quad \text{and} \quad J_f[x]\, =\int_0^\infty dk~k^2 \log\left[1+e^{-\sqrt{k^2+x}}\right] .
\end{equation}
Since the difference in $J_b$ and $J_f$ is small,
we will for simplicity set $n=0$ for the fermionic CFT \emph{d.o.f.}~in Eq.~(\ref{eq:v1Tloop}). 
For the bosonic CFT {\it d.o.f.}, we then choose the normalization
\be\label{eq:fixf}
\sum_{\text{CFT bosons}} \hspace{-.2cm} n \, = \, \frac{45 N^2}{4} 
\ee
which ensures that the free energy \eqref{eq:fcft} in the deconfined phase at $\chi=0$ is reproduced. 
Moreover, the thermal correction from the CFT \emph{d.o.f.}~becomes strongly suppressed for ${\chi\gtrsim T/g_\chi}$, thereby resolving the previously assumed step function.
This produces a barrier in the potential, whose height increases with $N$. Corresponding to this, the critical temperature \eqref{eq:CriticalTemperature} decreases with $N$. The strength of the phase transition is thus strongly dependent on $N$. 
In order to ensure that electroweak baryogenesis is possible, the temperature at which the phase transition takes place needs to be somewhat below the electroweak scale. This also ensures that a valley along $h \sim \chi$ as discussed in the last section for the zero-temperature potential is not too strongly modified by the temperature corrections. From \eqref{eq:CriticalTemperature}, we see that such a low phase-transition temperature can always be achieved by taking $N$ sufficiently large.

\begin{figure}[t]
\centering
\includegraphics[width=6.5cm]{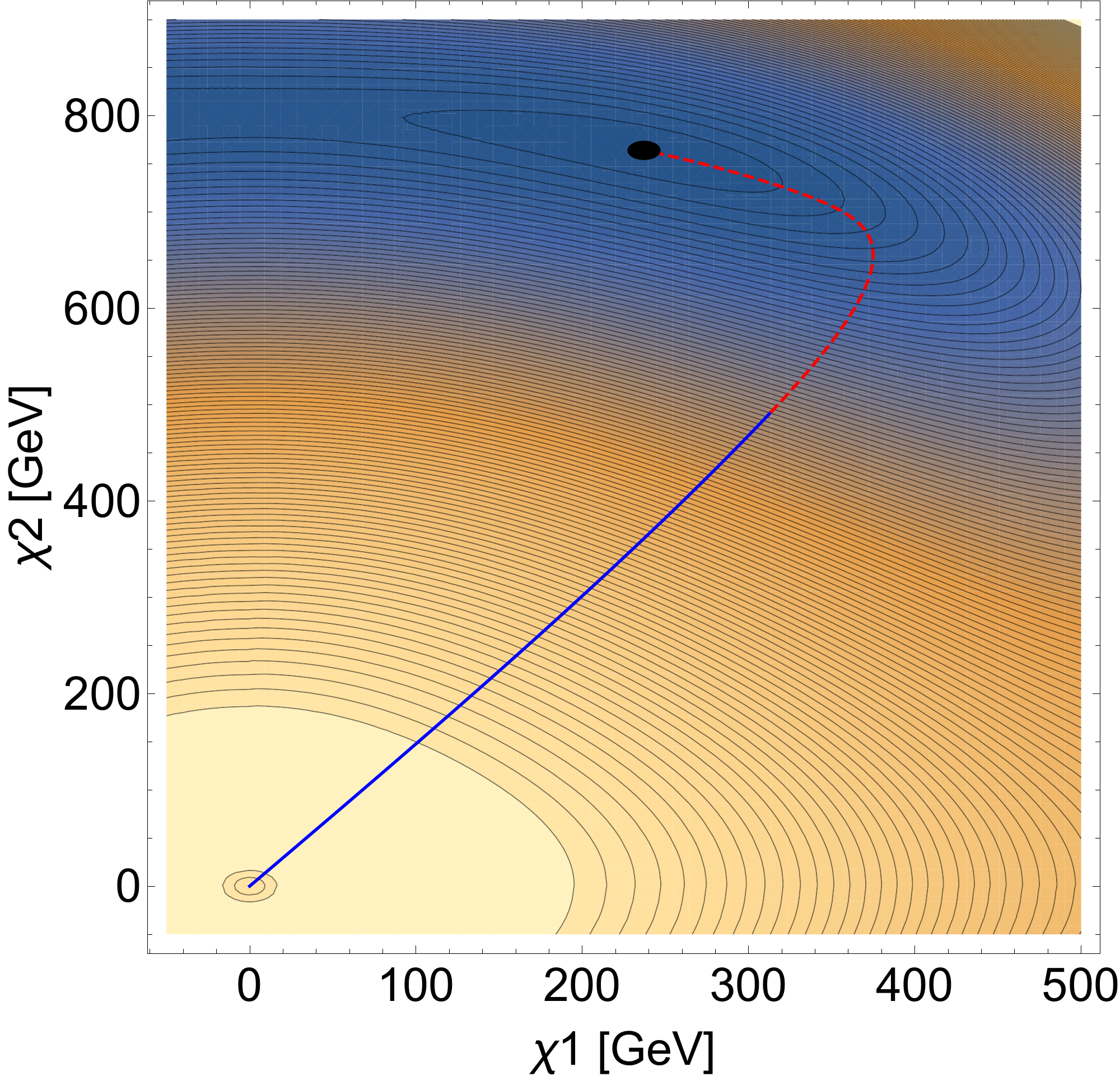}
\hspace{.3cm}
\centering\includegraphics[width=9cm]{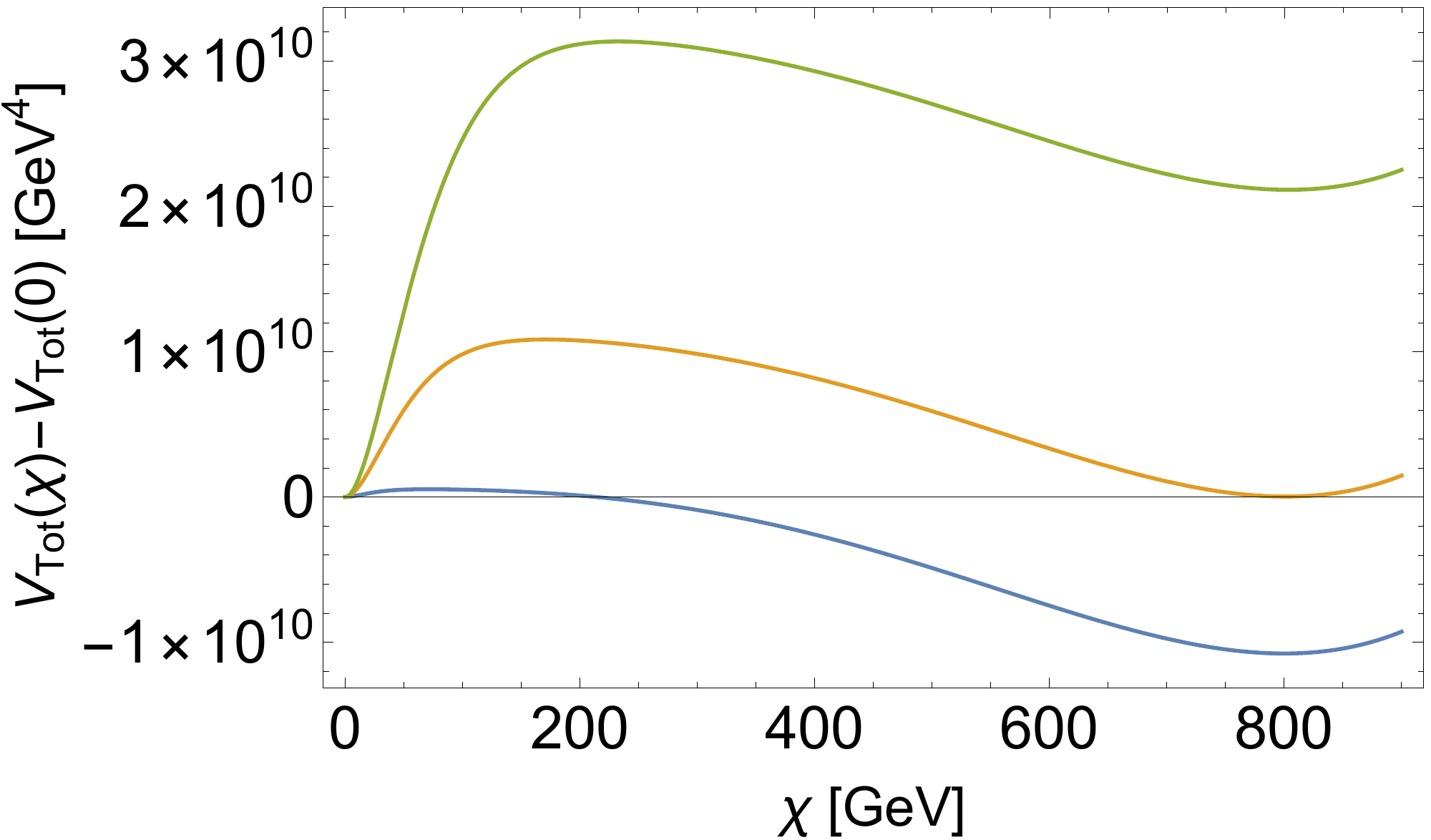}
\caption{\label{tunnellingPath2DFigure} \small \it Left: Potential as a function of $\chi_1$ and $\chi_2$ for a meson-like dilaton with mass ${m_\chi=500\,\text{GeV}}$ and $N=5$ and evaluated at the nucleation temperature $T_n\simeq 65.6\,$GeV. The other parameters are as in Table \ref{tab:bench}.
The solid blue line shows the tunnelling path to the release point, while the red dotted line indicates the subsequent rolling trajectory towards the minimum of the potential. Right: Potential at, from bottom to top, the nucleation temperature $T_n\simeq 65.6\,$GeV, the critical temperature $T_c\simeq 138.3\,$GeV and ${T=180\,\text{GeV}}$. The potential is plotted along the straight line parametrized by $\chi$ connecting the minimum at the origin with the second minimum at $\{\chi_1,\chi_2\} \simeq \{238 \, \text{GeV}, 764 \, \text{GeV}\}$, $\{175\, \text{GeV},782 \, \text{GeV}\}$ and ${\{ 0 \, \text{GeV}, 804 \, \text{GeV}\}}$, respectively.}
\end{figure}

We emphasize that while the procedure outlined above to estimate the potential in the regime between $\chi \sim 0$ and $\chi \sim T/g_\chi$ carries a certain amount of speculation, we expect that this does not affect the reliability of our main results (the dynamics of the phase transition and the induced CP asymmetry).  
Note also that we do not include daisy resummation in our analysis. It is a subdominant  effect in the region of validity of our description.

In summary, the potential that we will use for our study is
\be
\label{eq:totalpotential}
V_{\rm tot}[h,\chi] \, = \, V_{h}[h,\chi] \, + \,  V_{\chi}[\chi] \, + \, \Delta V^{\text{1-loop}}_T[h,\chi]\, ,
\ee
where $V_{h}$,  $V_{\chi}$ and $ \Delta V^{\text{1-loop}}_T$ are respectively given by Eqs.~(\ref{eq:Vhpotential}), (\ref{eq:Vchipotential}) and (\ref{eq:v1Tloop}). In the left panel of Fig.~\ref{tunnellingPath2DFigure}, we plot the combined potential $V_{\rm tot}$ as a function of the canonically normalized fields $\chi_1$ and $\chi_2$ (\emph{cf.}~Eq.~\eqref{chi1chi2def}) for a meson-like dilaton with mass ${m_\chi=500\,\text{GeV}}$ and $N=5$ and evaluated at the nucleation temperature $T_n\simeq 65.6\,$GeV (for which $S_3/T \approx 140$). The other parameters are as in Table \ref{tab:bench}. In the right panel, we plot the potential along straight lines connecting the two minima for three different temperatures (see caption for details). The aforementioned thermal barrier is clearly visible.

Before moving on with the analysis of the phase transition using this potential, we now discuss sources of CP violation arising in our construction which are relevant for electroweak baryogenesis.

\section{CP violation from varying Yukawa interactions}
\label{sec:cpv}
In electroweak baryogenesis, the baryon asymmetry is produced during charge transport in the vicinity of the Higgs bubble
walls that form during a first-order electroweak phase transition. 
In Ref.~\cite{Bruggisser:2017lhc}, it was shown that a new CP-violating source arises if the Yukawa couplings vary across the Higgs bubble wall and that this new source can allow for enough CP-violation to generate the observed baryon asymmetry. The kinetic equations incorporating the variation of the Yukawa
couplings across the Higgs bubble wall were derived and the induced CP-violating force was extracted. The resulting produced baryon asymmetry was predicted for a large set of parametrizations of the Yukawa variation. 
It was in particular shown that successful electroweak baryogenesis can
be realised from the variation of SM Yukawa couplings using only the top and charm. In the present work, we will apply these results using the precise Yukawa variation obtained in composite Higgs models.

The CP-violating source due to varying Yukawa couplings across the Higgs bubble wall which can enable electroweak baryogenesis reads~\cite{Bruggisser:2017lhc}
\be
\label{CPviolatingsource}
S_\text{\rm CPV} \sim \text{Im}[V^\dagger m^{\dagger \prime \prime} m V]_{ii} \, ,
\ee
where $m$ is the mass matrix of up- or down-type quarks (the leptons will not be important in the following), $V$ is the unitary matrix which diagonalizes $m^\dagger m$, \emph{i.e.}~$V^\dagger m^{\dagger} m V = {\text{diagonal}}$, the derivative is taken along the direction perpendicular to the bubble wall and the index $ii$ stands for the diagonal elements of the corresponding matrix. Using this expression one can single out two distinct ways of sourcing CP violation which we discuss in the following.

\subsection{CP violation with hierarchical quark mixings}

Let us assume that the elementary-composite mixings $y$ remain hierarchical and that only the top Yukawa is of order one in the entire interval $\chi \in (0, \chi_0)$. 
In this case the dominant contribution to $S_\text{\rm CPV}$ is expected to arise from the top mass:
\be
S_\text{\rm CPV} \, \sim \, \text{Im}[m_t^{\dagger \prime \prime}m_t] \, .
\ee
In order for this to be non-vanishing, the top mass needs to have a complex phase which varies along the bubble profile. 
The top mass as a function of the Higgs \emph{vev} is given by 
\be
m_t \, \simeq \, \frac{(y_{L})_{11} (y_{R})_{11}}{g_\star} \, h \,,
\ee
where we have set $g_\psi= g_\star$ as discussed in Sec.~\ref{sec:hssb} and assigned the index $j=1$ for the top for definiteness. 
In general we expect the mixings $y$ and the coupling $g_\psi$ to have constant phases so that obtaining a varying phase of the top mass is nontrivial. The subdominant corrections from the light quarks, on the other hand, are known to be insufficient to produce a large enough $S_\text{\rm CPV}$. We can consider two options to generate a varying phase of the top mass.

First,  we can assume that one of the top quark chiralities couples to two different composite operators. This produces the elementary-composite mixing~\cite{Bruggisser:2018mus}
\be\label{eq:irpc2}
f \bar q_t U (y_{t 1} \psi_{t1} + y_{t2} \psi_{t2})  \,.
\ee 
An overall phase change of $m_t$ can then be caused by a relative change of $y_{t1}$ versus $y_{t2}$, with constant but different phases. This can be sizeable if the two mixings have a comparable size.

A second possibility is that the phase changes because of an additional scalar field $S$ which enters the mixing,
\be\label{eq:irpc3}
y_t f \bar q_t U (1+ i S) \psi_t  \, ,
\ee 
and which undergoes a phase transition together with the Higgs. This option was proposed in~\cite{Espinosa:2011eu} but will not be considered further in this work as the nature of the electroweak phase transition in this case differs significantly from what we focus on. 

\subsection{CP violation with non-hierarchical quark mixings}

\begin{table}[t]
\centering
\begin{tabular}{V{2.5}c|c|c|c|cV{2.5}cV{2.5}}
\clineB{1-6}{2.5}
${\mathbb{I}}_{L}$\,&\,$\#(y_L)$\,&\,$g_\star^{-1}$\,&\,$\#(y_R)$\,&\,${\mathbb{I}}_{R}$\,&\, condition on $d_{ij}$ for $S_\text{\rm CPV}=0$\\
\clineB{1-6}{2.5}
$d$&2&$d$&$2$&$d$&always\\
&1& &$1$& &always\\
\hline
$d$&2&$\slashed d$&$2$&$d$&\;\;$d_{11}+d_{22}=d_{12}+d_{21}$\\
 &1& &$2$& & always \\
 &2& &$1$& & always \\
 &1& &$1$& & always \\
\hline
$\slashed d$&2&$d$&$2$&$d$& always \\
 &1& &$1$& & always \\
\hline
$\slashed d$&2&$\slashed d$&$2$&$d$& \;\;$d_{11}=d_{12}$,  $d_{22}=d_{21}$\\
 &1& &$2$& & always\\
 &2& &$1$& & $d_{11}=d_{21}$ \\
 &1& &$1$& & always \\
\hline
$\slashed d$&2&$d$&$2$&$\slashed d$&$d_{11}=d_{12}=d_{21} = d_{22}$\\
 &1& &$1$& & always \\
\hline
$\slashed d$&2&$\slashed d$&$2$&$\slashed d$&$d_{11}=d_{12}=d_{21} = d_{22}$\\
 &2& &$1$& &$d_{11}=d_{21}$\\
 &1& &$2$& &$d_{11}=d_{12}$\\
 &1& &$1$& &always\\
 \clineB{1-6}{2.5}
\end{tabular}
\caption{\it \small List of the possible choices for the components of the quark mass matrices written as ${m_{ij} \sim  
({\mathbb{I}}_L)\, (y_L) \,(g_\star^{-1}) \, (y_{R})^\dagger  \,({\mathbb{I}}_{R})}$, and conditions under which the resulting CP-violating source vanishes. These conditions are given in terms of the quantities $d_{ij}$ defined in Eq.~(\ref{eq:dmatrix}), where an index 1 (2) is for the top (charm). The symbols $d$ and $\slashed d$ stand respectively for a diagonal and anarchic structure of the corresponding matrices. Furthermore, $\#(y_{L/R})$ denotes the number of non-vanishing components of $(y_{L/R})_{ii}$.  We do not present the cases $\{{\mathbb{I}}_{L},g_\star^{-1},{\mathbb{I}}_{R}\} = \{d, d, \slashed d\},\{d, \slashed d, \slashed d\}$ which are analogous to respectively $\{\slashed d, d, d\},\{\slashed d, \slashed d, d\}$.
}
\label{tab:tab1}
\end{table}

We have seen that with the top quark alone, one necessarily needs a varying phase in the elementary-composite mixing to obtain non-vanishing CP violation from the new source. Such a varying phase requires non-trivial assumptions. It is therefore interesting to see if one can obtain CP violation with the help of other quarks, without invoking varying phases of the mixing parameters $y$. 
For the effects of other quarks to be relevant, their Yukawa couplings and thus their mixings need to grow when going from $\chi=\chi_0$ towards $\chi=0$. As we have seen in Sec.~\ref{sec:PartialCompositeness}, this is possible if the anomalous dimensions $\gamma$ of the corresponding operators which enter the RG equation \eqref{eq:yrun} are negative in this energy interval. 
Let us consider the minimal realization where only the charm mixings can become sizeable and grow to a comparable size as the top mixings somewhere in the interval $\chi \in (0, \chi_0)$. We can then restrict ourselves to the mass matrix for these two quarks. Using that elements $(y_{L,R})_{ij}$ of the mixing matrices with the same index $j$ have the same dependence on $\chi$ (as it results from the same anomalous dimension of the corresponding operator ${\cal O}_j$, \emph{cf.}~Eq.~\eqref{eq:uvirpcflav}), we can write $(y_{L,R})_{ij}= ({\mathbb{I}}_{L,R})_{ij} (y_{L,R})_{jj}$ with constant matrices $({\mathbb{I}}_{L,R})_{ij}$ and vectors $(y_{L,R})_{jj}$. The $(2 \times 2)$ mass matrix can then be written as
\be\label{eq:my2f}
m_{ij} \, \simeq \, (y_L)_{ik} \,(g_\star^{-1})_{kl} \, (y_R)_{lj}^\dagger \,h \, = \, 
({\mathbb{I}}_L)_{ik}\, (y_L)_{kk} \,(g_\star^{-1})_{kl} \, (y_{R})_{ll}^\dagger  \,({\mathbb{I}}_{R})_{lj} \,h \, . 
\ee 
The matrices ${\mathbb{I}}_{L,R}$ depend on the flavour structure of the model and can be either approximately diagonal or anarchic with all entries of order one (see \emph{e.g.}~\cite{Csaki:2008eh,Barbieri:2012tu, Matsedonskyi:2014iha}). Similarly, the matrix $g_\star^{-1}$ can be either approximately diagonal or anarchic.
Using this expression for the mass matrix, we can analyse under which conditions a sizeable $S_\text{\rm CPV}$ can arise. To this end, we make the simplifying assumption that some of the components of $(y_{L})_{ii}$ and $(y_{R})_{ii}$ vanish exactly. We then give the conditions under which $S_\text{\rm CPV}$ is non-vanishing. Having one of the two components of $(y_{L})_{ii}$ or $(y_{R})_{ii}$ vanish is an approximation to the situation where the corresponding charm mixing stays always much smaller than the top mixing and is thus negligible. Two non-vanishing components of $(y_{L})_{ii}$ or $(y_{R})_{ii}$, on the other hand, correspond to the respective charm mixing being of comparable size as the top mixing. The charm mixing then has to change sizeably with $\chi$, as it needs to grow relative to its small value at $\chi\sim \chi_0$.
It is convenient to give the conditions in terms of a $(2\times 2)$ matrix defined as 
\be
\label{eq:dmatrix}
d_{ij} \, = \, {(y_{Li} y_{Rj}^\star h)^{\prime\prime}}/{(y_{Li} y_{Rj}^\star h)}
\ee
for non-vanishing $y_{Li} y_{Rj}^\star$ and $d_{ij}=0$ otherwise. 
The results are presented in Table~\ref{tab:tab1}.
Here $d$ and $\slashed d$ stand for a matrix being diagonal and anarchic, respectively, and $\#(y_{L/R})$ denotes the number of non-vanishing components of $(y_{L/R})_{ii}$. 
From the information in the table we conclude that symmetry patterns which forbid significant off-diagonal flavour mixing in ${\mathbb{I}}_{L,R}$ and $g_\star^{-1}$ would not allow for sizeable $S_{\rm CPV}$ at all. All the other patterns require at least either ($y_{L})_{ii}$ or $(y_{R})_{ii}$ to have two sizeable components. A minimal scenario to obtain large CP violation, corresponding to $\{\#(y_L),\#(y_R)\} = \{2,1\}$ or $\{1,2\}$ in the table, then requires one charm mixing to grow to a comparable size as the top mixings, while the other charm mixing can stay small. Note that in this case the charm mass still remains significantly below the top mass. The required flavour patterns for this minimal scenario are $\{{\mathbb{I}}_{L},g_\star^{-1},{\mathbb{I}}_{R}\} = \{\slashed d, \slashed d, d\}$, $\{d, \slashed d, \slashed d\}$ or $\{\slashed d, \slashed d, \slashed d\}$.
The conditions to obtain large CP violation can also be satisfied in less minimal scenarios, with all four mixings being sizeable, and/or two and more mixings changing with $\chi$.

We have identified the conditions for creating non-vanishing CP violation from the running mixings. For concreteness, in the following part of this paper we will concentrate on the minimal option where only one charm mixing reaches the size of the top mixings when the dilaton $\emph{vev}$ is sent to zero, $\chi \rightarrow 0$. This can naturally happen if the operator $\mathcal{O}_i$ from the strong sector to which the elementary charm couples has a negative anomalous dimension $\gamma_i$ at energy scales somewhere below $\sim \chi_0$ (\emph{cf.}~Eq.~\eqref{eq:solmixing}). On the other hand, in order to reproduce the small charm mixings today, starting from order-one values in the far UV, the anomalous dimension of this operator at energies far above $\sim \chi_0$ should be positive.
This means that the anomalous dimension needs to become energy-dependent and in particular change its sign.
In \cite{vonHarling:2016vhf}, this was achieved in a dual description based on a Randall-Sundrum model \cite{Randall:1999ee}. To this end, the scalar field which stabilizes the extra dimension in the Goldberger-Wise mechanism \cite{Goldberger:1999uk} was coupled to the bulk fermions to source their masses. Since the \emph{vev} of the Goldberger-Wise scalar changes along the extra dimension, so do then the masses for the bulk fermions. Via AdS/CFT, this is dual to anomalous dimensions of the operators in partial compositeness which are energy-dependent (and which can in particular change their signs). 
In this paper, we will only assume that 
the anomalous dimensions become energy-dependent by virtue of the sources which explicitly break the conformal invariance and will otherwise be agnostic about the details. We will then assume that an anomalous dimension associated with the charm changes its sign somewhere above $\sim \chi_0$ and stays approximately constant at lower energies.

An advantage of varying charm mixings is that we do not need to rely on variations of the top mixings which have a certain subtlety. Recall that in Sec.~\ref{sec:hssb} we have made the simplifying identification $g_\psi = g_\star$, despite the fact that certain explicit models prefer $g_\psi<g_\star$ for the top. The price to pay for this assumption is that after fixing the top Yukawa with $\lambda_t\sim y_L[f] y_R[f]/g_\star$ rather than with $y_L[f] y_R[f]/g_\psi$, we obtain a larger $y_L[f] y_R[f]$ than is actually needed to obtain the right top mass. 
Also, the detuned part of the scalar potential proportional to $y_L^2, y_R^2$ will then be larger than it should be. Therefore the realistic scalar potential would differ from the one that we use by a less pronounced $y$-dependence, and also a smaller tuning (which was the original reason for taking $g_\psi < g_\star$). 
In principle this mismatch can be somewhat weaker than we have just shown. For instance if only $y_L$ varies with $\chi$, then $y_R$ has no effect on the dynamics. So we can implicitly assume that it is redefined to a smaller value to decrease $y_L[f] y_R[f]$. This, however has limits as $y_R$ can not be made smaller than 1, as otherwise the top mass can not be reproduced. 
To conclude, the assumption that $g_\psi = g_\star$ decreases the accuracy of our description in what concerns the $y$-dependence of the potential in the case of varying top mixings. 
This does not happen for varying charm mixings, as the charm partners exhibit no preference for a significant $g_\psi/g_\star$ separation.

\section{Numerical study}
\label{sec:res}

With all the necessary knowledge about the potential we are now in the position to examine numerically the dynamics of the confinement and electroweak phase transitions. One of the main features of the phase transition that we would like to test is whether it is sufficiently strong for successful electroweak baryogenesis, \emph{i.e.}~whether
\be
\label{eq:PTStrengthCriterion}
h[T_n]/T_n \, \gtrsim \, 1,
\ee
where $h[T_n]$ is the  Higgs \emph{vev} after the phase transition and $T_n$ is the temperature at which the transition occurs -- the nucleation temperature. 
As was discussed in the previous sections, the scalar potential in the presence of varying mixings $y[\chi]$ possesses non-trivial features, such as valleys. These may significantly influence the tunnelling trajectory in the two-dimensional field space and make it deviate from the line connecting the two minima.  
We therefore have to compute the tunnelling path in the two-dimensional field space. This is a problem which is hard to attack analytically and which we therefore have to solve numerically.

In the following subsections, we first summarize the parameters that determine the Higgs-dilaton potential (\ref{eq:totalpotential}) and discuss our choices for these parameters in the numerical study. We then give details about the numerical study and present our results for the features of the phase transition and the produced baryon asymmetry.

\subsection{Parameter space}
\label{sec:ParameterSpace}

We now summarize the effects of the different parameters on the potential and on the phase transition. Some of the parameters  are redundant and we mention explicitly when this happens.

\paragraph{$\bf f$} is the Goldstone decay constant which controls the overall size of the Higgs potential today. It is related to the current dilaton \emph{vev} $\chi_0$ as $g_\star f = g_\chi \chi_0$. It determines the required tuning of the Higgs potential, of order $v^2/f^2$, and therefore can not be  too large. On the other hand, the experimental data constrains $v^2/f^2 \lesssim 0.1...0.2$. For our analysis, we choose $f=800$~GeV (corresponding to $v^2/f^2\simeq 0.1$). For this value, one typically needs to impose a flavour symmetry in order to satisfy constraints from flavour- and CP-violating processes. We discuss one suitable flavour symmetry which also allows for a non-vanishing CP-violating force in Sec.~\ref{sec:numresBAU}.

\paragraph{$\bf g_\star, g_\chi$} are coupling constants of the strong sector. $g_\star f$ ($=g_\chi \chi_0$) sets the characteristic mass scale of the composite resonances, which are now excluded below 1~TeV~\cite{Matsedonskyi:2015dns,Matsedonskyi:2014lla,DeSimone:2012fs,Panico:2017vlk}. In our scan, we fix these couplings to the large-$N$ estimates $g_\star=4 \pi/\sqrt{N}$ and ${g_\chi=4 \pi/\sqrt{N}}$ (meson-like dilaton) or $g_\chi=4 \pi/N$ (glueball-like dilaton).

\paragraph{$\bf N$} is the number of colors of the underlying $SU(N)$ gauge theory which we use as a reference UV completion. It influences the potential through the coupling constants $g_\star, g_\chi$, and also determines the depth of the minimum at $\chi=0$. As the couplings crucially determine the potential, 
we perform a scan over a large range of values for $N$.

\paragraph{$\bf y_i$} are the mixings of the elementary fermions with their composite partners. In the numerical study of the phase transition, we focus on the case where only the mixing of one charm chirality (either left- or right-handed) changes significantly below the scale $f$. Let us denote this mixing as $y$. As we have discussed in Sec.~\ref{sec:cpv}, this case allows for a large CP-violating force if $y$ grows for $\chi \rightarrow 0$. The running of $y$ depends on the two parameters $\gamma_y$ and $c_y$ in the RG equation \eqref{eq:yrun} 
and on an integration constant which we trade for the mixing today $y[\chi_0]$. 
The more negative $\gamma_y$ is, the faster does the mixing grow when the dilaton \emph{vev} changes from $\chi=\chi_0$ to $0$, which results in a larger CP-violating force~\cite{Bruggisser:2017lhc}.
On the other hand, $\gamma_y$ is constrained to be $\gamma_y>-1$ and can therefore not be arbitrarily large and negative. We choose $\gamma_y=-0.735$.
Furthermore, we set $y[\chi_0]=\sqrt{\lambda_c g_\star}$ with $\lambda_c$ being the charm Yukawa and $c_y=1.5$. This then gives $y[0]=0.7 g_\star$ for the mixing in the unbroken phase, which does not exceed $g_\star$ so that the expansion in $y/g_\star$ in the RG equation \eqref{eq:yrun} remains applicable. 
The growth rate of $y$ also determines how quickly the Higgs potential becomes detuned for $\chi$ away from $\chi_0$ and therefore how strongly the tunnelling trajectory is shifted to the valley $h \sim \chi$. 
On the other hand, this does typically not affect the strength $h[T_n]/T_n$ of the phase transition as the Higgs potential is much smaller than the pure dilaton potential in most regions of parameter space.

\begin{figure}[t]
\centering
\includegraphics[width=14cm]{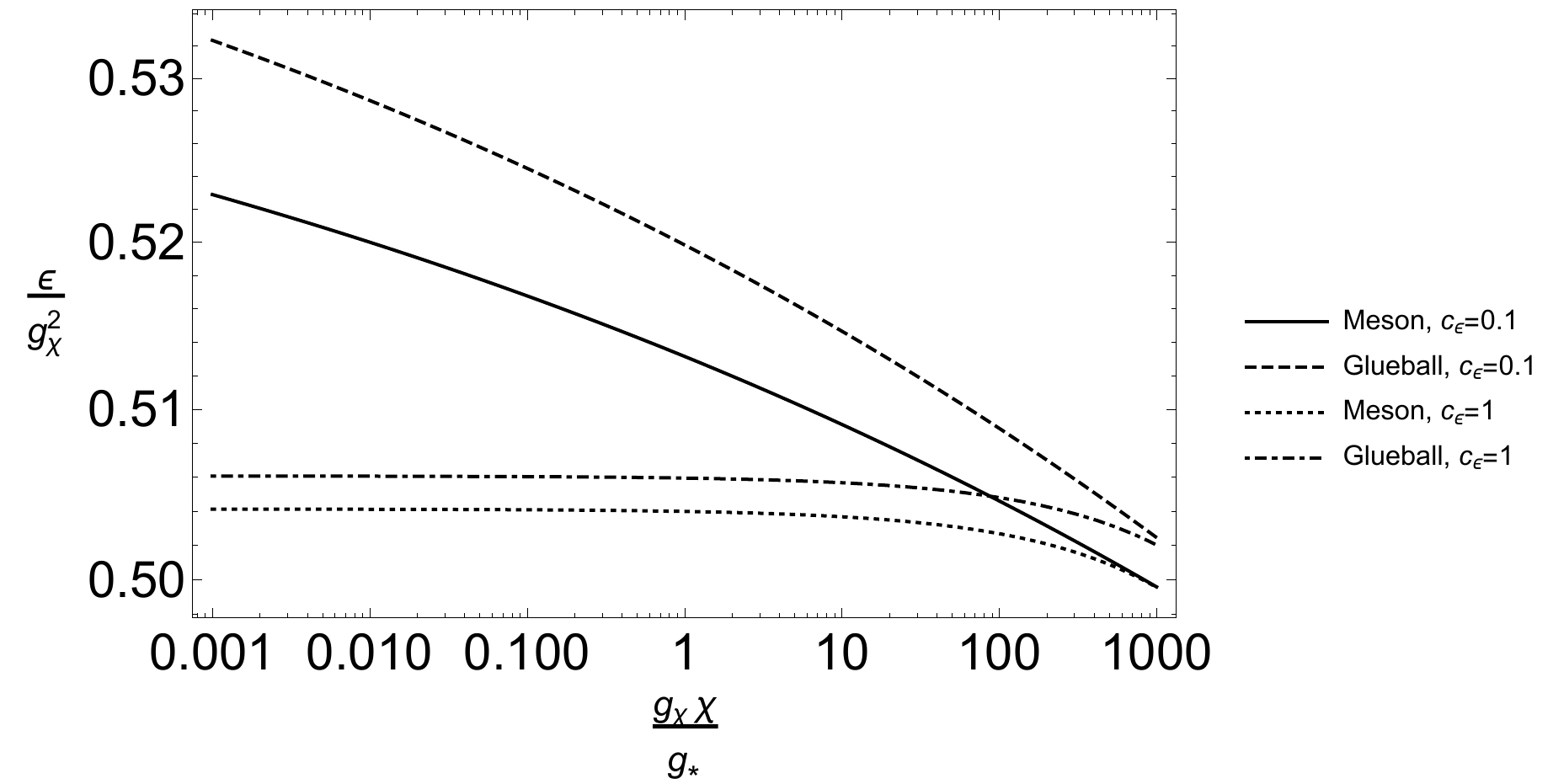}
\caption{\small \it The function $\epsilon[\chi]$ for a glueball-like and a meson-like dilaton with ${m_\chi=500\GeV}$, $N=5$ and different values of $c_\epsilon$.  The variable $g_\chi \chi/g_\star$ on the x-axis was chosen to ensure that the functions for the meson-like and glueball-like dilaton fit on the same plot.}
\label{fig:epsPlot}
\end{figure}

\paragraph{$\bf m_\chi$} -- the dilaton mass -- greatly influences the shape of the dilaton part of the potential. A smaller dilaton mass makes the potential more shallow and can hence delay the phase transition. If we choose a too small dilaton mass, on the other hand, the mass mixing between the dilaton and the Higgs becomes very large and hence problematic. The currently available analyses of experimental data place the bound $m_\chi\gtrsim 100$~GeV, see the discussion in Sec.~\ref{ExperimentalTests}. Another effect of a small dilaton mass is that the part of the potential which depends on the mixings becomes more important relative to the pure dilaton part. 

\paragraph{$\bf \epsilon$} is the coupling of the operator which gives the main contribution to the explicit breaking of conformal symmetry, and induces the minimum of the dilaton potential. 
The evolution of $\epsilon$ is determined by the two parameters $\gamma_\epsilon$ and $c_\epsilon$ in the RG equation \eqref{epsilonRGeq} and by an integration constant. We trade $\gamma_\epsilon$ and the integration constant for the dilaton mass $m_\chi$ and its \emph{vev} today $\chi_0$ ($=g_\star f/g_\chi$). For small values of the dilaton mass, the terms involving the varying mixings can sizeably affect the global minimum. In order to correct for this, we solve numerically for the values of $\gamma_\epsilon$ and the integration constant which reproduce the correct Higgs and dilaton masses and \emph{vev}s.
The parameter $c_\epsilon$ influences the size of $\gamma_\epsilon$ for a given $m_\chi$. We choose $c_\epsilon=0.1$ which ensures that $|\gamma_\epsilon|$ is typically much smaller than $1$ for the range of $m_\chi$ of interest. 
In Fig.~\ref{fig:epsPlot}, we plot $\epsilon$ as a function of $\chi$ for different parameter choices. We find that $\epsilon/g_\chi^2$ almost always stays smaller than 1 for the whole range of $\chi$ of interest which ensures perturbativity. The only exception is the corner of the largest $m_\chi$ and $N$ that we consider later (\emph{cf.}~fig.~\ref{fig:MNplots}) where $\epsilon/g_\chi^2$ becomes sightly larger than 1 near $\chi=0$. Given that the range in $\chi$ for which this happens is very small, we expect that this does not substantially affect our results.

\paragraph{$\bf c_\chi$} controls the scale-invariant quartic term in the dilaton potential. A small value makes it easier for $\epsilon$ to remain below $g_\chi^2$ to preserve perturbativity. We set $c_\chi=0.5$ for our study. 

\paragraph{$\bf c_{\chi y}$} controls the $y$-dependent contribution to the dilaton potential. This contribution in general has a small effect on the overall potential at large dilaton masses, whereas its effect can become comparable to the $y$-independent part for small dilaton masses. %or small $N$. 
Depending on the sign of $c_{\chi y}$, the $y$-dependent contribution can either facilitate or delay the phase transition. We set $c_{\chi y}=-1$.

\begin{table}[t]
\centering
%\begin{tabular}{V{2.5}c|c|c|c|c|c|c|cV{2.5}cV{2.5}}
%\clineB{1-10}{2.5}
\begin{tabular}{|c|c|c|c|c|c|c| c| c| c| c| c| c| c|}
\hline
 $c_\epsilon$ & $c_\chi$  & $c_{\chi y}$ & $c_\alpha$ & $c_\beta$ & $p_\beta$ & $\gamma_y$ & $c_y$ & $y[\chi_0]$ & $f$ \\
\hline
0.1 & 0.5& -1 & -1 & 1 & 0 & -0.735 & 1.5 & $\sqrt{\lambda_c g_\star}$ & 800 GeV \\
\hline
\end{tabular}
\caption{\it \small Definition of our benchmark point. Here $y$ refers to the one charm mixing (either for the left- or right-handed chirality) that is chosen to vary and $\lambda_c$ is the charm Yukawa.}
\label{tab:bench}
\end{table}

\paragraph{$\bf c_\alpha, c_\beta$} are order-one parameters, which influence the valley in the potential. These two parameters therefore can have an influence on the strength of the phase transition as well as on the direction of the tunnelling. However, the Higgs-dependent part of the potential is rather small compared to the pure dilaton part. As we will see in Sec.~\ref{sec:numresPT}, this means that the effect of $c_\alpha$ and $c_\beta$ on the strength of the phase transition is almost negligible in most regions of parameter space. We choose $c_\alpha= -1$ and $c_\beta=1$ for our study which for $p_\beta=0$ results in a valley along $\theta=\pi/4$.

\paragraph{$\bf p_\beta$} can take the values $0$ or $2$ depending on the embedding of the fermions in the symmetry group. The effect of $p_\beta$ on the potential is to further suppress the influence of $c_\beta$ if $p_\beta=2$ as compared to $p_\beta=0$. For simplicity, we fix $p_\beta=0$.

\medskip
\medskip

To summarize, most of the parameters of our model are dimensionless coefficients, whose absolute values are expected to be of order one. Many of their signs are constrained if we demand successful electroweak baryogenesis. The couplings $g_\star$ and $g_\chi$ are determined by the number of colors $N$ of the underlying gauge theory and whether the dilaton is meson-like or glueball-like. The current values of the running couplings $y$ and $\epsilon$, on the other hand, can be (partly) fixed from the quark masses and the condensation scale $f$. Furthermore, $\gamma_y$ is constrained by the requirement of sufficient CP violation during the phase transition. This leaves $\gamma_\epsilon$ and $N$ as free parameters, the first of which we trade for the dilaton mass $m_\chi$. In our numerical study, we then perform scans over $m_\chi$ and $N$ for both a meson-like and a glueball-like dilaton. 
In Table~\ref{tab:bench}, we summarize our choices for the remaining parameters which determine the Higgs-dilaton potential.

\subsection{Details of the numerical study}
\label{sec:num}

\begin{figure}[t]
\centering
\includegraphics[width=7.7cm]{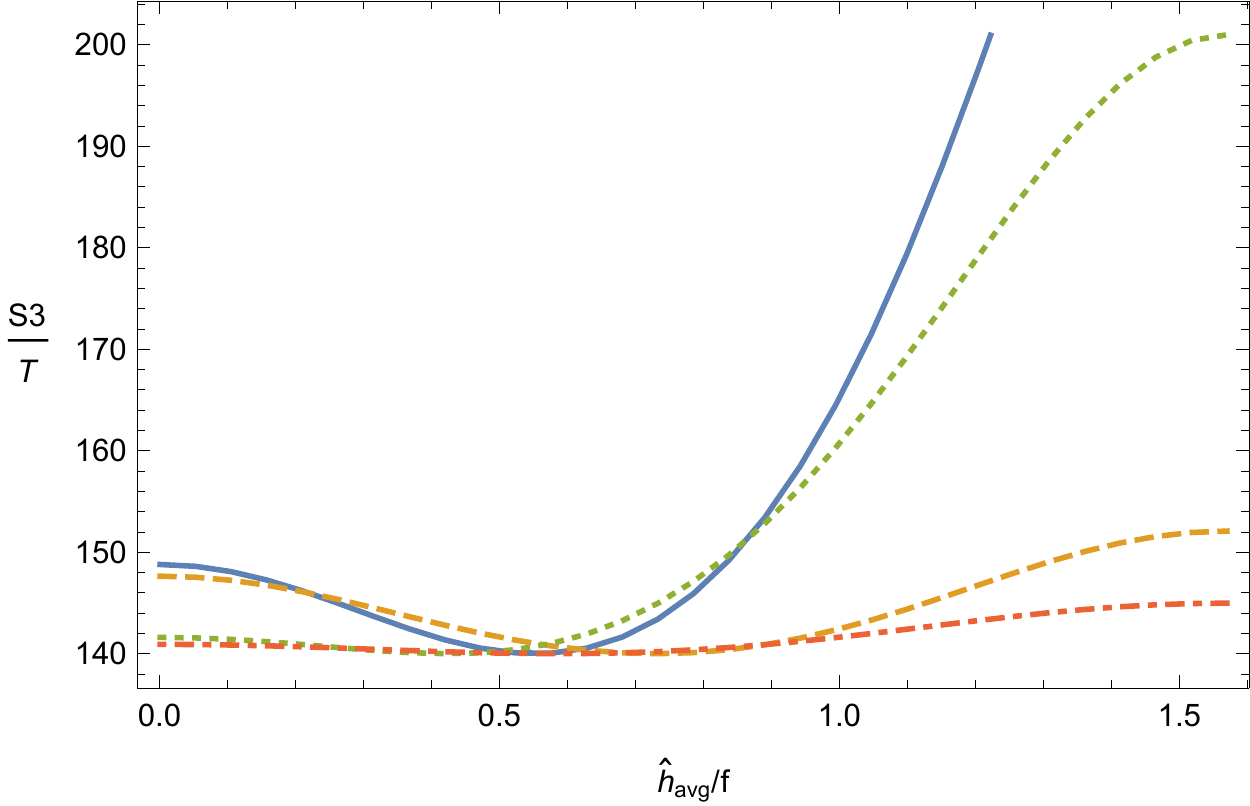}
\hspace{.3cm}
\includegraphics[width=7.7cm]{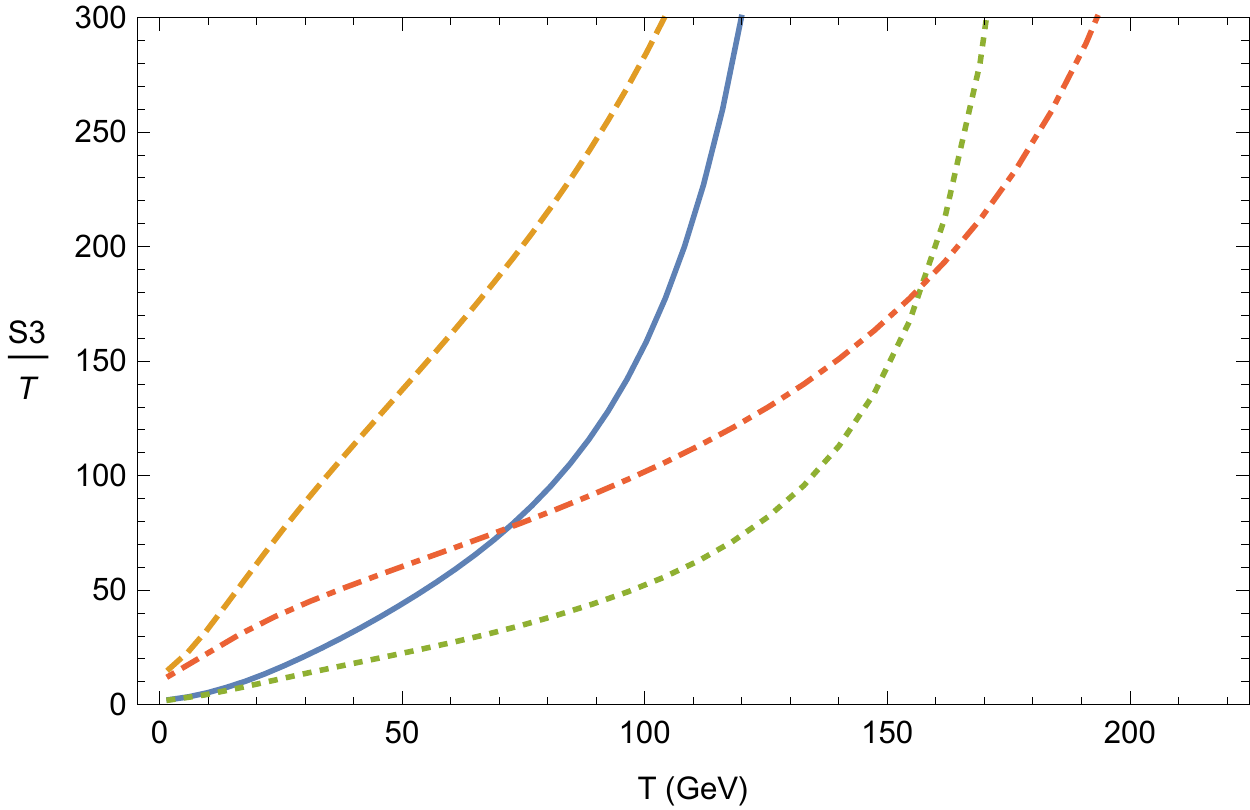}
\caption{\small \it The action $S_3/T$ of $O(3)$-symmetric bubbles for a meson-like dilaton with $m_\chi=600\,\text{GeV}$, $N=4$ (blue solid) and $m_\chi=1000\,\text{GeV}$, $N=4$ (green dotted) and a glueball-like dilaton with $m_\chi=600\,\text{GeV}$, $N=4$ (orange dashed) and $m_\chi=1000\,\text{GeV}$, $N=4$ (red dot-dashed). Left: At the nucleation temperature and as a function of $\hat{h}_{\rm avg}$ parametrizing the tunnelling path. 
%Notice that for the last case, $S_3/T$ depends only very weakly on $\hat{h}_{\rm avg}$. The minimum for this case is at $\hat{h}_{\rm avg}\approx 0$. 
Right: At the minimum with respect to 
$\hat{h}_{\rm avg}$ and as a function of the temperature.
The point of intersection with the critical action $S_3/T=140$ gives the nucleation temperature.}
\label{fig:procedure}
\end{figure}

We first discuss the calculation of the phase-transition properties. 
The formalism for calculating the tunnelling path and action in potentials for one and more fields is reviewed in Appendix~\ref{ap:vacuumtunnelling}. 
We have found that the tunnelling path for the canonically normalized fields $\chi_1$ and $\chi_2$ is typically well approximated by a straight line (see \emph{e.g.}~Fig.~\ref{tunnellingPath2DFigure}). Since the calculation of the exact tunnelling path in a two-dimensional potential is computationally quite intensive, we therefore only consider tunnelling along straight lines. These emanate from the false minimum at $\chi_1=\chi_2=0$. Using Eq.~\eqref{chi1chi2def}, we parametrize these straight lines in terms of the angular variable $\theta$ or $\hat{h}_{\rm avg} \equiv \theta f$ and the radial variable $\chi$.  
We calculate the corresponding tunnelling actions for both $O(3)$- and $O(4)$-symmetric bubbles. In the left panel of Fig.~\ref{fig:procedure}, we plot the action of $O(3)$-symmetric bubbles as a function of $\hat{h}_{\rm avg}$ for a meson-like and a glueball-like dilaton, different values of $m_\chi$ and $N$ (see caption for details; $O(3)$-symmetric bubbles dominate for these choices, at least near the
nucleation temperature and for the tunneling direction which minimizes the action) and the other parameters as given in Table~\ref{tab:bench}. In our approximation of tunnelling along straight lines, the most likely tunnelling trajectory is then found by minimizing this function with respect to $\hat{h}_{\rm avg}$. For ease of comparison, we have set the temperature to the nucleation temperature for the different cases shown in the plot. For other temperatures, the most likely tunnelling trajectory can be found in the same way. In the right panel of Fig.~\ref{fig:procedure}, we plot the action for the resulting trajectories as a function of the temperature for the same cases as in the left panel. The nucleation temperature is then found as the point where this function equals the critical action $S_3/T=140$. 
Finally, the trajectory is continued from the point of exit from the tunnelling, assuming that the fields subsequently follow the direction of steepest descent towards the global minimum of the potential (\emph{cf.}~Fig.~\ref{tunnellingPath2DFigure}). 
In our scan, we repeat the above steps for each analysed point in the parameter space.

\begin{figure}[t]
\centering
\includegraphics[width=14cm]{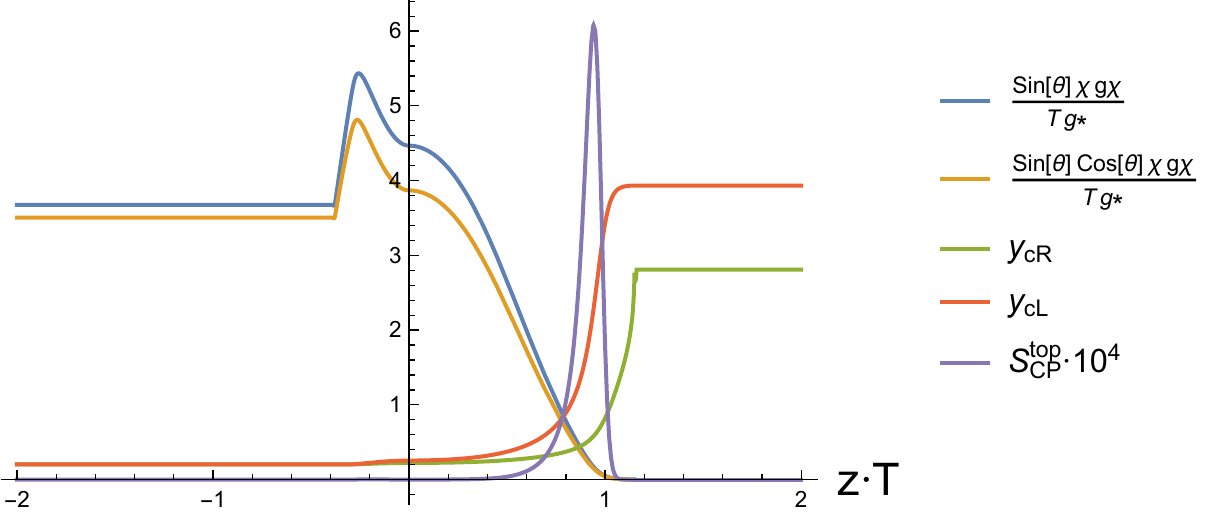}
%\vspace{-.03cm}
\caption{\small \it Typical profiles (here for a meson-like dilaton with mass ${m_\chi=500\,\text{GeV}}$ and $N=5$) along the direction $z$ perpendicular to the bubble wall of various dimensionless functions which are relevant for the produced baryon asymmetry assuming the flavour structure discussed in Sec.~\ref{sec:numresBAU}: In blue, the quantity which sets the gauge boson masses and the fermion masses in non-minimal composite Higgs models normalized to the nucleation temperature $T_n$ (see Sec.~\ref{sec:prelimAnalysis}); in yellow, the quantity which sets the fermion masses in minimal composite Higgs models normalized to the nucleation temperature; in green and red, the two charm mixings; in purple, the CP-violating force normalized to the nucleation temperature (see Sec.~\ref{sec:numresBAU}).}
\label{fig:baryoPlot}
\end{figure}

We next discuss the calculation of the baryon asymmetry. For more details see \cite{Bruggisser:2017lhc}. 
After a critical bubble is formed during the phase transition, it expands in the hot plasma and reaches a steady-state velocity when the pressure due to the latent-heat release balances the friction from the particles scattering off the bubble wall. During this bubble expansion, the baryon asymmetry is created. The profile of the steady-state bubble may in principle differ from the one of the critical bubble. However, we expect the two profiles to be similar and set them equal for our calculation.
On the other hand, a source of uncertainty comes from the wall velocity. Computing its precise values requires a dedicated analysis which is beyond the scope of this paper. We expect that the wall velocity in particular depends on the amount of supercooling of the phase transition, \emph{i.e.}~the ratio $T_c/T_n$ of the critical to the nucleation temperature. If there is a lot of supercooling and the temperature and density of the plasma surrounding the bubble is correspondingly very low, the friction of the particles in the plasma is not sufficient to significantly slow down the bubble wall. It can then accelerate to supersonic speeds for which electroweak baryogenesis can no longer work. However, we are mostly interested in regions of parameter space where $T_c/T_n$ is at most a few (\emph{cf.}~Fig.~\ref{fig:MNplots}) in which case we expect that the bubble wall is at subsonic speeds. In the following, we simply fix the wall velocity to $v_{\text{wall}}=10^{-1}$. We have checked 
that our results do not depend very sensitively on its exact value.
Note, on the other hand, that for a very strong phase transition, $T_c/T_n \gg 1$, there can be a few e-folds of inflation resulting from the large amount of supercooling, and the relevant baryogenesis mechanism may instead be {\it cold} baryogenesis rather than the standard charge transport mechanism \cite{Servant:2014bla}.

For illustration, in Fig.~\ref{fig:baryoPlot} we show typical profiles along the bubble wall of functions which are relevant for the produced baryon asymmetry. 
As can be seen, the bubble wall in the models that we study tends to be rather thin compared to the inverse nucleation temperature $T_n^{-1}$. This has two important consequences. 
Firstly, the baryon asymmetry roughly scales as $\eta_B\sim L_w^{-1}$, where $L_w$ is the bubble wall width. It thus grows when the bubble wall becomes thinner. On the other hand, the derivative expansion used in~\cite{Bruggisser:2017lhc} to derive the diffusion network and the CP-violating source is valid only for sufficiently thick bubble walls ($L_w\cdot T_n\gtrsim 1$). The profiles that we find are close to the limit of validity of this expansion, with larger values of $\hat{h}_\text{avg}/f$ becoming less reliable. 
Overall, the values of the baryon asymmetry presented in the following should only be taken as an indication of the order of magnitude.

\subsection{Results: Phase transition}
\label{sec:numresPT}

\begin{figure}[t]
\centering
\includegraphics[width=7.cm]{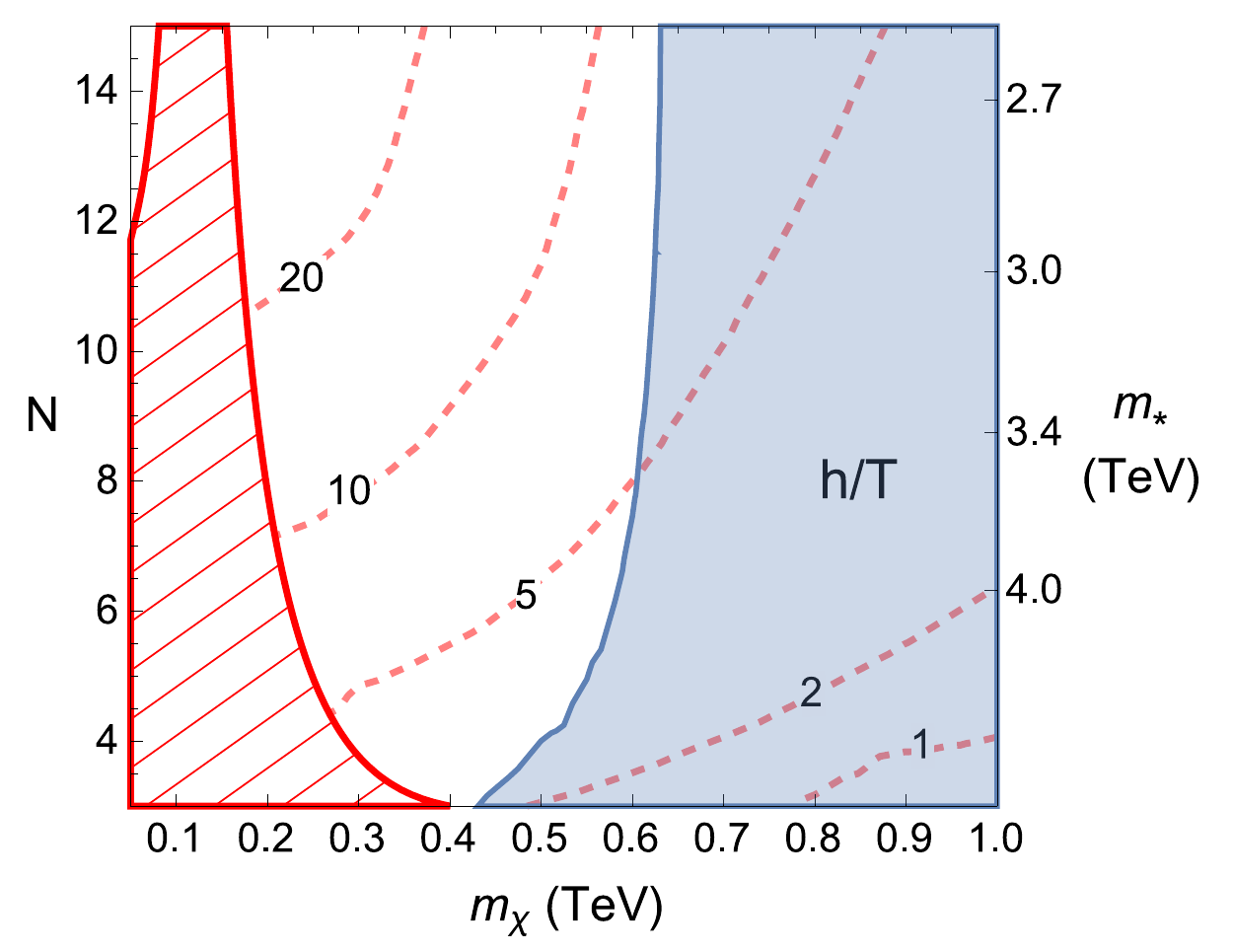}
\hspace{0.3cm}
\includegraphics[width=7.cm]{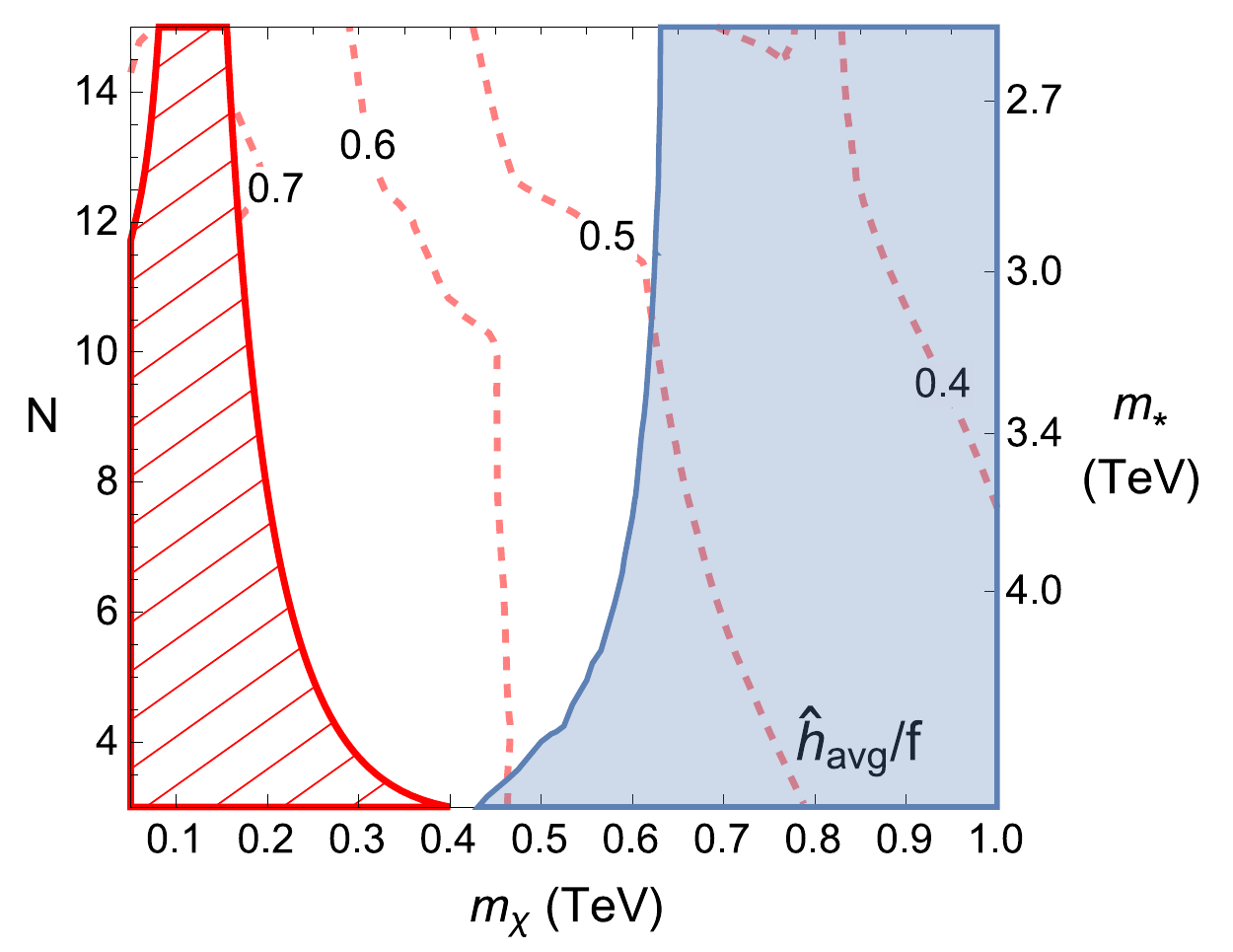} \\ \vspace{.4cm}
\includegraphics[width=7.cm]{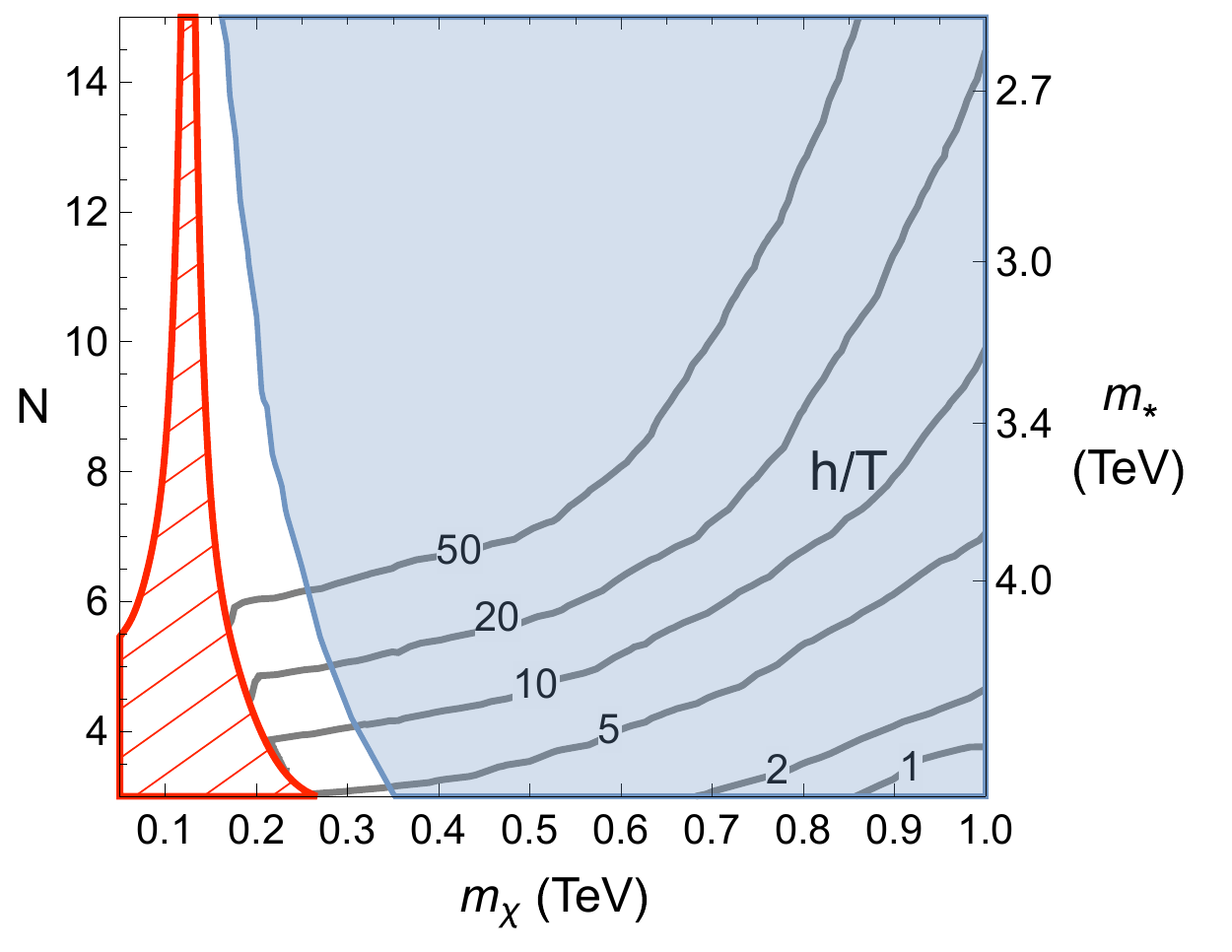}
\hspace{0.3cm}
\includegraphics[width=7.cm]{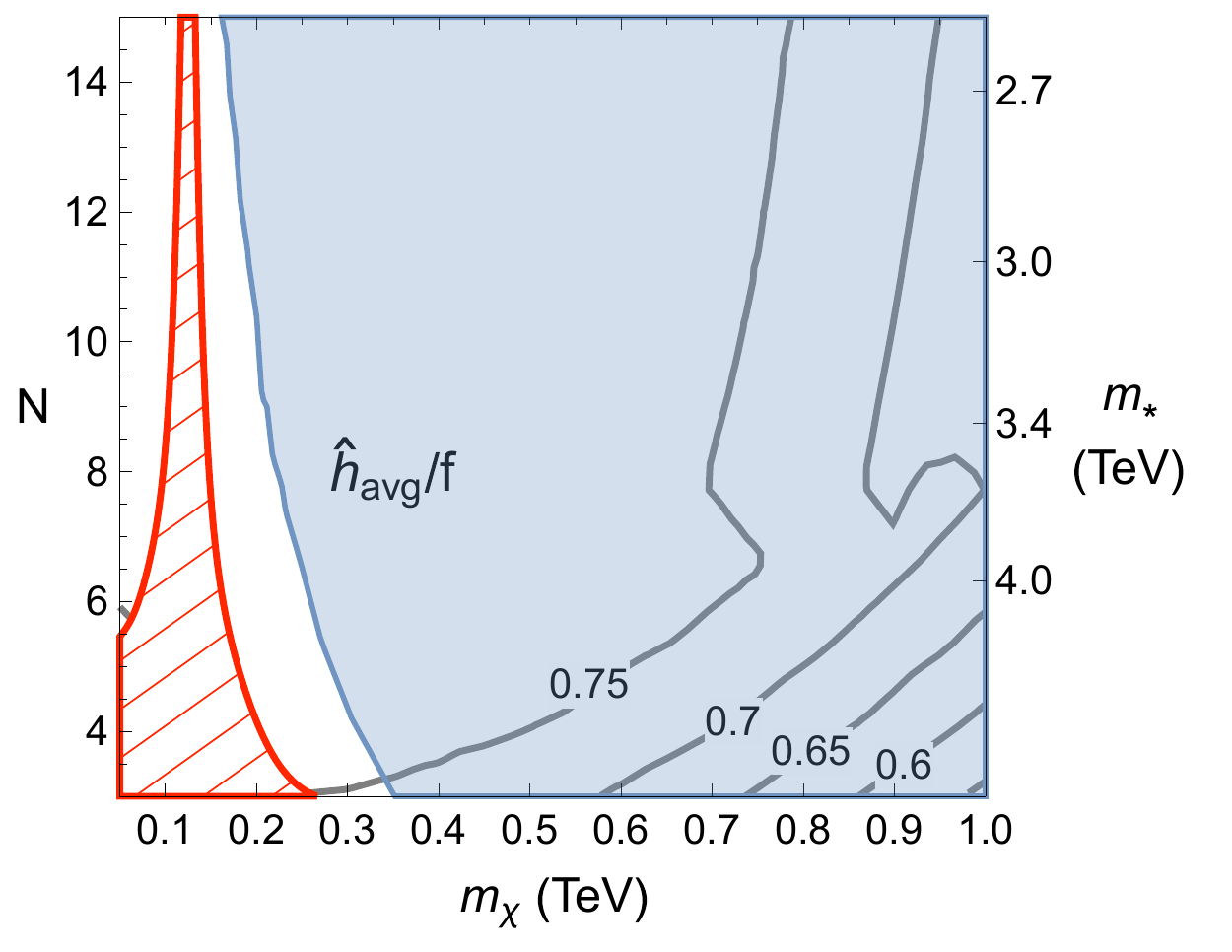}
\caption{\small \it{Results of our numerical study for a meson-like dilaton (upper panels; red dashed) and a glueball-like dilaton (lower panels; black solid). In the red dashed region, there is no phenomenologically viable electroweak minimum. The blue shaded region shows where $h[T_R]/T_R<1$, with $T_R$ being the reheating temperature, and where therefore the baryon asymmetry is washed out. Left panels: The strength $h[T_n]/T_n$ of the phase transition, with $T_n$ being the temperature where the phase transition completes, as a function of $m_\chi$ and $N$. Right panels: The average direction $\hat{h}_{\rm avg}$ of the tunnelling trajectory as a function of $m_\chi$ and $N$. We also show the cutoff $m_\star = g_\star f$ with $g_\star=4\pi/\sqrt N$, where the other composite states appear.}}
\label{fig:MNplots}
\end{figure}

\begin{figure}[t]
\centering
\includegraphics[width=7.8cm]{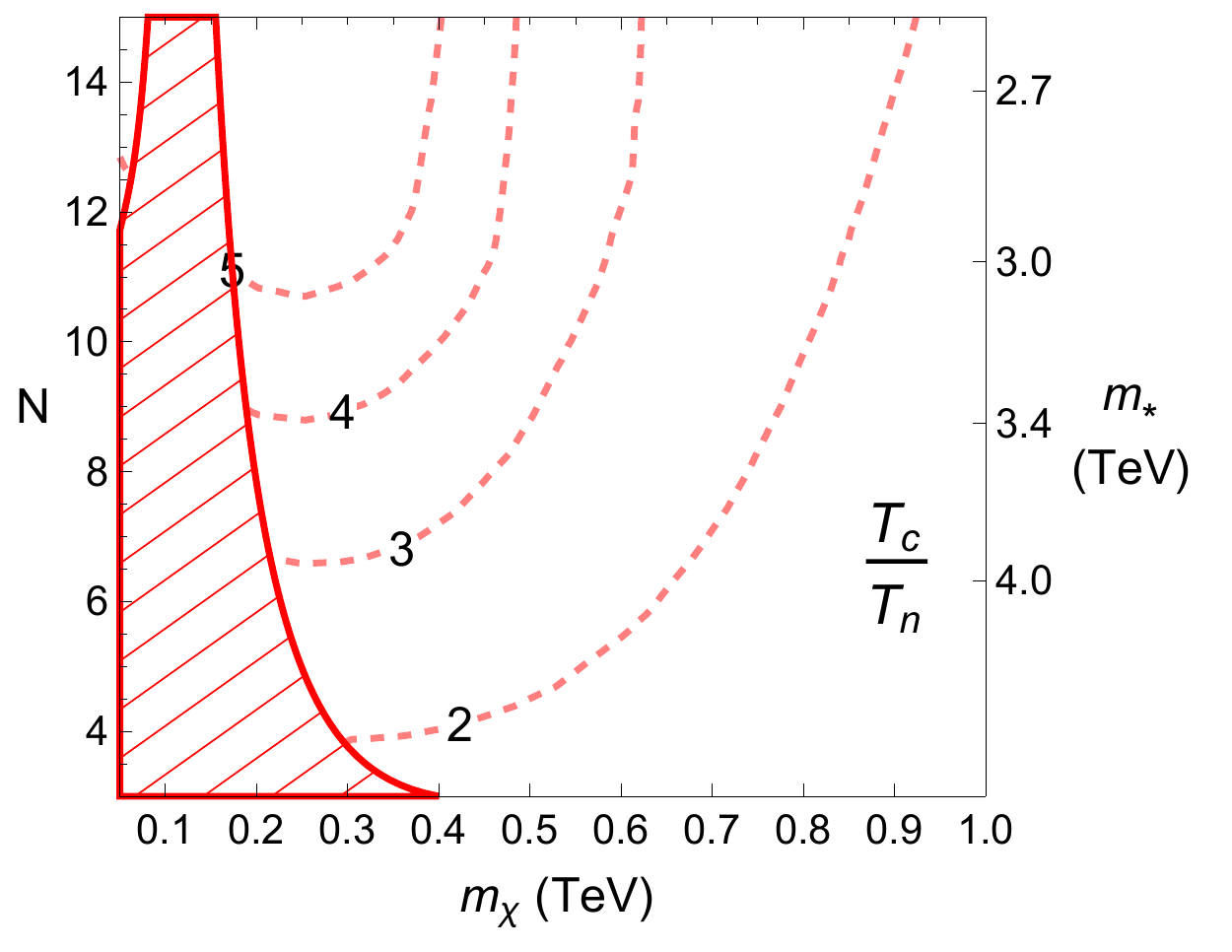}
\hspace{.2cm}
\includegraphics[width=7.8cm]{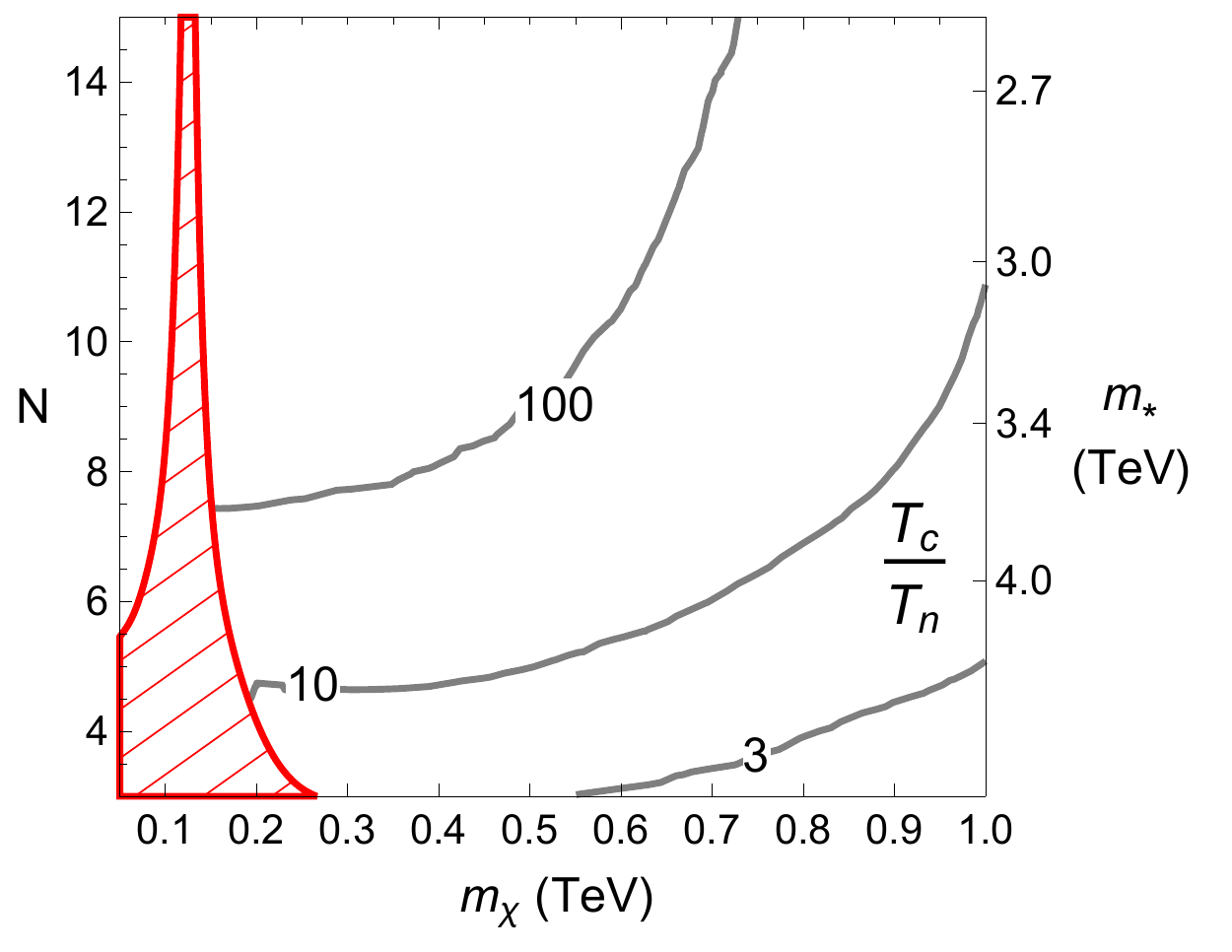}
\caption{\small \textit{The ratio $T_c/T_n$ of the critical to the nucleation temperature which measures the amount of supercooling of the phase transition for a meson-like dilaton (left panel; red dashed) and a glueball-like dilaton (right panel; black solid). In the red dashed region, there is no phenomenologically viable electroweak minimum.
%Right: The strength $h[T_n]/T_n$ of the phase transition for a glueball-like dilaton for the two cases where $f=0.8\TeV$ (black solid) and $f=2\TeV$ (black dot-dashed). For a better comparison, we report the strength as a function of the dimensionless ratios $m_\chi/f$ and $m_\star/f$. The reason for the former choice is that $\gamma_\epsilon$ is approximately determined by $m_\chi/f$ as follows from Eq.~\eqref{GammaEpsilonRel}. The black shaded area shows the excluded region for $f=2\TeV$, where the global minimum is shifted away from the one reproducing a correct Higgs \emph{vev}.
}}
\label{fig:MNCompareF}
\end{figure}

In Fig.~\ref{fig:MNplots}, we show the results of our numerical study for the phase-transition properties.
In the left panels, we plot the strength $h[T_n]/T_n$ of the phase transition as a function of $m_\chi$ and $N$ for a meson-like and a glueball-like dilaton. %
Notice that in both cases, the strength increases with smaller $m_\chi$ and larger $N$. This is due to two effects. Firstly, the critical temperature, where the phase transition becomes energetically possible, decreases with increasing $N$ as follows from Eq.~\eqref{eq:CriticalTemperature}. Similarly, one finds that $\gamma_\epsilon$ and with it the critical temperature decreases with decreasing $m_\chi$. A lower critical temperature implies a lower nucleation temperature and thus a stronger phase transition. Secondly, the amount of supercooling, \emph{i.e.}~the ratio $T_c/T_n$ of the critical to the nucleation temperature, also increases with smaller $m_\chi$ and larger $N$. This can be seen in Fig.~\ref{fig:MNCompareF}, where we plot $T_c/T_n$ as a function of $m_\chi$ and $N$. Larger values of $T_c/T_n$ lead to stronger phase transitions. The increase with $N$ is due to the fact that the tunnelling action for $O(3)$- and $O(4)$-symmetric bubbles scales like $N$ to a positive power. To see this, note that the $N$-dependence in the pure dilaton part of the Lagrangian, given by the dilaton kinetic term plus the potential in Eqs.~\eqref{eq:vcft} and \eqref{epsilonexpression} (ignoring the additional term in Eq.~\eqref{eq:Vchipotential} which typically only gives a small correction), enters via an overall factor of $g_\chi^2$ in the potential (up to corrections of order $g_*^{\gamma_\epsilon},g_\chi^{\gamma_\epsilon}$ which enter via $\chi_0$ but are small). 
Using this, one can show the aforementioned scaling of the tunnelling action (\emph{cf.~e.g.}~\cite{vonHarling:2017yew}).
This then delays the phase transition for larger values of $N$. Furthermore, a smaller $m_\chi$ corresponds to a smaller $\gamma_\epsilon$ which makes the pure dilaton potential flatter. This in turn also increases the tunnelling action and thereby makes the phase transition more supercooled. Notice also that the strength $h[T_n]/T_n$ of the phase transition increases much faster for the glueball-like dilaton compared to the meson-like dilaton. This can be understood from the different $N$-scalings of the coupling $g_\chi$ in the two cases. 

Apart from requiring $h[T_n]/T_n \gtrsim 1$, another important constraint for electroweak baryogenesis comes from reheating following the completion of the phase transition after bubble percolation. In order to ensure that reheating does not wash out the baryon asymmetry and the sphalerons are frozen in the broken electroweak phase, the ratio $h/T$ needs to remain larger than 1 also at the reheating temperature.
The energy stored in the plasma of the symmetric phase as well as its potential energy is transferred into the energy of particles in the confined phase. The reheating temperature $T_R$ can be derived from the equality (including the contribution in the symmetric phase from the fundamental degrees of freedom which in the confined phase mix with the composite states)
\be
\frac{\pi^2 g_c}{30} T_R^4 \, \simeq \, \Delta V  \, + \, \frac{3 \pi^2 N^2}{8} T_n^4  \, + \, \frac{\pi^2 g_c}{30} T_n^4 \,, 
%\Delta V(T_n) \, = \, \frac{\pi^2 g_c}{30} T_R^4 \, ,
\ee
where $g_c$ is the number of SM relativistic degrees of freedom in the confined-phase plasma and $\Delta V$ is the difference between the potential energies in the false and the true minimum (at zero temperature). In the approximation of Eqs.~(\ref{eq:vchimin}) and (\ref{GammaEpsilonRel}):
\be
\Delta V \, \simeq \, \frac{g_*^2}{g_\chi^2} \frac{m_\chi^2 f^2}{16} \, .
\ee
%The first term
% in the {\it rhs} 
%is the depth of the effective scalar potential at the global minimum, which comprises the energy density of the CFT plasma, 
%evaluated at the nucleation temperature. 
%(for which we used approximations (\ref{eq:vchimin}), (\ref{GammaEpsilonRel}))
% In the above expression we have neglected the energy density of elementary states in the symmetric phase. 
Thus for successful electroweak baryogenesis, both $h[T_n]/T_ n \geq 1$ and  $h[T_R]/T_R\geq1$ are required. In Fig.~\ref{fig:MNplots}, we have shaded the region, where the latter condition is not fulfilled, in blue.

Restricting the number of colors to reasonable values, say $N<15$ as in Fig.~\ref{fig:MNplots} (or equivalently restricting the cutoff to $m_\star\gtrsim2.6\, $TeV), we conclude from these plots that successful electroweak baryogenesis then implies a dilaton which is lighter than $\sim 0.65\, (0.35)\,$TeV in the case of a meson (glueball) dilaton. On the other hand, the fast increase of the amount of supercooling with decreasing $m_\chi$ for the glueball-like dilaton means that it can in this case not be too light either. Indeed, as we have discussed in the last section, with too much supercooling the bubble walls accelerate to wall velocities larger than the sound speed in the surrounding plasma and electroweak baryogenesis is no longer possible. The precise amount of supercooling for which this happens and the resulting lower bound on $m_\chi$ for the glueball-like dilaton would require a dedicated analysis which is beyond the scope of this work. However, the constraint from the reheating temperature pushes the phase transition for the glueball-like dilaton very much into the supercooled region. We therefore conclude that in our scenario, standard electroweak baryogenesis from charge transport is unlikely to work for a glueball-like dilaton. The possibility of cold baryogenesis nevertheless remains very attractive~\cite{Konstandin:2011ds,Servant:2014bla} and deserves further investigation. We will however continue to show our results also for the glueball case, as the study of the phase transition is interesting even in the absence of the possibility of standard electroweak baryogenesis.

In the right panels of Fig.~\ref{fig:MNplots}, we plot $\hat{h}_{\rm avg}$ as a function of $m_\chi$ and $N$ for a meson-like and a glueball-like dilaton. As discussed in the last section, $\hat{h}_{\rm avg}=\theta f$ enters our parametrization of the tunnelling trajectories along straight lines using Eq.~\eqref{chi1chi2def} and sets their angle in the two-dimensional field space. It determines the average Higgs \emph{vev} during the phase transition and thereby is in particular important for the amount of CP violation that can be generated from the varying mixings (\emph{cf.}~Eq.~\eqref{CPviolatingsource}). %The larger $\hat{h}_{\rm avg}$, the better. 
As can be seen from the plot, for the meson-like dilaton $\hat{h}_{\rm avg}$ increases slowly with decreasing $m_\chi$. For the glueball-like dilaton, $\hat{h}_{\rm avg}$ increases as well with decreasing $m_\chi$, but somewhat faster. In both cases, the tunneling angle is far from zero, therefore sufficient CP-violation for baryogenesis can be generated. Note, however, that as visible in Fig.~\ref{fig:procedure}, the minimum in the tunnelling action which determines $\hat{h}_{\rm avg}$ can be very shallow. This means that tunnelling in directions with somewhat different $\hat{h}_{\rm avg}$ may not be much less likely than in the direction with $\hat{h}_{\rm avg}$ at the minimum. The amount of CP violation can then be larger than what is naively expected from the right panel of Fig.~\ref{fig:MNplots}. This is discussed in more detail in the next section.

\subsection{Results: Baryon asymmetry}
\label{sec:numresBAU}

Our choice $f=800\, \text{GeV}$ for the Goldstone decay constant is relatively low.  In order to ensure that flavour-changing and CP-violating processes are within their experimental bounds,  one then typically needs to impose flavour symmetries. But as we have seen in Sec.~\ref{sec:cpv}, too large flavour symmetries can lead to the vanishing of the CP-violating force from the varying mixings. For definiteness, in this section we make use of a construction from Ref.~\cite{Csaki:2008eh} which is based on $U(1)$ flavour symmetries. The model in~\cite{Csaki:2008eh} is formulated in a warped extra dimension and features a U(1) symmetry acting on the bulk up-type fermions (the down-type fermions are irrelevant for our purpose hence we ignore them) and only broken by an IR-brane localised interaction. Such a structure can be reflected in our 4D construction in the following way. First, the elementary-composite mixings are diagonal. This is enforced by a U(1) symmetry acting on elementary quarks and the operators they couple to, such that different SM flavours have different U(1) charges. This symmetry is explicitly broken by the strong sector condensation, which is reflected in the anarchic form of the matrix $g_\star^{-1}$. The mass matrix then reads
\begin{equation}
\label{eqn:u1mass}
m_{ij} \, \simeq \, 
%\frac{1}{g_\star} \,
\left(
\begin{array}{cc}
  y_{tL} & 0 \\ 
  0 & y_{cL}(z)
% \end{array}\right)_{kk} \,({\mathbb{I}}_{S})_{kl} \, \left(
\end{array}\right)_{ik} \,(g_*^{-1})_{kl} \, \left(
\begin{array}{cc}
  y_{tR} & 0 \\ 
  0 & y_{cR}(z)
 \end{array}\right)_{lj}^\dagger \,h \, ,
\end{equation}
where 
%${\mathbb{I}}_{S}$ 
$(g_*^{-1})_{kl}$ is a matrix with order-$1/g_*$ \emph{complex} entries. From Table~\ref{tab:tab1} we see that a non-vanishing CP-violating source can arise for this flavour structure provided that at least the two charm mixings change with $\chi$ (which is why above they are written as functions of the direction $z$ perpendicular to the bubble wall). Note that having two varying charm mixings may also affect the properties of the phase transition compared to the case with only one varying charm mixing considered before.
However, as we have seen, the Higgs potential is small compared to the total potential in most regions of parameter space. In the regions where it plays a role, the second varying charm mixing just enhances the effect of the Higgs potential, hence making the tunnelling trajectory more strongly influenced by the valley along $h \sim \chi$. In order to reduce the dimensions of the parameter space, we introduce the second varying charm mixing only for the calculation of the CP-violating force and the baryon asymmetry. The properties of the phase transition are still computed with only one varying charm mixing which we expect to be a good approxmation. 

We fix the parameters for the first varying charm mixing and the other parameters as in Table \ref{tab:bench}. For the second charm mixing, which in a slight abuse of notation we again denote as $y$, 
we set ${y[\chi_0]=\sqrt{\lambda_c g_\star}}$ and $y[0]=0.5g_\star$. This fixes two of the three parameters in the RG equation \eqref{eq:yrun}. We then scan the produced baryon asymmetry over the third parameter which we choose to be the anomalous dimension $\gamma_y$ of the operator which is associated with the mixing. 
As we have seen in Fig.~\ref{fig:procedure}, for many points in parameter space the tunnelling action depends only relatively weakly on the direction $\hat{h}_{\rm avg}$ of the tunnelling trajectory. In the most extreme shown case, for a glueball-like dilaton with $m_{\chi} =1000\, \text{GeV}$ and $N=4$, the tunnelling action is almost flat as a function of $\hat{h}_{\rm avg}$. 
Since the rate of bubble nucleation scales as $\sim e^{-S_E}$ with $S_E$ being the bubble action, there are then bubbles nucleating in all the directions for which the action is approximately constant. We therefore scan the produced baryon asymmetry also over a range of values of $\hat{h}_{\rm avg}$. 
We consider a meson-like dilaton with $m_\chi=500 \GeV$ and $N=5$. Furthermore, we choose ${\mathbb{I}}_{S}\approx\left\{\left\{0.64 + 0.67 i, -0.14 + 0.14 i\right\},\left\{-1.11+0.034 i, -1.03 + 0.22 i\right\}\right\}$ for the order-one matrix in \eqref{eqn:u1mass}. We have checked that this choice reproduces the correct top and charm masses (to $10\%$) in the broken phase. Note, however, that the produced baryon asymmetry depends somewhat on this choice. 
Given that different matrices $\mathbb{I}_S$ can reproduce the correct quark masses, our results should not be taken at face value but rather as an indication of the order of magnitude.

\begin{figure}[t]
\begin{center}
\includegraphics[width=7.8cm]{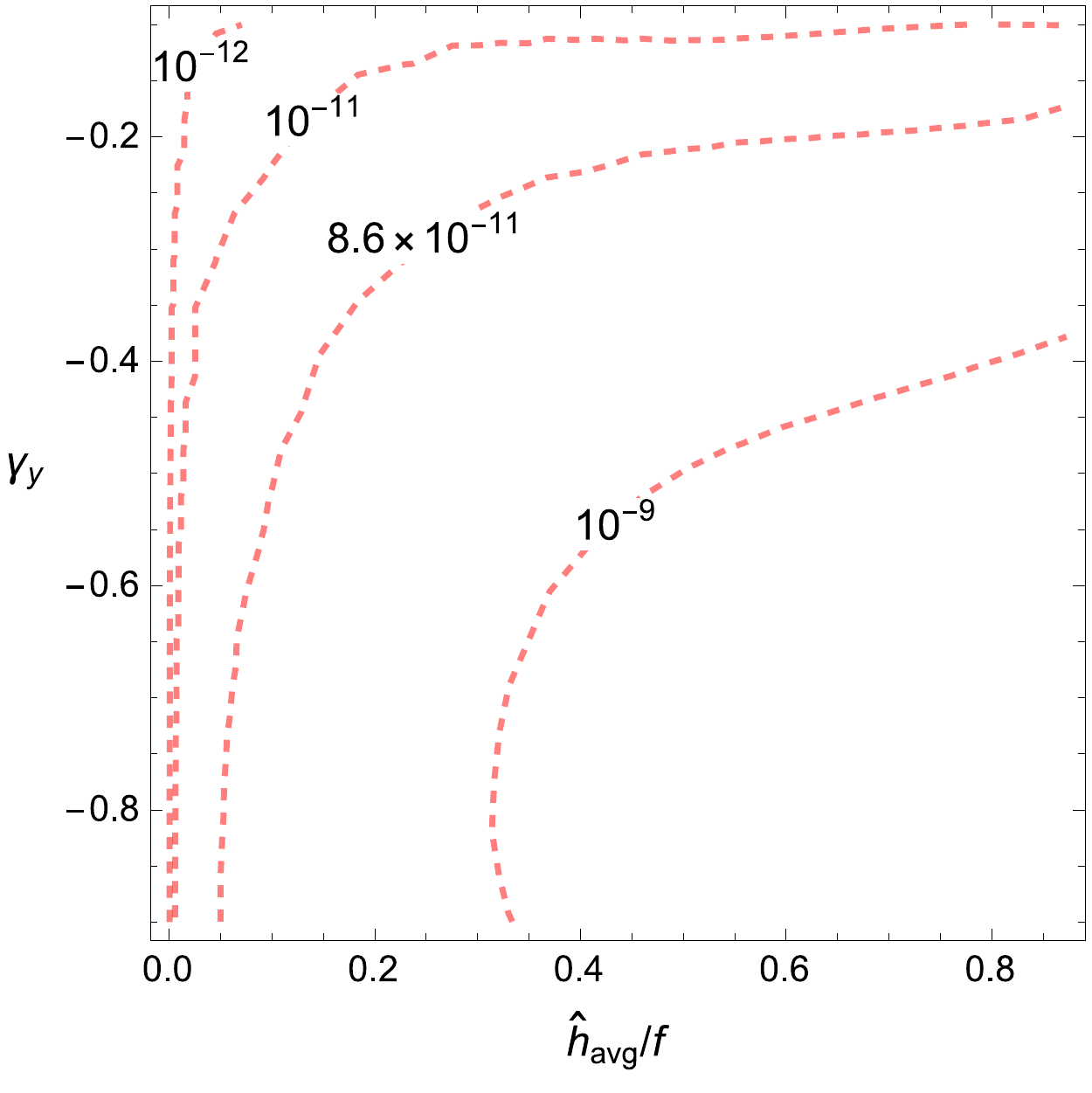}
\end{center}
%\hspace{0.3cm}
%\includegraphics[width=7.8cm]{alphaVsGammaPlotU1Glueball}
\caption{\small \it{The produced baryon asymmetry assuming the $U(1)$ flavour symmetries. As the reheating temperature is generally too high for a glueball dilaton, we focus here on the meson case, for which we have set $m_\chi=500\GeV$ and $N=5$. The measured baryon asymmetry $\eta_B= n_B/s\simeq 8.6\times 10^{-11}$~\cite{Ade:2015xua} is shown as a contour line.}}
\label{fig:etaBplots}
\end{figure}

In Fig.~\ref{fig:etaBplots}, we plot the resulting baryon asymmetry as a function of $\gamma_y$ and $\hat{h}_{\rm avg}$.
We see that in a large portion of parameter space a sufficient baryon asymmetry can be obtained. Note, however, that in this particular flavour model there is an effect that suppresses the baryon asymmetry. Indeed, all entries of the mass matrix (\ref{eqn:u1mass}) have a constant complex phase (the phases in the mass matrix are $z$-independent, the phases of the eigenvalues, on the other hand, are not). As pointed out in \cite{Bruggisser:2017lhc}, this implies that ${\rm Im} [V^\dagger {m^\dagger}'' m V]$ is traceless which in turn means that the CP violation from the top and the CP violation from the charm cancel each other to a very large degree. Obviously this suppresses the yield in baryon asymmetry that can be generated with this flavour structure. This effect might be absent in other flavour structures.

\section{Collider bounds and other experimental tests}
\label{ExperimentalTests}

A key question is about the experimental tests of our scenario.
How do we probe experimentally the nature of the electroweak phase transition in composite Higgs models?
More precisely, how are we able to distinguish between the three scenarios displayed in Fig.~\ref{fig:phtr} and to probe that Yukawa couplings have varied during the phase transition?

\subsection{Dilaton production}

\begin{figure}[t]
\centering
\includegraphics[width=8.cm]{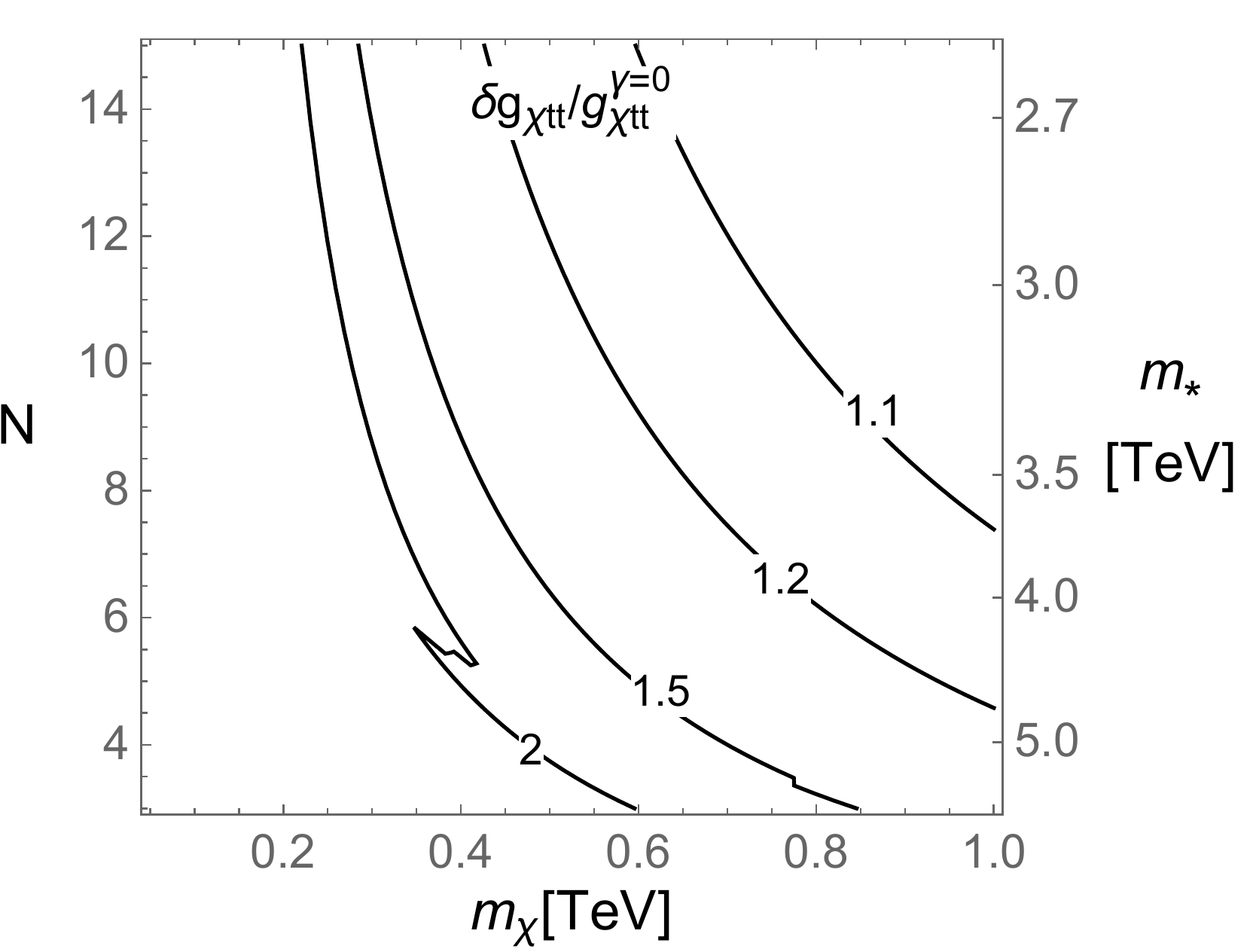}
\caption{\small \it{Strength of the dilaton-top coupling~(\ref{eq:coll1}) in the benchmark model with a varying top mixing, normalized to the analogous strength for a constant top mixing. For the thus defined relative strength the difference between the glueball and meson dilaton is negligible. 
}}
\label{fig:collider}
\end{figure}

An important step would be to detect the dilaton.
The general way of analysing it once it has been detected would be to fit the parameters of the effective potential~(\ref{eq:totalpotential}) to the available data. 
There is however one particularly pronounced effect that we can single out, residing in the couplings of the dilaton to the massive SM states, and whose existence can be traced back to the sizeable energy dependence of the elementary-composite fermion mixings.   
Let us first consider the dilaton-fermion couplings. To this end, we define the mass eigenstate fields $\tilde h$ and $\tilde \chi$, with $\langle \tilde h \rangle = \langle \tilde \chi \rangle = 0$ and rotated by an angle $\delta$ with respect to the fields
\be
\chi = \chi_0  + c_\delta \tilde \chi - s_\delta \tilde h\,, \quad \hat{h} = v + c_\delta \tilde h + s_\delta \tilde \chi \, ,
\ee
where $s_\delta=\sin \delta, c_\delta=\cos \delta$ and $\delta$ is defined by ${\tan \delta = m_{\hat{h} \chi}^2/m_\chi^2}$ in the limit of large dilaton mass.
The leading dilaton-fermion coupling can be obtained from the part of the effective action that gives the fermion masses\footnote{The dependence of the fermion masses on the Higgs can be more complex and here we have chosen the simplest option.}
\be\label{eq:qchi}
\Gamma_{qq \tilde \chi} \, \sim \, \frac{\delta}{\delta \bar q \, \delta q\, \delta \tilde \chi } (\lambda_q v_{\text{SM}} \bar q q)  \, =  \, (\partial_{\tilde \chi} \lambda_q) v_{\text{SM}}  + \lambda_q (\partial_{\tilde \chi} v_{\text{SM}}) \, = \,
\beta_{\lambda_q}  c_\delta \frac{v}{\chi_0} + \lambda_q (\partial_{\tilde \chi} v_{\text{SM}}) \, ,
\ee
where $\beta_{\lambda_q}$ is the $\beta$-function of the Yukawa coupling $\lambda_q$ and $v_{\text{SM}} \equiv (g_\chi \chi/g_\star) \sin \theta$.
Notice that in our scenario both terms on the {\it r.h.s.} of this expression can significantly differ from the typical composite Higgs case. 
The first term is explicitly sensitive to the running of the mixing $y_q$ of the fermion $q$. In particular, it can then be enhanced for the top quark, if the latter is chosen to have a varying Yukawa, \emph{e.g.}~as discussed in Sec.~\ref{sec:cpv}. In addition, this term may allow to test the sign of the $\beta$-function of the varying mixings, which is crucial for the phase transition.    
As for the second term, let us first rewrite $v_{\text{SM}}$ in terms of the mass eigenstates $\tilde h$ and $\tilde \chi$:
\be
v_{\text{SM}} \,\simeq \, \left(f+ \frac{g_\chi}{g_\star} \left(c_\delta \tilde \chi- s_\delta \tilde h\right)\right) \left(\sin[v/f] + \cos[v/f] \frac{ (c_\delta \tilde h + s_\delta \tilde \chi)}{f}\right)  \, .
\ee   
Using this expression in the last term of Eq.~(\ref{eq:qchi}), we get  
\be\label{eq:coll1}
\Gamma_{qq\tilde \chi} \, \sim \, c_\delta  \beta_{\lambda_q} \frac{v}{\chi_0} + c_\delta  \lambda_q  \frac{v}{\chi_0}  + s_\delta \lambda_q\,.
\ee
Therefore, not unexpectedly, this interaction is sensitive to the Higgs-dilaton mixing, which in turn 
depends on the running of $y$ and the coefficient $c_\alpha$ (see Eq.~(\ref{eq:hchimix2})).  
Notice that it is mostly sensitive to the largest varying mixing $y$, \emph{i.e.}~not necessarily the one corresponding to the fermion $q$. 
These $y$ and $c_\alpha$ are crucial for ensuring that the tunnelling trajectory goes far from the $(h=0)$-line. Hence measuring the dilaton-fermion couplings can give an important information for assessing the possibility of electroweak baryogenesis in this scenario. In the case that all mixings are constant or small, we have $\beta_{\lambda_q} \simeq s_\delta \simeq 0$, and the expression~(\ref{eq:coll1}) reduces to $\Gamma_{qq\tilde{\chi}} \sim \lambda_q v /\chi_0$ as expected.

We emphasize that large Higgs-dilaton mixing is only expected for varying top mixings. For varying charm mixings, on the other hand, a significant effect does typically not arise as $\beta_{y_c} \sim y_c$, and the value of $y_c$ in the global minimum is relatively small. In this section, we will therefore consider a benchmark model with a varying top mixing. 
We assume that the top Yukawa contains a constant and a dilaton-dependent contribution,
\be
\lambda_t[\chi] = \lambda_t^{\text{SM}} + \left[y_{tL}[\chi] \, y_{tR}-y_{tL}[\chi_0] \, y_{tR} \right]/g_\star
\ee 
with 
\be
y_{tR}=\sqrt{\lambda_t^{\text{SM}} g_\star} ,\;\;\;
y_{tL}[\chi]=0.3\sqrt{\lambda_t^{\text{SM}} g_\star} \, (\chi/\chi_0)^{\gamma_{y_{tL}}} e^{i \pi/2}
\ee
and the anomalous dimension of the running mixing $y_{tL}$ given by $\gamma_{y_{tL}}=-0.3$. Moreover, the remaining parameters are chosen as in Table~\ref{tab:bench}. This is similar to the benchmark model considered in \cite{Bruggisser:2018mus} which was shown to produce enough CP violation for successful electroweak baryogenesis. In Fig.~\ref{fig:collider} we show how the dilaton-top coupling for this benchmark model deviates from the case with a constant top Yukawa. Note that in the case of a varying charm mixing the relative deviation of the dilaton-charm coupling is also expected to be sizeable, though the overall coupling size is of course small.

The discussed behaviour of $\delta_{\tilde \chi} v_{\text{SM}}$ also shows up in the couplings of the dilaton to the massive SM gauge bosons. These couplings can again be derived from the corresponding mass terms,
\be\label{eq:wchi}
\Gamma_{WW\tilde\chi}\sim\frac{\delta}{\delta  W_\mu \, \delta W^\mu\, \delta \tilde\chi} (g^2 v_{\text{SM}}^2 W_\mu W^\mu)   =  (\delta_\chi g^2) v^2 c_\delta + 2 g^2 v_{\text{SM}} (\delta_{\tilde \chi} v_{\text{SM}}),
\ee
and contain a contribution which is sensitive to the Higgs-dilaton mixing. The couplings of the dilaton to the top and the massive SM gauge bosons also induce dilaton-photon and dilaton-gluon couplings, which thus carry the dependence on the same parameters.  

Let us now briefly discuss the possible ways to test these couplings. The dilaton phenomenology is in many respects similar to the phenomenology of the SM Higgs, with differences mainly caused by the different masses, and the different couplings which are reduced roughly by a factor of $v/\chi_0$ for the dilaton~\cite{Blum:2014jca,Efrati:2014aea} (a more precise estimate of the couplings in our case is given by the expressions above). The dilaton couplings to the massive SM gauge bosons~(\ref{eq:wchi}) control the vector boson fusion channel for dilaton production, which can be tagged by forward jets and thus be well tested even if the dominant $\chi$ production channel is different, {\it i.e.}~gluon-gluon fusion. The same couplings also control the less significant $W$ or $Z$ associated production. The effect of the direct dilaton-fermion couplings~(\ref{eq:qchi}) on dilaton production is expected to be much weaker than that of the dilaton-vector couplings. They control for instance $t\bar t$-associated $\chi$ production.

For what concerns decays, the most stringent current bounds on the dilaton \cite{Blum:2014jca,Efrati:2014aea} were obtained from searches for $\chi \to VV$ decays~\cite{Aad:2015agg,Chatrchyan:2013yoa}, directly sensitive to the couplings~(\ref{eq:wchi}).  If the dilaton is sufficiently heavy, the decays into $t \bar t$ can also become sizeable, though still below the rate of decays into vectors.

The bound on the dilaton mass is around $m_\chi \gtrsim 100$~GeV~\cite{Blum:2014jca}. As discussed in Sec.~\ref{sec:numresPT}, this is not far from the mass range which is preferable for electroweak baryogenesis. So one can expect the near future experiments to probe the interesting dilaton mass range. In this regard, it would be important to conduct an updated recast of the latest existing experimental searches based on the 13~TeV dataset, such as~\cite{Aaboud:2017gsl,CMS:2017sbi} into bounds on the dilaton. A naive comparison of the exclusion reaches of these new searches to the ones obtained with the full 7~TeV and 8~TeV data sets does not show a dramatic improvement. We therefore start our plots at $m_\chi = 50$~GeV, leaving a more thorough study of the experimental bounds for future work.

\subsection{Flavour violation}
\label{sec:FCNC}

There is another important type of experimental constraints that our scenario has to face -- the bounds on flavour-changing four-fermion operators. It is well known that these bounds bring severe constraints on composite Higgs models. A set of solutions has been proposed in the literature, with additional symmetries which can suppress these unwanted effects.  
We have discussed one such solution, which makes use of $U(1)$ flavour symmetries, in Sec.~\ref{sec:numresBAU}. 
One may also investigate whether $U(2)$ symmetric constructions~\cite{Redi:2012uj,Barbieri:2011ci,Barbieri:2012tu} can be incorporated into our scenario, or 
a proposal \cite{Redi:2011zi} to impose a CP symmetry on the strong sector and most of the elementary-composite mixings, with the exception of those of the third generation.
This and a more rigorous study of flavour constraints in general deserve a separate analysis, which we leave for future work. Additionally, we should mention that the scenario with a varying top mixing (see \cite{Bruggisser:2018mus}) can accommodate any of the flavour or CP symmetries mentioned above.    
   
\subsection{Higgs couplings and CP violation}

\begin{figure}[t]
\centering
\includegraphics[width=7.85cm]{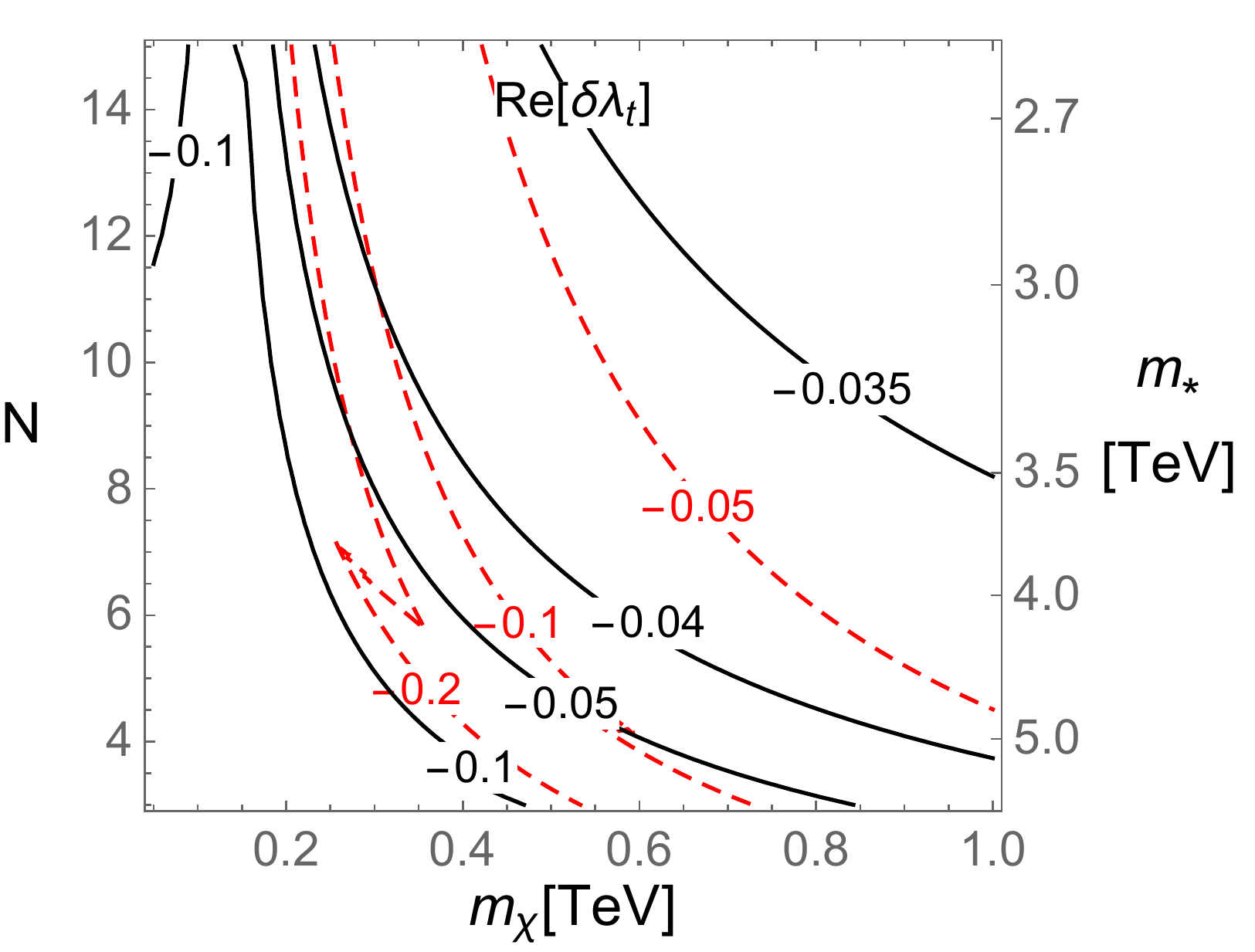}
\hspace{0.4cm}
\includegraphics[width=7.85cm]{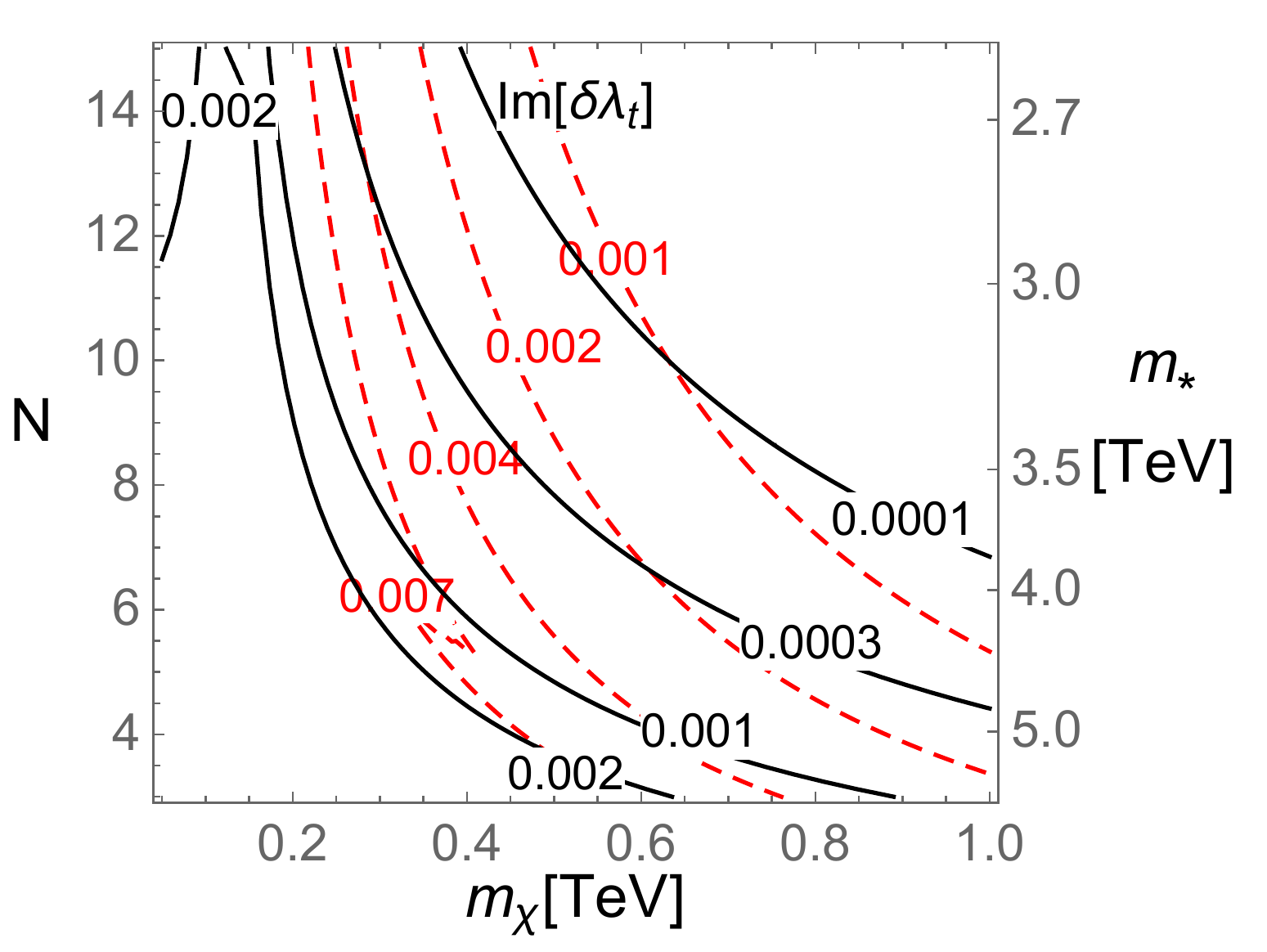}
\caption{\small \it{Real (left panel) and imaginary (right panel) parts of the top Yukawa modification in the benchmark model with a varying top mixing for a meson-like dilaton (red dashed) and a glueball-like dilaton (black solid). The real part can be tested by CLIC at the 4\% level at $1\sigma$~\cite{Abramowicz:2013tzc}, and a pure composite Higgs contribution to it (with no Higgs-dilaton mixing) is -0.05. For the tests of the imaginary part see text.}}
\label{fig:collider2}
\end{figure}

Last but not least, information about the dilaton sector can come from Higgs physics. As was discussed in Sec.~\ref{sec:prelimAnalysis}, the deviations of the Higgs couplings depend explicitly, and potentially sizeably, on the dilaton-Higgs mixing. In particular, one of the smoking guns of our scenario with a varying top Yukawa would be CP-violating top-Higgs couplings. To see this, let us derive the top-Higgs couplings from the term giving rise to the top mass:
\be
\begin{aligned}
\label{eq:qh}
\Gamma_{qq \tilde h} &\sim  \frac{\delta}{\delta \bar q \, \delta q\, \delta \tilde h} (\lambda_q v_{\text{SM}} \bar q q)  =  (\partial_{\tilde h} \lambda_q) f \sin[v/f]  + \lambda_q (\partial_{\tilde h} v_{\text{SM}})  = 
- \beta_{\lambda_q}  s_\delta \frac{v}{\chi_0} + \lambda_q (\partial_{\tilde h} v_{\text{SM}}) \\
&  =  - \beta_{\lambda_q}  s_\delta ({g_\chi}/{g_\star}) \sin[v/f] + \lambda_q \left(c_\delta \cos[v/f] - s_\delta ({g_\chi}/{g_\star}) \sin[v/f]\right) \,.
\end{aligned}
\ee
First of all, we notice that the leading deviations are proportional to $\sim s_\delta v/f$. By the end of the LHC operation, these will have been tested to a precision of at most $10\%$ at $1\sigma$, and could be tested up to $4\%$ at future linear colliders. The predictions of the modification of the $tth$ coupling for our benchmark model with a varying top mixing are shown in the left panel of Fig.~\ref{fig:collider2}. 

Secondly, we see that this coupling may carry an observable CP-violating phase, coming from the $\beta$-function. The latter is in general complex, with a phase which is different from the phase of the Yukawa coupling itself. This follows provided that the phase of the Yukawa coupling changes with the dilaton \emph{vev}:
\be
\begin{aligned}
0 & \, \ne \, \arg[\lambda_t[\chi_0]] - \arg[\lambda_t[\chi_0+\delta \chi]] \, = \, 
\arg[\lambda_t[\chi_0]] - \arg[\lambda_t[\chi_0]+\beta_t \frac{\delta \chi}{\chi_0}]  \\
 &\, \Rightarrow \, \arg[\beta_t] \, \ne \,  \arg[\lambda_t[\chi_0]]\,.
\end{aligned}
\ee
Therefore, when we perform a complex rotation of the fermions to make the Yukawa coupling real, the contribution to $\Gamma_{qq \tilde h}$ which is proportional to $\beta_{\lambda_q}$ remains complex:
\be
\text{Im} [\Gamma_{qq \tilde h}] \, \sim \, \text{Im} [\beta_{\lambda_q}]  s_\delta ({g_\chi}/{g_\star}) \sin[v/f]\,.
\ee
Analogously, one can show that the dilaton-top coupling carries a complex phase. There are several ways of testing these phases. One is looking for effects induced by electric and chromoelectric dipole moments (EDMs)\footnote{Here we only focus on the contributions to the EDMs caused by the running of the mixings. For other contributions which can potentially arise in composite Higgs models, see \emph{e.g.}~\cite{Panico:2017vlk}.}. The strongest bounds result from the former~\cite{Cirigliano:2016nyn}, which affect the electron EDM which is currently bounded to be $d_e/e < 8.7 \times 10^{-29} \, \text{cm @ 90\% CL}$~\cite{Baron:2013eja} (see also~\cite{Afach:2015sja,PhysRevLett.116.161601,Yamanaka:2017mef}). This gives the constraint~\cite{Cirigliano:2016nyn}
\be
\text{Im} [\Gamma_{qq \tilde h}] \,\lesssim \, 0.018 \;\;@ \, 90\% \, \text{CL}\, .  
\ee
Future experiments are expected to improve the bound on $d_e/e$ by a factor of $\sim 10$, with a similar rescaling of the constraint on the imaginary coupling~\cite{Cirigliano:2016nyn}. For comparison, we present in the right panel of Fig.~\ref{fig:collider2} some typical values of the CP-violating Higgs-top coupling obtained for the glueball and meson case. For the glueball, these values do not exceed $\sim10^{-3}$, which means that they satisfy current constraints, but some part of the parameter space can be probed by future experiments. For the meson, the values can instead reach the current experimental sensitivity for the lowest dilaton masses, while future experiments are expected to have a good sensitivity to a large fraction of the parameter space.  
Notice that in the case where only the charm mixings vary, the imaginary part of the charm Yukawa is suppressed by both the small charm Yukawa itself and by the small Higgs-dilaton mixing, therefore we do not expect that the resulting CP violation~\cite{Sala:2013osa} can be testable in the near future.

\begin{figure}[t]
\centering
\includegraphics[width=7.87cm]{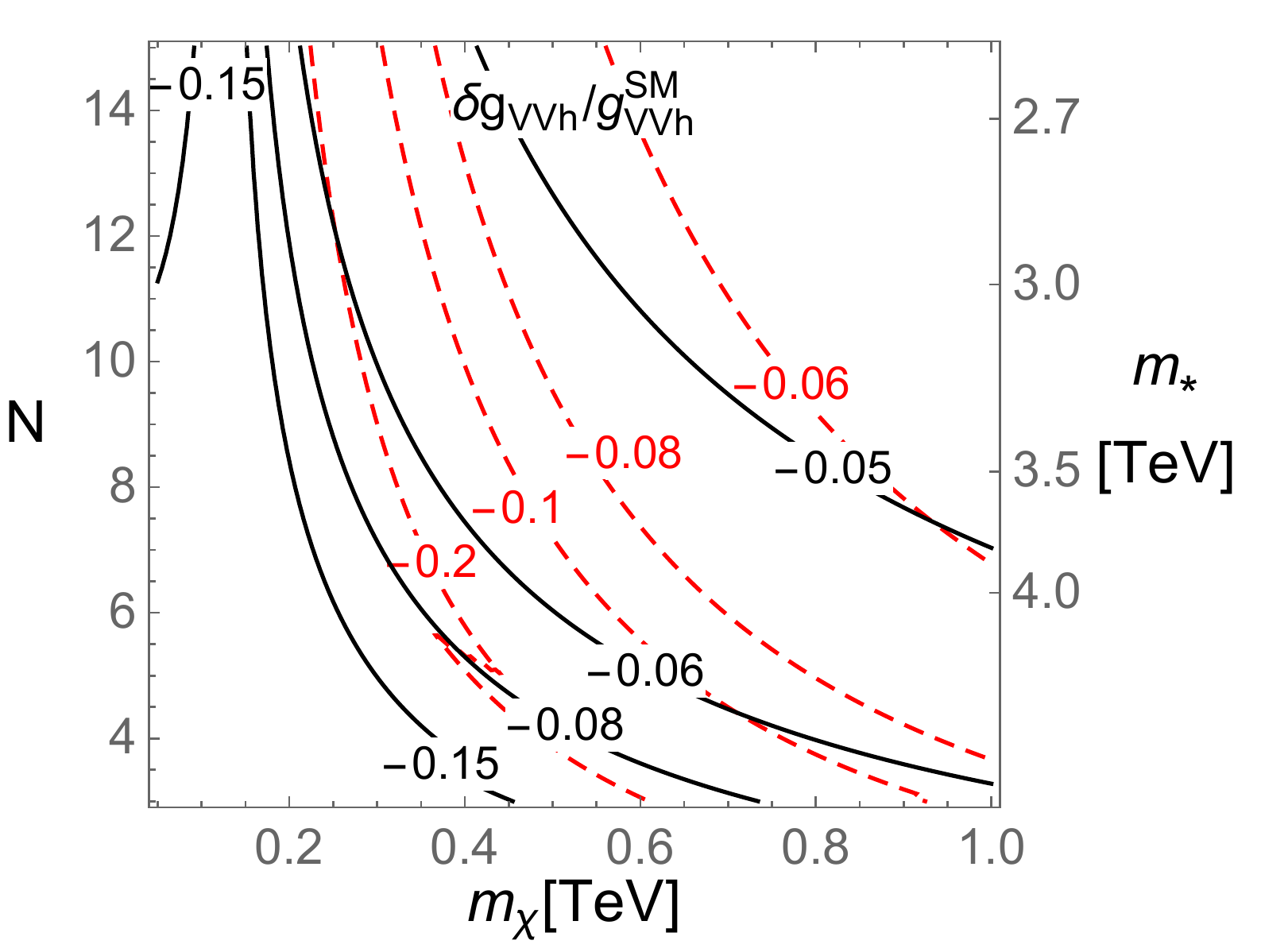}
\hspace{0.4cm}
\includegraphics[width=7.8cm]{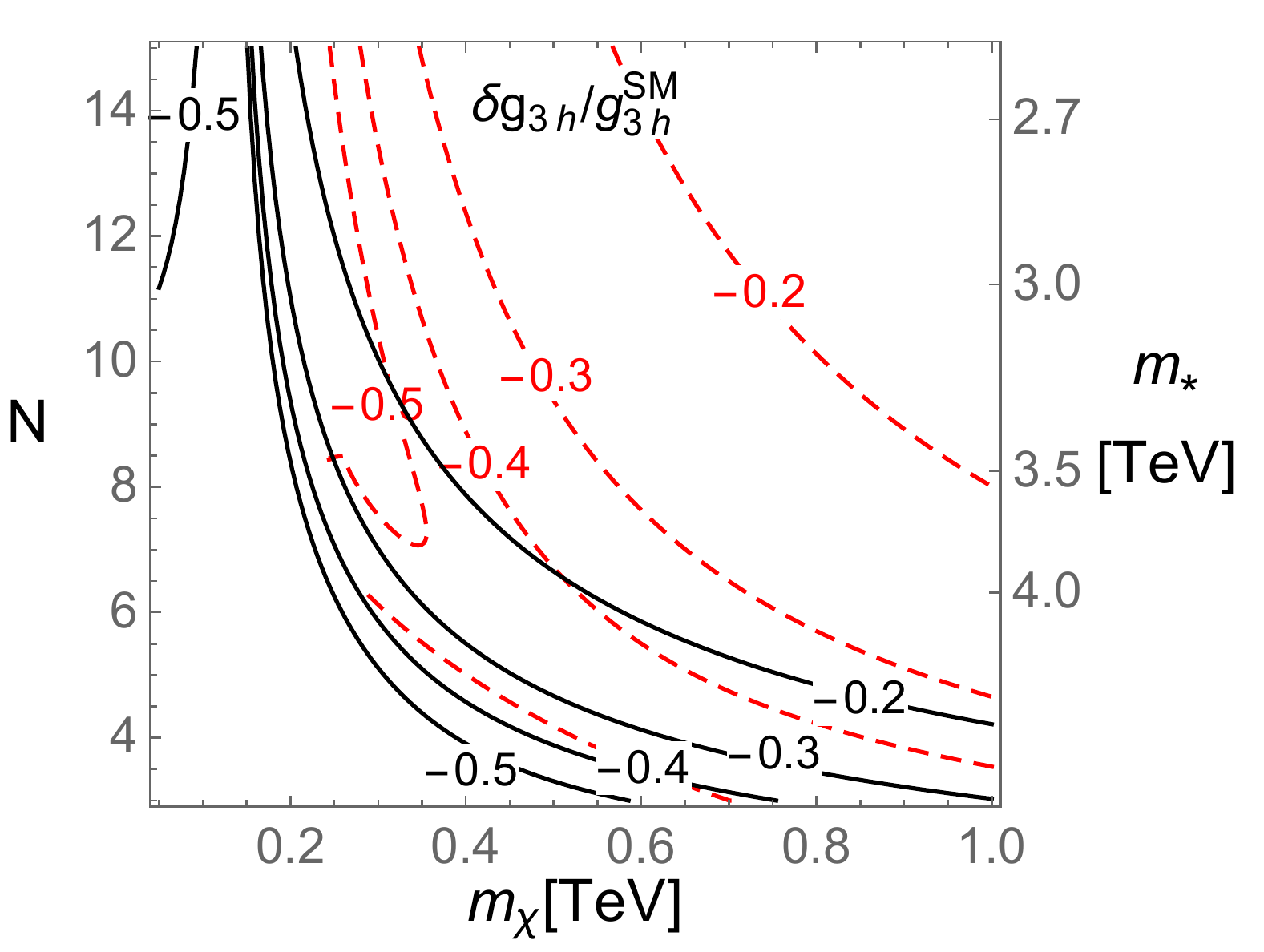}
\caption{\small \it{Relative deviation of the Higgs couplings to electroweak gauge bosons (left panel) and of the triple Higgs coupling (right panel) for a meson-like dilaton (red dashed) and a glueball-like dilaton (black solid). Future ($1\sigma$) sensitivity to the former is expected to be 0.8\% at CLIC~\cite{Abramowicz:2016zbo} and 0.15\% at FCC~\cite{Gomez-Ceballos:2013zzn}, while the expected precision for the latter is order-one at the high-luminosity LHC~\cite{DiVita:2017eyz} and $10-40\%$ at future leptonic colliders~\cite{Abramowicz:2013tzc,DiVita:2017vrr}.}}
\label{fig:collider3}
\end{figure}

Furthermore, the CP-violating Higgs-top interactions can be measured directly at the LHC. These are, however, expected to give much weaker sensitivity by at least one order of magnitude. This situation will not improve significantly even at the high-luminosity LHC~\cite{Rindani:2016scj,Buckley:2015vsa}. Therefore the first signs of CP-violation in Higgs-top interactions arising in a scenario with varying top mixings are expected to come from EDM experiments. 
For completeness, in Fig.~\ref{fig:collider3} we also show the predicted deviations of the Higgs couplings to the electroweak gauge bosons, and in the triple Higgs couplings.

\subsection{Gravitational waves}
\label{sec:GW}

Cosmological first-order phase transitions can lead to a stochastic background of  gravitational waves (GWs)~\cite{PhysRevD.30.272,Hogan:1986qda,Kamionkowski:1993fg,Grojean:2006bp}. Towards the end of the phase transition, the bubbles take up a large fraction of space and start to collide. During this collision, some of the free energy released during the phase transition is converted into GWs. The GWs hence created are then present today as a stochastic background characterized by its energy-frequency spectrum. It turns out that a strong first-order phase transition happening around the electroweak scale generates a spectrum of GWs that lie in the observable frequency bands of the Laser Interferometer Space Antenna (LISA) (see~\cite{Caprini:2015zlo,Weir:2017wfa} for reviews). The relevance of a confinement phase transition at the TeV scale for LISA was stressed in~\cite{Randall:2006py}.

The spectrum of GWs is controlled by two main parameters~\cite{Caprini:2007xq,Huber:2008hg,Hindmarsh:2015qta}: $\alpha$ roughly corresponds to the latent heat which is released during the phase transition and $\beta^{-1}$ measures the duration of the phase transition. These two parameters are given by
\begin{equation}
 	\alpha=\frac{\epsilon}{\rho_{\rm rad}}\simeq\frac{(V_{\rm tot}[0,0]-V_{\rm tot}[v,\chi_0])_{T_n}}{3\pi^2N^2T_n^4/8}~, \qquad \frac{\beta}{H}\simeq T_n \left.\frac{{\rm d} S_E}{{\rm d} T}\right|_{T_n},
 \end{equation}
 where $\epsilon$ is the latent heat, $\rho_{\rm rad}$ the radiation energy density of the unbroken, deconfined phase, $H$ the Hubble rate at the time of the phase transition and $S_E$ the euclidean action of the bounce. Since $\beta/H\gg1$ for the parameter region of interest, the nucleation temperature and the temperature when the phase transition completes are equal to a good approximation. We therefore evaluate the expressions above at the nucleation temperature $T_n$.  For $\alpha$,  we also use Eq.~\eqref{eq:fcft}. 

\begin{figure}[t]
\centering
\includegraphics[width=7.5cm]{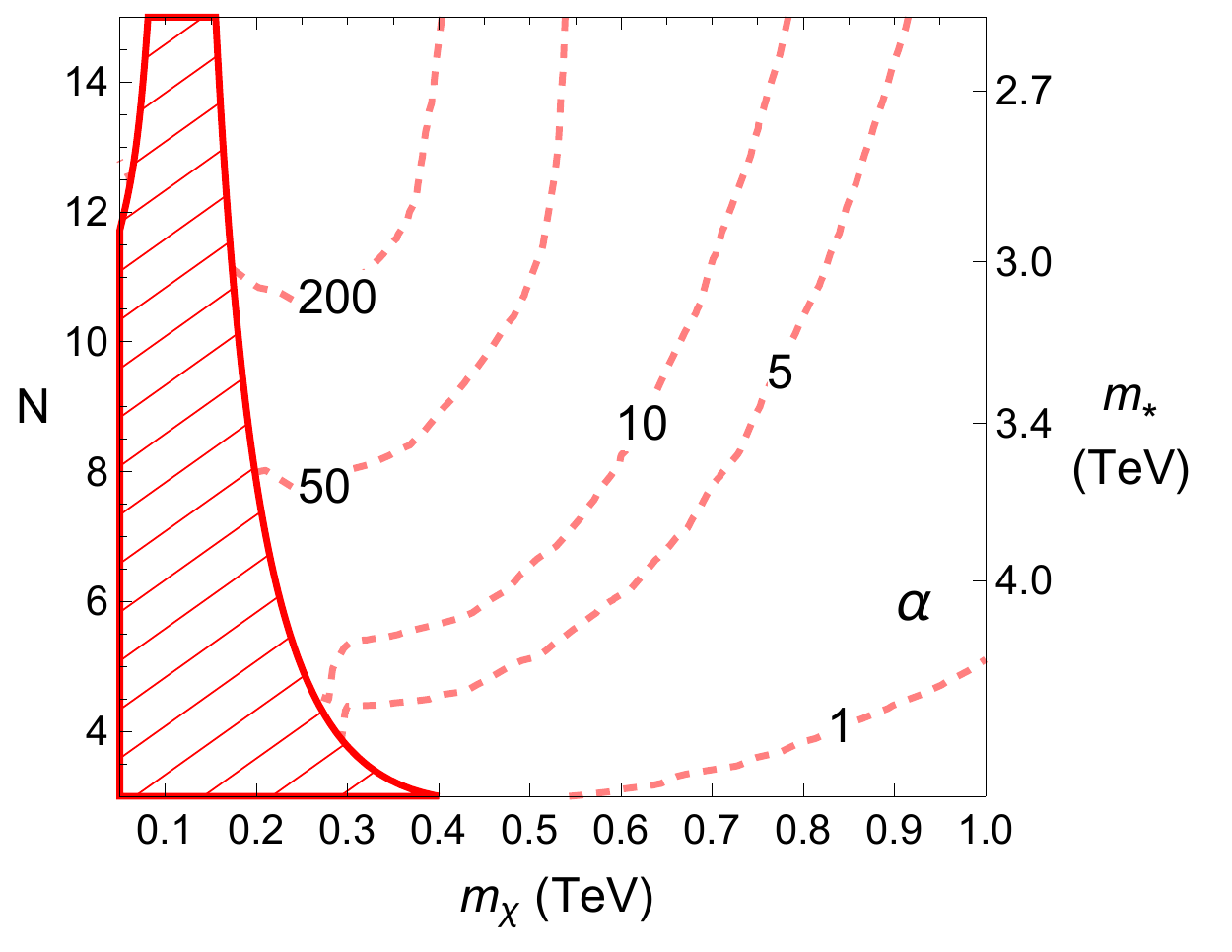}
\hspace{0.2cm}
\includegraphics[width=7.5cm]{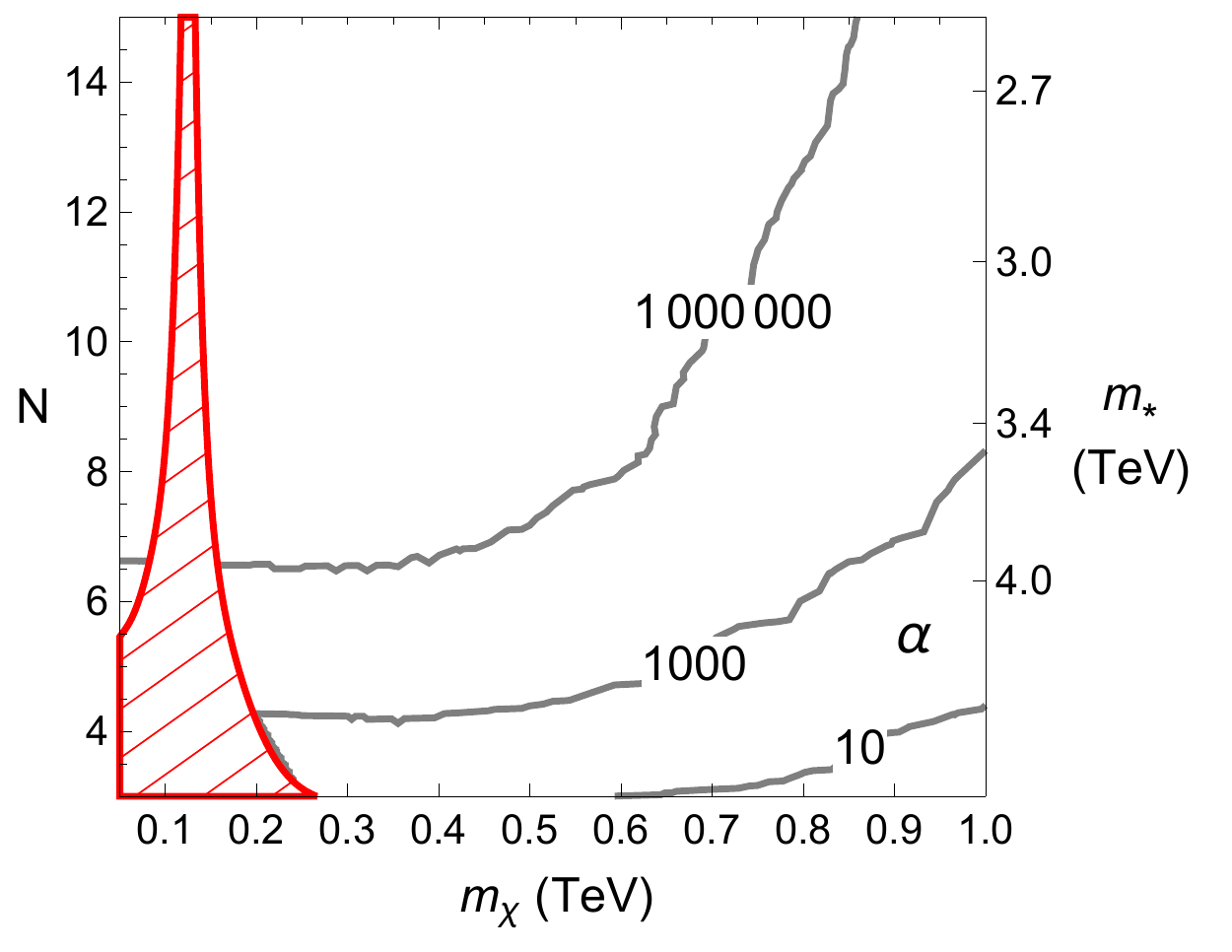} \\ \vspace{.4cm}
\includegraphics[width=7.5cm]{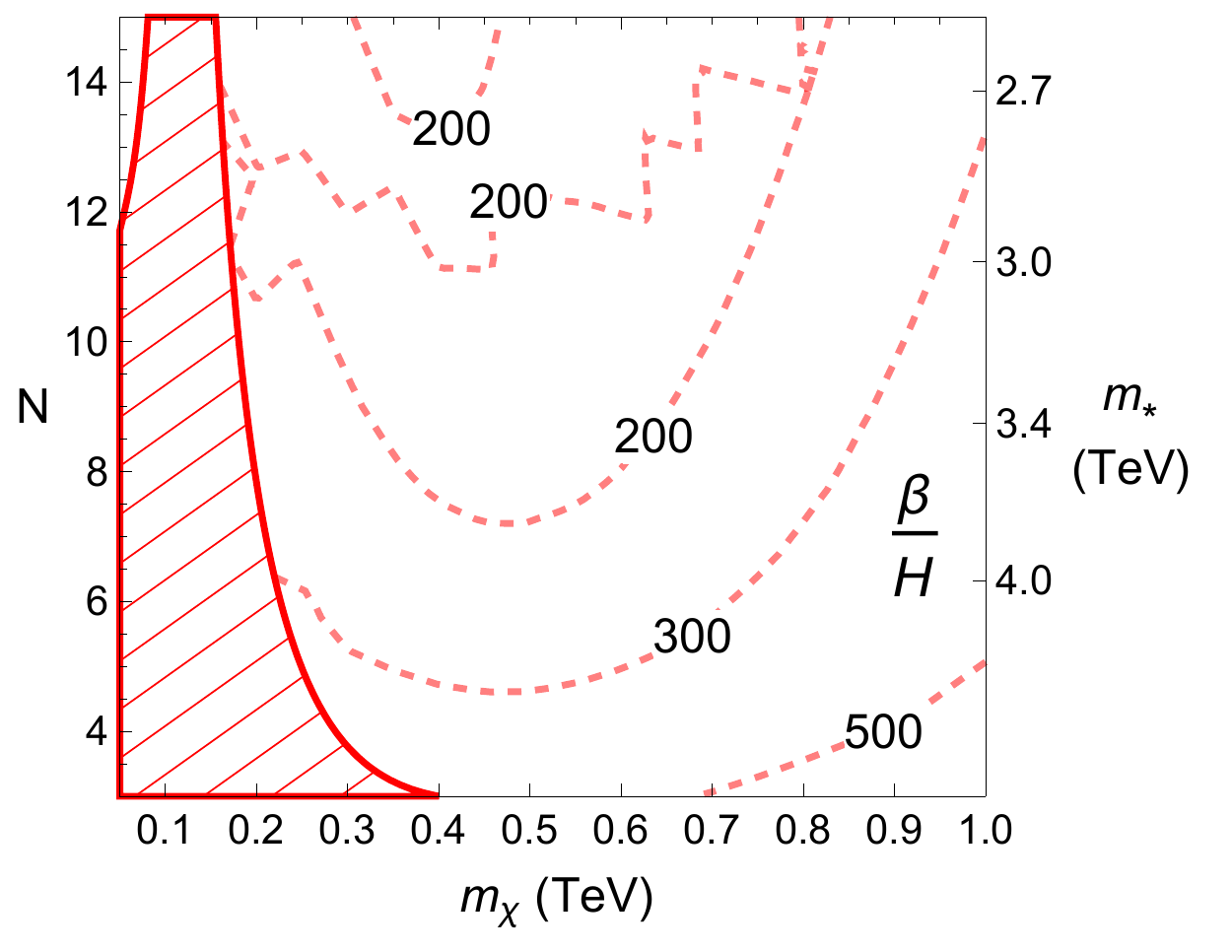}
\hspace{0.2cm}
\includegraphics[width=7.5cm]{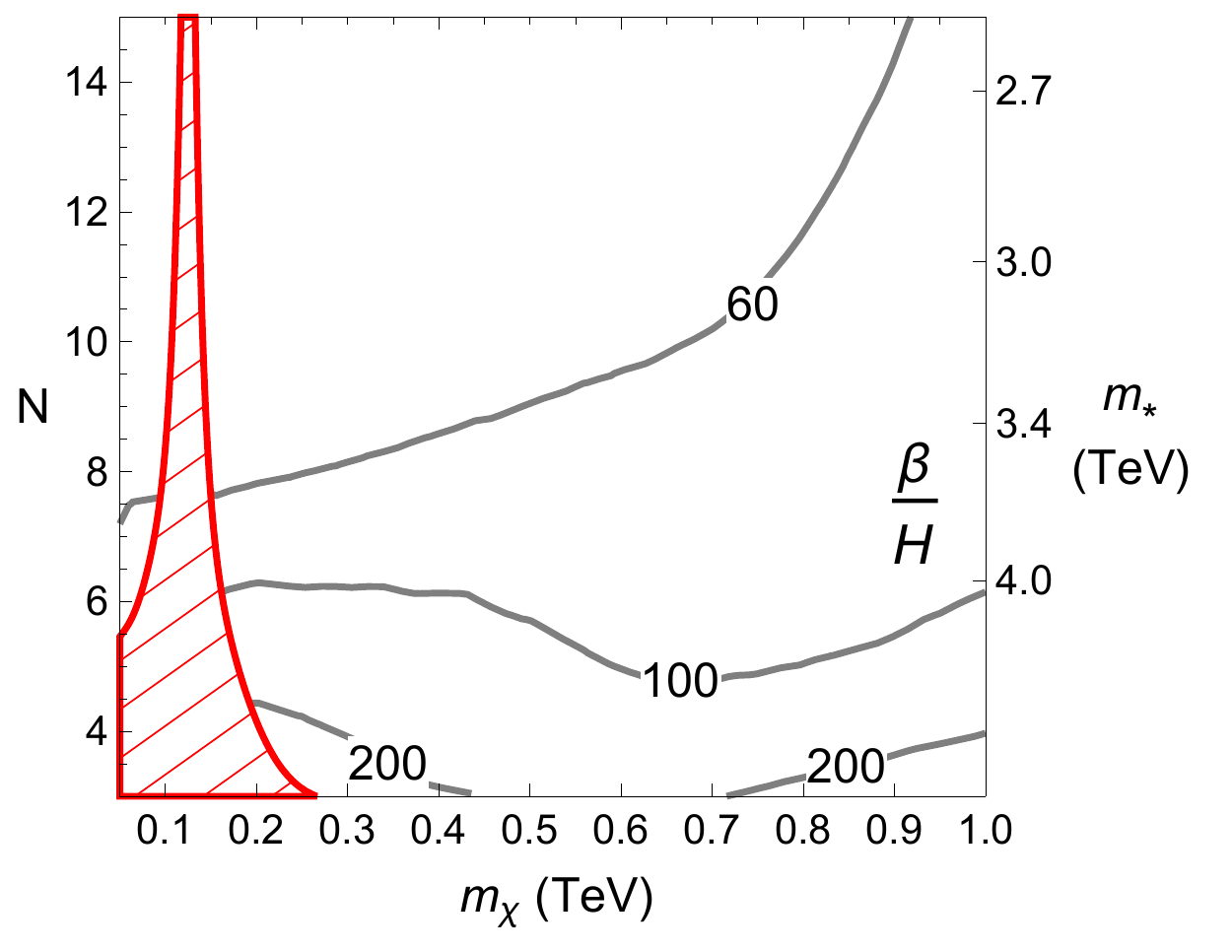}
\caption{\small \textit{The parameters $\alpha$ (upper panels) and $\beta/H$ (lower panels) which determine the GW signal as a function of $m_\chi$ and $N$ for a meson-like dilaton (left panels; red dashed) and a glueball-like dilaton (right panels; black solid). A signal in the LISA band is expected for $1\lesssim\beta/H\lesssim 10^4$ and $\alpha\gtrsim 0.1$~\cite{Grojean:2006bp,Caprini:2015zlo}. In the red dashed region, there is no phenomenologically viable electroweak minimum. 
}}
\label{fig:GWParameters}
\end{figure}

We have computed both $\alpha$ and $\beta/H$ and report the results in Fig.~\ref{fig:GWParameters}. A signal from a phase transition at $T_n=100\GeV$ will be detectable by LISA if $1\lesssim\beta/H\lesssim 10^4$ and $\alpha\gtrsim 0.1$~\cite{Grojean:2006bp,Caprini:2015zlo}.
We see that in a large portion of the parameter space, a strong GW signal can be expected, giving a testable prediction of our scenario. Particularly unique to this class of nearly-conformal potentials, $\alpha$ can be extremely large (such as $10^6$ in the glueball case). In turn, the absence of a GW signal at LISA will lead to non-trivial constraints on the parameters of our model, especially on $m_\chi$ and $N$.

Should GWs, compatible with a strong first-order electroweak phase transition, be observed by LISA, a more detailed study of their spectrum can help to discriminate our model against other models leading to such a phase transition. In particular, a separate evaluation of the contributions coming from the colliding bubble envelopes (see \emph{e.g.}~\cite{Kosowsky:1992vn,Huber:2008hg}), the sound waves (see \emph{e.g.}~\cite{Hindmarsh:2015qta}) as well as the turbulence (see \emph{e.g.}~\cite{Caprini:2009yp}) is needed to get a more refined prediction for the spectrum. But most importantly, one has to confront a GW signal with data from colliders and EDM experiments.

\section{Conclusions}
\label{sec:conc}

Composite Higgs models with partial compositeness feature a tight interplay between electroweak symmetry breaking and
flavour physics as the same interactions which generate the Higgs potential are also responsible for the Yukawa couplings. These interactions are mixings between the elementary and composite sectors whose sizes depend on the confinement scale which in turn is set by the dilaton \emph{vev} $\chi$. Therefore, a comprehensive investigation of the  electroweak phase transition requires to study the dynamical interplay between the Higgs and the dilaton. 
Extending the analysis in~\cite{Bruggisser:2018mus}, we have studied this interplay and have shown that the running mixings can play a key role in the nature of the phase transition. Not only do they control the Higgs potential, but they also represent an additional source of conformal symmetry breaking and thus affect the dilaton potential as well. In addition, the running mixings can provide a new source of CP violation, allowing for successful electroweak baryogenesis.

In~\cite{Bruggisser:2018mus}, we have focussed on a particularly simple case with a sizeably varying top quark mixing. In the present work, we have found that other options are available too. We have systematically determined the conditions under which a large CP-violating source is active during the phase transition due to varying mixings.
In particular, another simple option is CP violation from the interplay of varying top and charm quark mixings.
We have studied the possible tunnelling trajectories in the Higgs-dilaton field space for this case as a function of the model parameters and have identified the features of the scalar potential that lead to a strong first-order electroweak phase transition. 
We have then calculated the produced baryon asymmetry and found that the observed baryon asymmetry can be obtained for natural choices of the parameters.
Our work thus strengthens the conclusion that PNGB composite Higgs models provide a natural realization of the flavoured electroweak baryogenesis mechanism of Ref.~\cite{Bruggisser:2017lhc}.

Our description is based on a simplified four-dimensional effective field theory in the confined phase, reflecting  symmetries such as spontaneous $G\to H$ breaking and spontaneous breaking of the scale invariance, together with the sources of explicit breaking of these symmetries. We have determined the joint potential for the Higgs and the dilaton which is determined by these symmetries and their explicit breakings as well as dimensional analysis and the large-$N$ expansion.    
We have then performed a numerical study of the potential and a scan over several of the most relevant parameters.
This has allowed us to identify interesting regions in the parameter space where the phase transitions of the electroweak and the strong sector happen simultaneously and where electroweak baryogenesis can be successful.

This work together with~\cite{Bruggisser:2018mus} shows that the resulting properties of the phase transition significantly depend on the properties of the dilaton, namely whether it obeys the $N$-scaling of a meson or a glueball. Answering the question of which of these two options is preferred by realistic UV completions is beyond the applicability of our approach, but our results add motivation to searching for the answer in calculable models of the strong dynamics.

In summary, the generic key feature responsible for a supercooled electroweak phase transition in composite Higgs models is the thermal barrier generated by the large number of degrees of freedom coupled to the dilaton, as captured  in Eqs.~(\ref{eq:fcft})
and (\ref{eq:fixf}). More precisely, what makes the EW phase transition strongly first-order is the following combination:

\begin{enumerate} 
\item A nearly conformal zero-temperature dilaton potential.
\item A large number of degrees of freedom in the strongly coupled sector which acquire a mass during the confinement phase transition.
\item A substantial part of the potential where $\chi_0>\chi>T$ (as required by the consistency of our approach).
\end{enumerate}

In the end, in this setup, the strength of the EW phase transition (i.e.~the value of the tunnelling action) is not too much affected by the varying Yukawas in the way discussed in \cite{Baldes:2016rqn}, especially for large $m_{\chi}$ and small Yukawas. The corresponding effects are subdominant compared to the ones listed above. However, the direction of the tunneling crucially depends on the varying Yukawas. So the nature of the EW phase transition is still very much affected by the varying Yukawas, when taking the tunnelling direction as being part of the nature of the phase transition. In the case of small $m_{\chi}$, the Yukawa-dependent parts of the potential actually have some influence on the tunnelling action. The cross terms ($c_{\chi y}$) also modify the potential.

The range of dilaton masses which is favored for obtaining a first-order electroweak phase transition while being compatible with constraints on dilaton-Higgs mixing  begins around one hundred GeV,  well within LHC reach. This offers an interesting opportunity to explore questions relevant to the cosmological history in collider experiments. 
The upper bound on the dilaton mass from the reheating/washout constraint is much stronger for the glueball than for the meson ($\sim \, 350 \GeV$ vs.~$\sim \, 650 \GeV$, respectively)  as the depth of the true minimum is controlled parametrically by
$m_{\chi}^2 f^2  [\times  N$ for glueball].
So for the same dilaton mass, the glueball will have more latent heat and reheat to a higher temperature, unless $N$ is 1.
Besides the prediction of the dilaton mass, our analysis suggests that precise measurements of its couplings to SM states may give constraints on the trajectory of the phase transition in the Higgs-dilaton plane, which in turn crucially affects the generated baryon asymmetry. The same concerns the couplings of the Higgs to SM states, which can also be linked to the phase transition properties.     

Data from flavour physics experiments can also be important for assessing the viability of electroweak baryogenesis in composite Higgs models. As we have shown, the amount of the induced CP asymmetry crucially depends on the possible (approximate) flavour symmetries of the theory. For instance, in models with large flavour symmetries, the simplest way of inducing a CP asymmetry is to have top quark mixings with a varying complex phase. The immediate consequence would be a quick detuning of the Higgs potential by the top quark mixings, once the dilaton changes from its \emph{vev} today. This detuning would then be reflected in the dependence of the dilaton couplings to SM states, through the quantity $\partial_\chi v$.       
 
The quick detuning of the Higgs potential discussed above, as well as a large relative size of the mixing-induced corrections to the scalar potential would also lead to constraints on the global symmetry structure of the elementary-composite mixings. Indeed these two features, once they are pronounced enough, can attract the tunnelling trajectory to the $\theta=\pi/2$ direction, which is only compatible with electroweak baryogenesis for certain non-minimal embeddings of the elementary fermions into the global symmetry $G$. This, in particular, can directly affect the collider signatures of the strong sector resonances. 

Finally, the strong first-order phase transition that we find generates a large stochastic  gravitational wave signal that would be detectable by LISA in most of the relevant parameter space. This provides another future independent way of testing our scenario. The very large amount of supercooling and therefore the very large predicted gravitational wave signal is particularly unique in the glueball case.

The results which we have obtained from the study within this simple framework give powerful non-trivial constraints on the possible underlying UV completions. The fact that we have identified viable regions of the parameter space motivates further studies in this direction using more complete models, such as deconstructed and 5D models of the composite Higgs and lattice simulations. 
It will also be interesting to study cold baryogenesis in this framework, instead of standard electroweak baryogenesis through charge transport.

\section*{Acknowledgments}
BvH thanks Fermilab for hospitality while part of this work
was completed. This visit has received funding/support from the European Union's Horizon 2020 research and innovation programme under the Marie Sk\l{}odowska-Curie grant agreement No 690575. BvH also thanks the
Fine Theoretical Physics Institute at the University of Minnesota for hospitality and partial support. OM thanks the Mainz Institute for Theoretical Physics (MITP) for its hospitality and support during the completion of this work.

\appendix
\section{Vacuum tunnelling}
\label{ap:vacuumtunnelling}
In order to calculate the probability of vacuum tunnelling from an unstable ground state in a potential $V[\phi]$ into another ground state, one needs to minimize the Euclidean action 
\be
S_E \, = \, \int d^4x \left[\frac{1}{2}(\partial_\mu\phi)^2\, +\, V[\phi]\right] 
\ee
for this transition. The tunnelling rate is then given by \cite{Coleman:1977py,Callan:1977pt}
\begin{equation}
\Gamma \, = \, A e^{-S_E} \, , 
\end{equation}
where $A$ is a factor that depends on the action as well as on determinant factors that are extremely difficult to compute. The exponential suppression is very sensitive to the value of the action and hence it is sufficient to estimate $A$ based on dimensional grounds. 
The phase transition completes once the tunnelling probability per Hubble time and Hubble volume becomes of order one. For a phase transition near the electroweak scale, this translates to the criterion on the Euclidean action (for a pedagogical review see \emph{e.g.}~\cite{Quiros:1999jp})
\begin{equation}
S_E \, \sim \, 140\, .
\end{equation}

Let us first consider a potential that depends only on one field. 
At zero temperature the field configuration that minimizes the action features an $O(4)$ symmetry. The formula for the action then greatly simplifies,
\begin{equation}
\label{eq:O4action}
S_4 \, = \, 2\pi^2 \int_0^\infty r^3 dr\left[\frac{1}{2}\left(\frac{d\phi}{dr}\right)^2+V[\phi]\right] , 
\end{equation}
where $r^2\equiv t_E^2+\vec{x}^2$. Shifting $\phi$ such that the false minimum is at $\phi=0$ and the potential such that $V[\phi=0]=0$,
the field configuration that minimizes this action follows from solving the equation of motion
\begin{equation}
\label{eq:O4equation}
\frac{d^2\phi}{d r^2} \, + \, \frac{3}{r}\frac{d\phi}{dr} \, = \, V'[\phi]
\end{equation}
subject to the boundary conditions
\be 
\label{eq:O4boundaryconditions}
\lim_{r\rightarrow \infty}\phi[r] \, = \, 0 \quad \text{and} \quad \left.\frac{d\phi}{dr}\right|_{r=0} \, = \, 0\, .
\ee
The equation of motion is equivalent to that of a classical particle moving in a potential $-V[\phi]$ with an $r$-dependent friction term $\frac{3}{r}\frac{d\phi}{dr}$, where $r$ plays the role of time. The boundary conditions then correspond to the particle being released with zero initial velocity and approaching $\phi=0$ in the limit $r\rightarrow \infty$.

This problem can be solved with the overshoot/undershoot method. To this end, one guesses an initial release point and then lets the particle evolve. If the release point was chosen too close to the true vacuum, the particle will stay close to this point for a long time until the $r$-suppressed friction term is completely negligible. The particle will subsequently start rolling towards the false minimum. Due to the negligible friction at this stage, it will however have too much energy and overshoot the hill in the potential $-V[\phi]$ at $\phi=0$ and end up rolling towards $\phi\rightarrow  -\infty$. On the other hand, if the particle is released too close to the false vacuum it will not have enough energy to reach the hill at $\phi=0$ and hence it will fall back and start oscillating around the minimum of the potential $-V[\phi]$. This situation is called undershoot. The numerical method to find the solution to Eqs.~\eqref{eq:O4equation} and \eqref{eq:O4boundaryconditions} then proceeds as follows:
Starting with a guess $\phi[r=0]=\phi_1$, one moves the initial value $\phi[r=0]$ to lower values $\phi_2<\phi_1$ if the solution overshoots and to larger values $\phi_2>\phi_1$ if it undershoots. This procedure is repeated until the right $\phi[r=0]$ is found such that the particle comes to rest at $\phi[r]=0$ as $r\rightarrow \infty$. Using this solution, one then computes the Euclidean action \eqref{eq:O4action}. The analogy with the classical particle shows also that a bubble has a typical size $R$. This can be seen by realising that the particle will come very close to the hill at $\phi=0$ after a finite time $r=R$ and then only very slowly move exactly to $\phi=0$.

The case of finite temperature $T\neq0$ is formally equivalent to a periodicity of $T^{-1}$ in imaginary time $t_E$. The field is then subject to the additional constraint
\begin{equation}
\phi[t_E,\mathbf{x}] \, = \, \phi[t_E+1/T,\mathbf{x}]\, .
\end{equation}
If $T$ is large, the solution that minimizes the action has an $O(3)$ symmetry. The Euclidean action in this case simplifies to
\begin{eqnarray}
S_E & \, = \, &S_3/T\\
S_3 & \, = \, &4\pi\int_0^\infty s^2 ds \left[\frac{1}{2}(d\phi/ds)^2+V[\phi]\right] \, ,
\end{eqnarray}
where $V[\phi]$ is the finite-temperature potential and $s^2 \equiv \vec{x}^2$. The field configuration that minimizes this action is found by solving the equation of motion 
\begin{equation}
\label{eq:O3equation}
\frac{d^2\phi}{d s^2} \, + \,\frac{2}{s}\frac{d\phi}{ds} \, = \, V'[\phi] 
\end{equation}
subject to the boundary conditions
\be 
\label{eq:O3boundaryconditions}
\lim_{s\rightarrow \infty}\phi[s] \, = \, 0 \quad \text{and} \quad \left.\frac{d\phi}{ds}\right|_{s=0}  = \, 0 \, . 
\ee
The equation of motion is again that of a classical particle moving in a potential $-V$ with an $s$-dependent friction term $\frac{2}{s}\frac{d\phi}{ds}$, where $s$ plays the role of time. Hence we can use the same method to find the solution as for the $O(4)$-symmetric case. 

In order to determine the nucleation temperature and evaluate whether the bubbles nucleate in the $O(4)$- or the $O(3)$-symmetric solution, we have to compute the Euclidean actions $S_4$ and $S_3/T$ for different temperatures. The nucleation temperature is the highest temperature for which $S_E\sim 140$. If the nucleation temperature is very low compared to the typical bubble size, $R\ll T^{-1}$, and the $O(4)$-symmetric solution has the lowest action, the bubbles nucleate in the $O(4)$-symmetric solution. For the potentials studied in this work, however, the bubbles usually nucleate in the $O(3)$-symmetric solution.

We next discuss what happens in a potential that depends on more than one scalar field. For simplicity, let us focus on $O(3)$-symmetric bubbles. For multiple fields, the equation of motion \eqref{eq:O3equation} becomes
\begin{equation}\label{eq:O3equationMultifield}
\frac{d^2\vec{\phi}}{d s^2} \, + \, \frac{2}{s}\frac{d\vec{\phi}}{ds} \, =\, \nabla V[\vec{\phi}]\, ,
\end{equation}
while the boundary boundary conditions \eqref{eq:O3boundaryconditions} now read
\begin{equation}\label{eq:O3boundaryconditionsMultifield}
\lim_{s\rightarrow \infty}\vec{\phi}[s] \, = \, \vec{0} \quad \text{and} \quad \left.\frac{d\vec{\phi}}{ds}\right|_{s=0}  = \, \vec{0} \, . 
\end{equation}

The overshoot/undershoot method does no longer work in higher-dimensional field space as the path in this case does not have to pass by $\vec{\phi}=\vec{0}$ but can also avoid this point by going around it in field space. In order to numerically find the path in field space that minimizes the action, a modified procedure is required. To this end, 
one starts with an initial guess $\vec{\phi}_g[x]$ for this path, where $x$ is a parameter that measures the distance along the path. If we normalize the path according to $\left|d\vec{\phi}_g/dx\right|=1$, the equation of motion \eqref{eq:O3equationMultifield} can be nicely separated into parts parallel and perpendicular to the path \cite{Beniwal:2017eik,Wainwright:2011kj}:
\begin{eqnarray}\label{eq:eomAlongPath}
	\frac{d^2x}{ds^2} \, + \, \frac{2}{s}\frac{dx}{ds} \, = \, &\partial_x V[\vec{\phi}_g[x]] \\
	\label{eq:eomPerpPath}
	\frac{d^2\vec{\phi}_g}{dx^2}\left(\frac{dx}{ds}\right)^2 \, = \, &\nabla_\perp V[\vec{\phi}_g]\, .
\end{eqnarray}
The first equation is just the usual bounce equation for a one dimensional potential which can be solved using the overshoot/undershoot method. The second equation, on the other hand, can be understood
as defining a normal force 
\begin{equation}\label{eq:normalForceAlongPath}
\vec{N} \, \equiv \, \frac{d^2\vec{\phi}_g}{dx^2}\left(\frac{dx}{ds}\right)^2 - \, \nabla_\perp V[\vec{\phi}_g]
\end{equation}
that acts perpendicularly on the path until the second equation is fulfilled. The procedure to find the path that minimizes the action is then the following:
\begin{enumerate}
\item Guess an initial path. Usually we will take this initial path to be a straight line defined by an angle $\alpha$ (we try to guess the angle that minimizes the action, see Sec.~\ref{sec:num}).
\item Calculate the bounce for $x[s]$ from the equation of motion  \eqref{eq:eomAlongPath}.\label{it:bounce}
\item Determine the normal force \eqref{eq:normalForceAlongPath} and deform the path in the direction of the force.
\item Go to step \ref{it:bounce} until the path does not get significantly deformed any more.
\end{enumerate}
Note that for certain potentials, with multiple valleys, this algorithm might not automatically converge to the right solution. In those cases there is still some manual adjusting needed.

In practice the force is only calculated at a finite number $n$ of points along the path and then those points are displaced in the direction of the force. In our case, with only two fields $\vec{\phi}[r]=\left(\phi^1[x], \phi^2[x]\right)$, the algorithm for the deformation of the path then is:
\begin{enumerate}
	\item Define points $\phi^{1,2}_i=\phi^{1,2}[x_i]$ along the path, where $x_i=x_{max}/n*i$ and $x_{\rm max}$ is the value of $x$ for which $\vec{\phi}[x_{\rm max}]=\vec{0}$.
	\item Displace the points, $\hat{\phi}^{1,2}_i = \phi^{1,2}_i+\rho\vec{N}[x_i]$, where $\rho$ is a small step size (typically we use $\rho=0.02$).
	\item Fit a function to the displaced points, $\phi^{1,2}[x]=\operatorname{Fit}[\{x_i,\hat{\phi}^{1,2}_i\},P_l]$ with $P_l$ being a polynomial of order $l$.
\end{enumerate}
We have found that $l=6$ is sufficient and adding higher orders does not significantly improve the result.

%\begin{thebibliography}{99}

\bibliographystyle{JHEP}  %was plainnat apsrev-new
\bibliography{biblio}

%\end{thebibliography}

\end{document}